\documentclass[twoside,british,english]{article}
\usepackage[T1]{fontenc}
\usepackage[latin9]{inputenc}
\usepackage[a4paper]{geometry}
\usepackage{babel}
\usepackage{verbatim}
\usepackage{refstyle}
\usepackage{float}
\usepackage{booktabs}
\usepackage{amsmath}
\usepackage{amsthm}
\usepackage{amssymb}
\usepackage{stmaryrd}
\usepackage{graphicx}
\usepackage{esint}
\usepackage{bbding}
\usepackage{placeins}

\usepackage[unicode=true,
 bookmarks=true,bookmarksnumbered=false,bookmarksopen=false,
 breaklinks=false,pdfborder={0 0 1},backref=false,colorlinks=false]
 {hyperref}
\hypersetup{pdftitle={Calibration to Vanilla and No-Touch options},
 pdfauthor={Matthieu Mariapragassam, Christoph Reisinger},
 pdfsubject={Calibration}}

\makeatletter

\AtBeginDocument{\providecommand\algref[1]{\ref{alg:#1}}}
\AtBeginDocument{\providecommand\algolineref[1]{\ref{algoline:#1}}}
\providecommand{\tabularnewline}{\\}
\floatstyle{ruled}
\newfloat{algorithm}{tbp}{loa}
\providecommand{\algorithmname}{Algorithm}
\floatname{algorithm}{\protect\algorithmname}
\RS@ifundefined{subref}
  {\def\RSsubtxt{section~}\newref{sub}{name = \RSsubtxt}}
  {}
\RS@ifundefined{thmref}
  {\def\RSthmtxt{theorem~}\newref{thm}{name = \RSthmtxt}}
  {}
\RS@ifundefined{lemref}
  {\def\RSlemtxt{lemma~}\newref{lem}{name = \RSlemtxt}}
  {}

\numberwithin{equation}{section}
\numberwithin{figure}{section}
\numberwithin{table}{section}

  \theoremstyle{remark}
  \newtheorem*{rem*}{\protect\remarkname}
\theoremstyle{plain}
  \theoremstyle{remark}
  
\theoremstyle{plain}

  \theoremstyle{plain}
  
  \theoremstyle{plain}
  
 \usepackage{algolyx}

\makeatother

\usepackage{eurofont}
\usepackage{graphicx}
\usepackage{epstopdf}
\usepackage{hyperref}
\usepackage{babel}
\usepackage[T1]{fontenc}
\usepackage{ae,aecompl,aeguill}
\usepackage{newfloat}
\usepackage{caption}
\usepackage{pdflscape}
\usepackage{mathtools}
\usepackage{tikz}
\usepackage{microtype}

\usepackage[justification=justified,singlelinecheck=false,font=small,labelfont=bf]{caption}
\usepackage{subcaption}
\usepackage{booktabs} 
\usepackage{tabularx}
\newcolumntype{Y}{>{\centering\arraybackslash}X} 
\allowdisplaybreaks[4]
\usepackage{dsfont}
\usepackage{bold-extra}
\usepackage{cases}
\usepackage{enumerate}
\usepackage{verbatim}
\usepackage{xcolor}
\usepackage{mathtools}

\usepackage{amssymb}
\usepackage{stmaryrd}
\usepackage{graphicx}
\usepackage{esint}

\allowdisplaybreaks[4]
\usepackage{hyperref}
\hypersetup{pdfpagemode=UseNone} 
\hypersetup{pdfstartview={XYZ null null 0.85}} 

\newcommand{\db}{{\, \rm d}b}

\newcommand{\rf}{r^{\rm f}}
\newcommand{\rd}{r^{\rm d}}
\newcommand{\rdt}{r^{\rm d}\!\!\!\!\;\;(t)}
\newcommand{\rft}{r^{\rm f}\!\!\!\!\;\;(t)}
\newcommand{\rdT}{r^{\rm d}\!\!\!\!\;\;(T)}
\newcommand{\rfT}{r^{\rm f}\!\!\!\!\;\;(T)}
\newcommand{\rdTm}{r^{\rm d}\!\!\!\!\;\;(T_m)}
\newcommand{\rfTm}{r^{\rm f}\!\!\!\!\;\;(T_m)}
\newcommand{\Ddt}{D^{\rm d}\!\!\!\!\;\;(t)}
\newcommand{\DdT}{D^{\rm d}\!\!\!\!\;\;(T)}
\newcommand{\DfT}{D^{\rm d}\!\!\!\!\;\;(T)}
\newcommand{\Dft}{D^{\rm f}\!\!\!\!\;\;(t)}
\newcommand{\Qd}{\mathbb{Q}^{\rm d}}
\newcommand{\EQd}{\mathbb{E}^{\mathbb{Q}^{\rm d}}}

\newcommand{\Bf}{B_{\rm first}}
\newcommand{\Bl}{B_{\rm last}}

\newcommand{\X}{{\bf X}}

\newcommand{\vol}{Y}

\newcommand{\dom}{{\mathcal{D}}}

\newcommand{\loc}{\sigma_{\mathrm{LV}}}
\newcommand{\lev}{\sigma}
\newcommand{\xiH}{\xi_{\mathrm{H}}}
\newcommand{\locM}{\sigma_{\mathrm{LMV}}}
\newcommand{\levM}{\sigma}
\newcommand{\sig}{\sigma}
\newcommand{\rhos}{\hat{\rho}}

\let\originalleft\left
\let\originalright\right
\renewcommand{\left}{\mathopen{}\mathclose\bgroup\originalleft}
\renewcommand{\right}{\aftergroup\egroup\originalright}

\numberwithin{equation}{section}
\numberwithin{figure}{section}
 \theoremstyle{definition}
 \theoremstyle{assumption}
 \newtheorem*{assn*}{Assumption}
 \theoremstyle{assumption}
 
  \theoremstyle{plain}
  \newtheorem*{fact*}{Fact}
  \theoremstyle{plain}

\usetikzlibrary{calc}
\tikzstyle{decision} = [diamond, draw, fill=blue!20, 
    text width=4.5em, text badly centered, node distance=3cm, inner sep=0pt]
\tikzstyle{block} = [rectangle, draw, fill=blue!20, 
    text width=10em, text centered, rounded corners, minimum height=4em, text centered,font=\bf]
\tikzstyle{output} = [rectangle, draw, fill=green!20, 
    text width=10em, text centered, rounded corners, minimum height=4em, text centered,font=\bf]
\tikzstyle{line} = [draw, -latex']
\tikzstyle{cloud} = [draw, ellipse,fill=black!10,node distance=5cm,
    minimum height=3em,text centered,font=\bf]

\author{
\scshape{alan bain}\thanks{\footnotesize\scshape{BNP Paribas, 10 Harewood Avenue, London, NW1 6AA, United Kingdom}} \and
\scshape{matthieu mariapragassam}\thanks{\footnotesize{\scshape{Mathematical Institute and Oxford-Man Institute of Quantitative Finance, University of Oxford,
Woodstock Road, Oxford, OX2 6GG, United Kingdom}}, 
matthieu.mariapragassam@gmail.com, christoph.reisinger@maths.ox.ac.uk \newline
\footnotesize{The second author gratefully acknowledges financial support from the \textsc{Oxford-Man
Institute} and \textsc{BNP Paribas}. } } \,\Envelope \and \scshape{christoph reisinger\footnotemark[2]}}

\date{}
\allowdisplaybreaks[0]
\usepackage{algorithmic}
\makeatother

  \addto\captionsbritish{\renewcommand{\corollaryname}{Corollary}}
  \addto\captionsbritish{\renewcommand{\lemmaname}{Lemma}}
  \addto\captionsbritish{\renewcommand{\remarkname}{Remark}}
  \addto\captionsbritish{\renewcommand{\theoremname}{Theorem}}
  \addto\captionsenglish{\renewcommand{\corollaryname}{Corollary}}
  \addto\captionsenglish{\renewcommand{\lemmaname}{Lemma}}
  \addto\captionsenglish{\renewcommand{\remarkname}{Remark}}
  \addto\captionsenglish{\renewcommand{\theoremname}{Theorem}}
  \providecommand{\corollaryname}{Corollary}
  \providecommand{\lemmaname}{Lemma}
  \providecommand{\remarkname}{Remark}
\addto\captionsbritish{\renewcommand{\algorithmname}{Algorithm}}
\providecommand{\theoremname}{Theorem}

\begin{document}


\title{Calibration of Local-Stochastic and Path-Dependent Volatility Models 
to Vanilla and No-Touch Options}

\maketitle%
\begin{abstract}
We propose a generic calibration framework to both vanilla and no-touch
options for a large class of continuous semi-martingale models. The
method builds upon the forward partial integro-differential equation
(PIDE) derived in 
\emph{
Hambly et al.\ (2016),
  {A forward equation for barrier options under the Brunick \& Shreve Markovian projection},
{Quant.\ Finance},
16 (6), {827--838}},
which allows fast computation of up-and-out
call prices 
for the complete set of strikes, barriers
and maturities.
It also utilises a novel
two-state 
particle method to estimate the Markovian projection of the variance
onto the spot and running maximum.
We 
detail a step-by-step
procedure for a 
Heston-type local-stochastic volatility model with local vol-of-vol,
as well as two path-dependent volatility models where the local volatility component
depends on the running maximum.
In numerical tests we benchmark these new models 
against standard models
for a set of EURUSD market data,
all three models are seen to
calibrate
well within 
the market no-touch bid--ask.
\end{abstract}

\section{Introduction}


For derivative pricing models to be useful in practice, they need to allow calibration to the market prices of
liquid contracts, as well as exhibit a dynamic behaviour consistent with that of the underlying
and with future options quotes.
Vanilla options prices provide a snapshot of the market implied distributions of the underlying which is the key ingredient for pricing European options, but they provide limited information about the joint law of the underlying observed at different times, which is needed for pricing path dependent options.
There is evidence (see, e.g., \cite{Ayache2004}) that the market prices of contracts with barrier features contain additional information 
on the dynamic behaviour of the volatility surface not already seen in vanilla quotes.
The topic of this paper is hence the simultaneous calibration of volatility models to European call and no-touch
(or, more generally, barrier) options.

The case of calibration to vanilla options, i.e.\ European calls and puts,
has been considered extensively in the literature. The seminal work
of Dupire \cite{Dupire} gives a
constructive solution to the calibration problem for local volatility (LV) models, which can perfectly match
call prices for any strike and maturity.
Nowadays,
local-stochastic volatility (LSV) models are in widespread use in financial institutions
because of their ability to calibrate exactly to vanilla options due to the local volatility component while
embedding a stochastic variance component, which improves the dynamic properties.
The calibration problem
of LSV models to vanilla quotes is reviewed already in \cite{piterbarg2006markovian}, and 
is addressed, more recently, in the works of Guyon
and Henry-Labord{\`e}re \cite{guyon2012being, GuyonLabordere2013} by a particle method, and in
\cite{Ren&Madan2007} by solution of a nonlinear Fokker-Planck PDE; see also
\cite[Section 6.8]{Clark2010}.

In addition to call options,
practitioners are increasingly interested in including the quotes
of touch options in the set of calibration instruments, which will improve the
pricing and risk-management of exotic contracts with barrier features. 
In some markets, for example in foreign exchange, the next most visible layer of option prices after the European vanilla prices are the American barrier options (for example products such as one-touch, double no-touch and vanilla knock-out options). Observation and model parameter adjustment derived from these prices is a well-established part of model calibration for short-dated FX options.  

A few published
works already address this question for different model classes: Crosby
and Carr \cite{Carr&Crosby2008} consider a particular class of jump models
which gives a calibration to both vanillas and barriers; Pironneau
\cite{Pironneau} proves that an adaptation of the Dupire equation is valid for
a given barrier level, under the local volatility model.

This paper addresses the simultaneous calibration to vanilla and barrier (specifically, no-touch)
options systematically for a wide class of volatility models.
The focus is less on the calibration of a particular model -- although we do calibrate three different new models
-- but on a methodology which allows the efficient calibration of \emph{any} volatility model.

We assume that interest rates are deterministic. 
We take the Brunick--Shreve mimicking
point of view from \cite{BrunickShreve2013} that the joint law of a stock price and its
running maximum, or equivalently, barrier prices for all strikes, barriers levels and maturities,
can be reproduced by a one-factor model with a deterministic volatility function of the spot, the running maximum and time.
This is a natural extension of Gy{\"o}ngy's result in \cite{Gyongy1986}
that the  stock price distribution, or equivalently,
call prices for all strikes and maturities, can be reproduced
with a deterministic volatility function of the spot and time.
In the latter case, this volatility function is the expectation of the variance process conditional on the
spot price, while in the path-dependent case the expectation is also conditional on the path-dependent quantity,
here the running maximum. 
These conditional expectations are often referred to as Markovian projections.

In the vanilla case, exact calibration is guaranteed if the Markovian projection of the instantaneous variance
onto the spot coincides
with the squared local volatility function derived from vanilla quotes by Dupire's formula.
Conversely, given the local volatility, model prices can be computed by the forward Dupire PDE,
formulated in strike and maturity.
The estimation of the local volatility from observed prices is an ill-posed inverse problem,
and regularisation approaches have been proposed, e.g., in 
\cite{jsh98}, \cite{Egger2005}, or \cite{Crepey2010}.
If the underlying model to be calibrated is not itself a local volatility model, the Dupire PDE
can still be used to compute the model prices by utilising the mimicking result, i.e.,
the Dupire PDE with the Markovian projection of the instantaneous variance onto the spot
as diffusion coefficient gives the correct model prices.
The conditional expectation of the stochastic variance under the desired model can be estimated, 
e.g., by the particle method in \cite{GuyonLabordere2013,guyon2012being}.


The natural extension of the forward Dupire PDE for calls to a forward equation for barrier option prices,
with strike, barrier level and maturity as independent variables, is the
forward PIDE derived in \cite{Hambly2016}.
It has as diffusion coefficient a volatility function of spot, running maximum, and time, which we view as a `code book' 
for barrier option prices, a name coined in \cite{carmona09} for local volatility and European options. 
We re-iterate that the underlying diffusion can be
a general continuous stochastic process. Specifically, the variance
process does not need to contain the running maximum in its parametrisation. The link between the original model and this path-dependent volatility is
given by Corollary 3.10 in \cite{BrunickShreve2013} (see also (\ref{proj}) below).

We investigate as example
a Heston-type LSV model with a local volatility component as well as a stochastic volatility
with simple parametric, spot- and time-dependent vol-of-vol (LSV-LVV).
Hence, we can perform a best-fit of the vol-of-vol function to no-touch options at each quoted
maturity while ensuring perfect calibration to vanilla options through
the local volatility function.
The tests show that the calibrated LSV-LVV model prices no-touch options well within the market
bid--ask spread for all barrier levels and maturities.
The approach can
easily be generalised to other types of stochastic volatility diffusions.

We also construct a 
``local maximum volatility'', i.e.\ a spot and running maximum dependent volatility function (LMV) consistent with market prices of
calls and no-touches by solving an inverse problem for the PIDE discussed above using regularisation.
The calibration of this maximum-dependent local volatility function using the forward PIDE
is inspired by, and extends,
the literature on the local volatility model calibration, see e.g.\ \cite{jsh98,Egger2005,Crepey2010}.
We then
consider an extension of the model in the spirit of LSV models,
i.e., a local maximum-dependent volatility function (LMSV) multiplied by
a stochastic volatility.
These two models fall into the class of
path-dependent volatility models and can be useful to replicate a market's spot-volatility
dynamics as explained in \cite{Guyon2014}.

In the calibration of the LSV-LVV and LMSV models, one can compute
the Markovian projection of the stochastic variance by
either extending the particle method introduced in
\cite{GuyonLabordere2013,guyon2012being} or by solving the Kolmogorov forward PDE for the joint density of 
$\left(S_{t},M_{t},V_{t}\right)$, the spot, maximum and volatility, numerically. In our approach, we rely on a two-dimensional
particle method (in $(S_t,M_t)$) as it offers 
a straightforward extension to additional stochastic factors.
The computationally most expensive part is, as often, retrieval of the neighbouring particles,
for which we propose a binary tree search, specifically on a $k$-d tree.
The use of $k$-d trees for particle method calibration is a novel approach which can easily be generalised to higher-dimensional state spaces.


The remainder of this paper is organised as follows. In Section \ref{sec:Model-definition},
we define the models and calibration condition for up-and-out barrier option quotes.
Section \ref{sec:NumericalPIDE} presents an efficient numerical
solution of the forward PIDE for barrier options, which is central for
the algorithms in this paper.
In Section \ref{sec:Calibration Brunick--Shreve}, we present a
possible calibration algorithm for the path-dependent volatility (LMV) model by
forward PIDE and regularised gradient-based optimisation. 
Then, in Section \ref{sec:LMSV calibration} specifically,
we use again a particle method to calibrate the LMSV model by Markovian projection.
Section \ref{sec:LSV local Xi calibration} makes use
of a Markovian projection with a two-dimensional conditional state of spot and running maximum
for the LSV-LVV model and combines it with the forward PIDE
for barrier options in order to best-fit no-touch quotes while perfectly
calibrating vanilla market prices. 
The calibration results for all
these models are presented and compared in Section \ref{sec:Calibration-results}.
Section \ref{sec:Conclusion} concludes with a brief discussion.


\section{Models and calibration conditions \label{sec:Model-definition}}
 \label{sec:Necessary-and-sufficient}


We consider a 
spot exchange rate $S_{t}$ associated with the currency pair $\text{FORDOM}$, which is the amount
of units of domestic currency $\text{DOM}$ needed to buy one unit
of foreign currency $\text{FOR}$ at time $t$. 
We assume the existence of a filtered probability space ({$\Omega$},
$\mathcal{F},\left\{ \mathcal{F}_{t}\right\} _{t\geq0},\Qd)$
with domestic risk-neutral measure $\Qd$, under which
$S$ follows the SDE
\begin{equation}
\cfrac{dS_{t}}{S_{t}}=\left(\rdt-\rft\right)\,dt+\vol_{t}\,dW_{t}\,,
\label{eq:Model Definition}
\end{equation}
where $W$ is a one-dimensional $\mathcal{F}_{t}$-adapted standard Brownian motion,
$\vol$ is a continuous and positive $\mathcal{F}_{t}$-adapted
semi-martingale, where
\begin{equation}
\EQd\left[\int_{0}^{t}Y_u^{2}S_u^{2}\, du\right] <\infty \,, \label{eq:bounded-second-moment}
\end{equation}
and the domestic and foreign short rates, $\rd$ and $\rf$, are deterministic functions of time,
such that 
\begin{eqnarray*}
\rdt = -\frac{\partial\ln P^{\rm d}\left(0,t\right)}{\partial t}, \qquad
\rft=-\frac{\partial\ln P^{\rm f}\left(0,t\right)}{\partial t},
\end{eqnarray*}
with $P^{\rm d/f}\!\left(0,T\right)$ the market zero-coupon bond
prices for the domestic and the foreign money market accounts, respectively (see Chapter 9.1 in \cite{Musiela2009}).
The domestic and foreign discount factors are then defined as 
\begin{eqnarray}
\label{discout}
\Ddt =\text{e}^{-\int_{0}^{t}r^{\rm d}\!\left(u\right) du},\quad \Dft=\text{e}^{-\int_{0}^{t}r^{\rm f}\!\left(u\right) du}\,.
\end{eqnarray}

A model widely used in the industry is the
Heston-type LSV model
\begin{equation}
\begin{array}{c}
\begin{cases}
\cfrac{dS_{t}}{S_{t}}=\left(\rdt-\rft \right)\,dt+\lev\left(S_{t},t\right)\sqrt{V_{t}}\,dW_{t}\\
dV_{t}=\kappa\left(\theta-V_{t}\right)\,dt+\beta\xi\sqrt{V_{t}}\,dW_{t}^{V},
\end{cases}\end{array}\label{eq:LSV-mixing=00003D0.55}
\end{equation}
where $W$ and $W^V$ are standard Brownian motions with constant correlation $\rho$. Moreover,
$v_{0}$, the a priori unknown initial value of $V$, and $\kappa,\theta,\xi, \beta$ are non-negative scalar parameters, while
the local volatility component
$\lev \colon \mathbb{R}^{+}\times\left[0,T\right]\rightarrow\mathbb{R}^{+}$, is
assumed to be bounded and locally Lipschitz in $S$. This ensures, alongside (\ref{eq:bounded-second-moment}),
the  necessary conditions to use the forward equation from \cite{Hambly2016} referenced below in (\ref{eq:Volettera-Type-PIDE}).
Here, the parameter $\beta$ is redundant (as only the product $\beta\xi$ appears) and can be set to 1 for the time being;
it will be used in Section \ref{sec:LSV local Xi calibration} to interpolate between the pure local volatility model ($\beta=0$, $\kappa=0$, $v_0=1$)
and the Heston model ($\beta=1$, $\lev= 1$). Similarly, although we will calibrate $\theta$ (alongside $\kappa, \xi, v_0, \rho$) to vanilla options using a pure Heston model, we note that the use of $\sigma$ makes $\theta$ a redundant parameter because of the scaling properties of the model; one could fix $\theta=1$ instead.

Extending (\ref{eq:LSV-mixing=00003D0.55}), we introduce a Heston-type local-stochastic volatility model with local vol-of-vol (LSV-LVV),
\begin{equation}
\begin{array}{c}
\begin{cases}
\cfrac{dS_{t}}{S_{t}}=\left(\rdt-\rft\right)\,dt+\lev\left(S_{t},t\right)\sqrt{V_{t}}\,dW_{t}\\
dV_{t}=\kappa\left(\theta-V_{t}\right)\,dt+\xi\left(S_{t},t\right)\sqrt{V_{t}}\,dW_{t}^{V}, 
\end{cases}\end{array}\label{eq:Heston LSV model}
\end{equation}
with 
$\xi \colon \mathbb{R}^{+}\times\left[0,T\right]\rightarrow\mathbb{R}^{+}$.
The motivation for this model is the freedom gained through the local volatility function $\lev$ and the vol-of-vol function $\xi$
for the calibration to two classes of options.
In particular, while $\lev$ plays a similar role to  that in (\ref{eq:LSV-mixing=00003D0.55}) and enables calibration to
calls with different strikes and maturities, we will use $\xi$ to match (no-)touch option quotes with different barrier levels and maturities.

The choice of $\xi$ as a function of $S$ reflects the fact that $S$ is an observable quantity.  
One could argue for other choices, for instance have the extra function $\xi$ depend on $V$, however,
absolute levels of $V$ lack financial interpretation (they will be adjusted for in the $\sigma(.,)$).
One could also have a spot-vol correlation that depends on $S$, however, local correlation models as in \cite{reghai2010breaking, langnau2010dynamic}
can be fragile as there are tight limits on the range of the correlation (for example, by requirements that the correlation matrix be positive semi-definite).

As calibration instruments, in addition to vanillas, we will consider one-touch options, which
pay $1$ at maturity, in one of the currencies, if the FX rate breaches the up-barrier $B$
during the product lifespan (with continuous monitoring).
We note that on the market, touch options paying either foreign or domestic notional are quoted.
We convert the market quotes
for foreign one-touches denominated in the foreign currency numeraire,
$\mathrm{FOT}$, 
to foreign no-touch options denominated in domestic currency numeraire, $\mathrm{FNT}$, with the following formula, 
\begin{eqnarray}
\mathrm{FNT}\left(B,T\right) 
= \DdT {\mathbb{E}^{\mathbb{Q}^{\rm d}}\left[S_{T}\mathbf{1}_{M_{T}<B}\right]}
  =  {S_{0}} \left(\DfT -\mathrm{FOT}\left(B,T\right) \right). \label{eq:foreign no touch definition} 
\end{eqnarray}
In the following, if no specification of notional currency is given, the price of a no-touch is defined as in (\ref{eq:foreign no touch definition}).


No-touches and vanilla calls are two special cases of barrier calls, and we therefore work under this more general framework.
In the remainder of this section, we give a calibration condition for 
up-and-out call prices 
under model (\ref{eq:Model Definition}). 
The up-and-out call price under model (\ref{eq:Model Definition}) for a notional of one unit of $\text{FOR}$ is 
\begin{eqnarray}
\label{barr}
C\left(K,B,T\right)=\DdT \, \mathbb{E}^{\mathbb{Q}^{\rm d}}\left[\left(S_{T}-K\right)^{+}\mathbf{1}_{M_{T}<B}\right]\,.
\end{eqnarray}

From \cite{BrunickShreve2013}, under the integrability condition (\ref{eq:bounded-second-moment}), any model of the form (\ref{eq:Model Definition}) can be ``mimicked'' by a 
one-factor, path-dependent volatility model.
More precisely, consider
\begin{equation}
\label{eq:bsv}
\begin{cases}
\cfrac{d\widehat{S}_{t}}{\widehat{S}_{t}}=\left(\rdt-\rft\right)\,dt+\sigma_{\mathrm{LMV}}\left(\widehat{S}_{t},\widehat{M}_{t},t\right)\,d\widehat{W}_{t}\\[10pt]
\widehat{M}_{t}=\underset{0\leq u\leq t}{\max} \widehat{S}_{u}\,,
\end{cases} 
\end{equation}
with a standard Brownian motion $\widehat{W}$ defined on a probability space 
({$\widehat{\Omega}$},
$\widehat{\mathcal{F}},\{\widehat{\mathcal{F}}_{t}\} _{t\geq0},\widehat{\mathbb{Q}}^{\rm d})$
and
$\sigma_{\mathrm{LMV}} \colon {\mathbb{R}^{+}}^2\times\left[0,T\right]\rightarrow\mathbb{R}^{+}$ a ``local maximum volatility'' function,
i.e.\ a function of the spot, its running maximum, and time.

Then, the joint law of the pair $(\widehat{S}_T, \widehat{M}_T)$ 
agrees with that of $(S_T, M_T)$ for all $T$ if, for all $T,K,B$,
\begin{eqnarray}
\label{proj}
\sigma^2_{\text{LMV}}\left(K,B,T\right) & = & \EQd \left[Y_{T}^{2}\,|\,S_{T}=K,\,M_{T}=B\right]\,.
\end{eqnarray}
Consequently, the prices of barrier options coincide under both models.


Furthermore, it is shown in \cite{Hambly2016}, that $C$ satisfies a Volterra-type PIDE, 
expressed as
an initial boundary value problem, for any $B\geq S_{0}$, $0\leq K\leq B$
and $T\geq0$, {\small{}
\begin{eqnarray}
\hspace{-0.5 cm} \frac{\partial C\left(K,B,T\right)}{\partial T}+\rfT C\left(K,B,T\right) & = & -\left(\rdT-\rfT\right)K\frac{\partial C\left(K,B,T\right)}{\partial K}\label{eq:Volettera-Type-PIDE}
  +  \frac{1}{2} \sigma^2_{\text{LMV}}\left(K,B,T\right)K^{2}\frac{\partial^{2}C\left(K,B,T\right)}{\partial K^{2}} \\
 & - & \frac{1}{2} \sigma^2_{\text{LMV}}\left(B,B,T\right)B^{2}\left(B-K\right)\frac{\partial^{3}C\left(B,B,T\right)}{\partial K^{2}\partial B}\nonumber \\
 & - & \int_{S_{0}\lor K}^{B}\frac{1}{2}K^{2}\frac{\partial^{2}C\left(K,b,T\right)}{\partial K^{2}}\frac{\partial \sigma^2_{\text{LMV}}\left(K,b,T\right)}{\partial b}\,\db\,,\nonumber 
\end{eqnarray}
where
\begin{eqnarray*}
C\left(K,B,0\right)=\left(S_{0}-K\right)^{+}\mathbf{1}_{S_{0}<B}, & \quad & 0 \le K \vee S_0 \le B \\
C\left(B,B,T\right)=0, & \quad & B\ge S_0, T>0 \\
C\left(K,S_{0},T\right)=0, & \quad & K\le S_0, T>0. 
\end{eqnarray*}
}

The equation is degenerate at $K=0$ due to the factors $K$ and $K^2$ in the first and third line of (\ref{eq:Volettera-Type-PIDE}),
and no boundary condition is needed (the process $\widehat{S}$ in (\ref{eq:bsv}) does not attain the zero boundary).
Moreover, due to the nature of the integral term, the solution $C(\cdot,B,T)$ is fully determined without any asymptotic boundary
condition for large $B$,
hence (\ref{eq:Volettera-Type-PIDE}) is solved up to the largest barrier level 
needed.

%
%

We describe the numerical solution of (\ref{eq:Volettera-Type-PIDE}) by finite differences in Section \ref{sec:NumericalPIDE}
and the estimation of (\ref{proj}) by particle method in Section \ref{subsec:particle}.

\subsection{Market data \label{sub:Market-Data}}

We hereby describe the available data that we use throughout the paper for the different calibration routines.
These are market quotes from $28/03/2013$ for the $\text{EURUSD}$ currency
pair\footnote{The call option prices and no-touch prices were provided by Markit. The zero-coupon rates for both EUR and USD curves
	were retrieved from Bloomberg.}.

We use at-the-money (spot or forward)
volatility, 10 and 25 delta smile-strangles and risk-reversals, i.e.\
for each maturity 5 volatilities on a delta scale (spot-delta convention up to 1Y included, forward-delta convention afterwards),
denoted as
$
10\text{D-Put},25\text{D-Put},50\text{D},\,25\text{D-Call},\,10\text{D-Call},
$
and
the following maturities, relative to 28/03/2013:
$
\text{3M, 6M, 1Y, 2Y, 3Y, 4Y, 5Y}.
$
{The implied volatility is plotted in Figure
	\ref{fig:markit-market-volatility-surface} on a strike scale for different maturities.
	
	\begin{figure}[h]
		\centering
		\begin{minipage}[c]{1\columnwidth}%
			\begin{minipage}[c]{1\columnwidth}
				\selectlanguage{english}
				\begin{center}
					\includegraphics[scale=0.55]{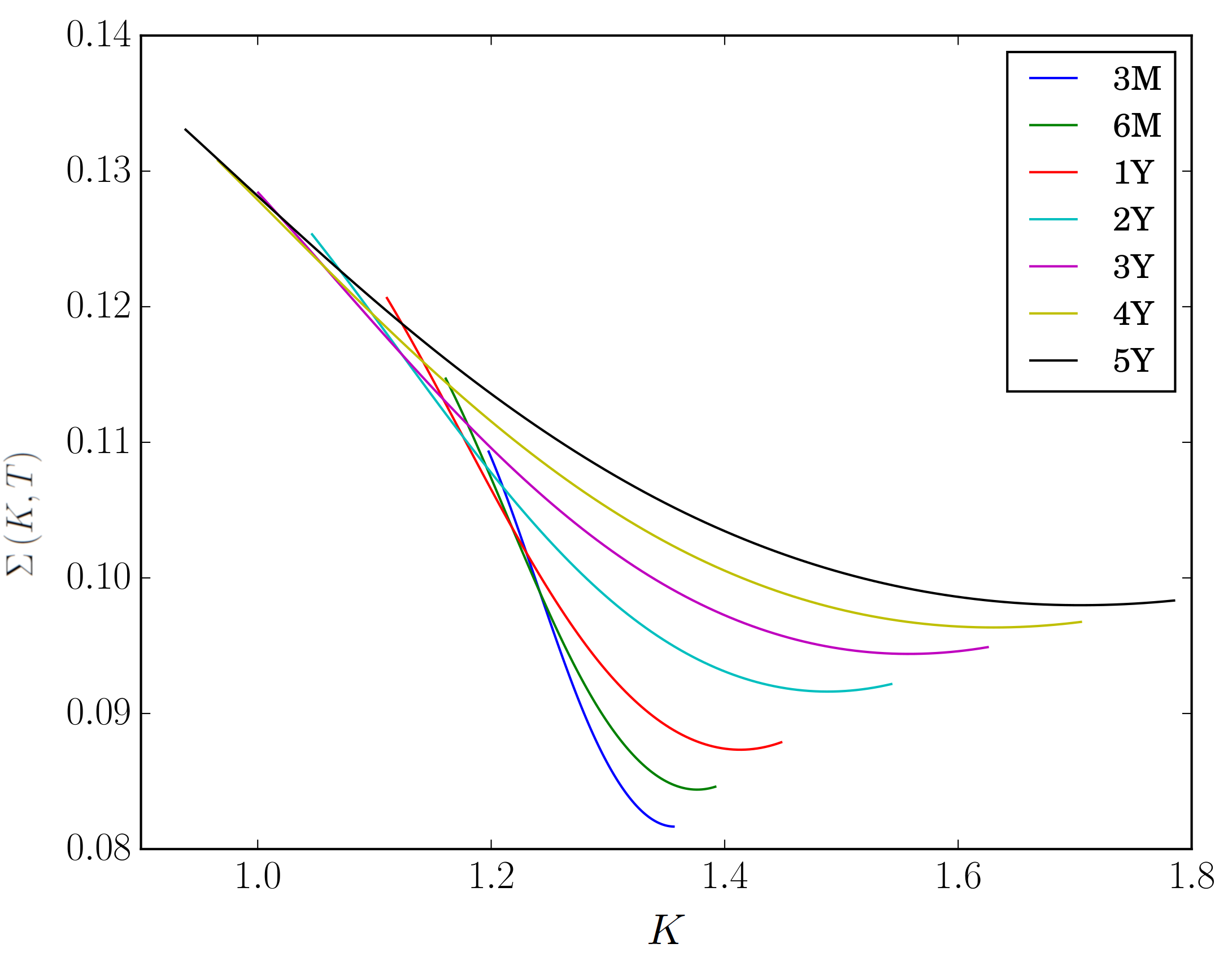}\foreignlanguage{british}{\captionof{figure}{Market volatility surface
							for $\text{EURUSD}$ on 28/03/2013.
							The spot value was $S_{0}=1.2837.$
						}\label{fig:markit-market-volatility-surface}}
					\par\end{center}\selectlanguage{english}%
			\end{minipage}~~~~%
		\end{minipage}
	\end{figure}

	%

	Additionally, we will perform calibration on 
	quotes for foreign one-touch options 
	for the following maturities, relative to 28/03/2013:
	$
	\text{3M, 6M, 1Y, 2Y, 3Y, 4Y, 5Y},
	$
	and
	barrier levels $B$ chosen such that the discounted foreign no-touch-up probabilities are approximately 
	$
	50\%,\,60\%,\,70\%,\,80\%,\,90\%.
	$

\section{A second order scheme for the PIDE}\label{sec:NumericalPIDE}


In this section, we introduce a second order accurate and empirically stable numerical scheme
for the PIDE (\ref{eq:Volettera-Type-PIDE}).
More specifically, we construct a tailored non-uniform spatial mesh, combined with finite differences for the derivatives and quadrature of the integral term,
and a backward differentiation formula (BDF) on a non-uniform time mesh, which 
is shown in tests to have better stability than the usual Crank-Nicolson scheme. In this section, for simplicity of notation, we drop the $\text{LMV}$ subscript from $\sigma_{\text{LMV}}$ in (\ref{eq:bsv}).


\subsection{Space discretisation}
\label{subset:space}

We define 
$M+1$ time points $T_m$, $N+1$ strike points $K_i$ and $P+1$ barrier points $B_j$,
leading to the following implicit definition of the non-uniform step sizes functions $\Delta_{T}, \Delta_{K}, \Delta_{B}$, respectively:
\begin{eqnarray*}
\begin{array}{lcll}
T_{m} & = & \sum_{m'=0}^{m-1}\Delta_{T}\left(m'\right), & 0\le m\le M, \\
K_{i} & = & \sum_{i'=0}^{i-1}\Delta_{K}\left(i'\right), & 0\le i \le N, \\ 
B_{j} & = & S_{0}+\sum_{j'=0}^{j-1}\Delta_{B}\left(j'\right), & 0\le j\le P.
\end{array}
\end{eqnarray*}
We denote by $N_{0}$ the node such that $K_{N_{0}}=S_{0}$. For simplicity,
we relate the step sizes functions $\Delta_{K}$ and $\Delta_{B}$ 
by $\Delta_{K}\left(i\right)=\Delta_{B}\left(i-N_0\right)$ for
any $i$ with $N_0\le i\le N$. 
This will ensure
that for any $B_{j}$, the corresponding mesh row $(\cdot, B_j,T_m)$ will contain
all $(B_{u},B_{j},T_m)$ for all $u<j$, which is
useful for the following algorithm. We denote the discrete solution
vector in such a row $(\cdot, B_j,T_m)$ by 
\[
\mathbf{u}_{.,j}^{m}=\begin{bmatrix}C\left(K_{0},B_{j},T_{m}\right),\ldots,C\left(B_{j},B_{j},T_{m}\right)\end{bmatrix}^{'} \in \mathbb{R}^{n_j},
\]
where $n_{j}=N_{0}+j+1$ and $'$ denotes the transpose. Define
$\mathbf{I}_{n}$ the identity matrix of size $n\times n$.

Derivatives are approximated by centered finite differences at each
space point except at $K=0$ and $K=B_{j}$, where they are computed,
respectively, by three-point forward and backward one-sided differences. We allow a non-uniform grid
and rely on the algorithm in \cite{Fornberg1988} to define
two finite difference operators $\delta_{K}\mathbf{u}_{i,j}^{m}$ and $\delta_{KK}\mathbf{u}_{i,j}^{m}$
(acting on the index $i$)
as well as the corresponding matrix derivative operators $\mathbf{D}$
and $\mathbf{D_{2}}$ respectively (see \cite{Hambly2016} for more details).


The integral term 
\begin{eqnarray*}
F\left(K_{i},B_{j},T_{m}\right) & = & \int_{S_{0}\lor K}^{B_{j}}g\left(K_{i},b,T_{m}\right)\db
\end{eqnarray*}
is computed using the trapezoidal quadrature rule
on the non-uniform grid,
with
\[
g\left(K,b,T\right)=-\frac{1}{2}K^{2}\frac{\partial^{2}C\left(K,b,T\right)}{\partial K^{2}}\frac{\partial\sigma^{2}\left(K,b,T\right)}{\partial b},
\]
and where we recall that we dropped for simplicity the subscript $\text{LMV}$ from $\sigma$.
We define
\[
\bar{g}\left(K_{i},B_{j},T_{m}\right)=\begin{cases}
-\frac{1}{2}K_{i}^{2}\delta_{KK}\mathbf{u}_{i,j}^{m}\frac{\partial\sigma^{2}\left(K_{i},B_{j},T_{m}\right)}{\partial B}\,, & K_{i}<B_{j}\,, \\
0\,, & K_{i}=B_{j}\,,
\end{cases}
\]
since (see \cite{Hambly2016})
\[
\frac{\partial^{2}C\left(B,B,T\right)}{\partial K^{2}}=0\,.
\]
Let $j\geq1$ and assume that we have an approximation to the solution of the
PIDE for the points $\left(\cdot,B_{j'},T_m\right)_{j'<j}$.
Hence, we can write
\begin{eqnarray}
\label{def:rhs}
\!F\left(K_{i},B_{j},T_{m}\right) \!\!\! &\!\!\!& \!\!\! \approx f\left(K_{i},B_{j},T_{m}\right)+\frac{1}{2}\Delta_{B}\left(j\right)\bar{g}\left(K_{i},B_{j},T_{m}\right), \\
f\left(K_{i},B_{j},T_{m}\right) \!\!\! &\!\!\!& \!\!\! :=
\frac{1}{2}\sum_{j'=1}^{j-1}\! \Delta_{B}\left(j' \right)\left(\bar{g}\left(K_{i},B_{j'},T_{m}\right)+\bar{g}\left(K_{i},B_{j'-1},T_{m}\right)\right)+
\frac{1}{2}\Delta_{B}\left(j\right) \bar{g}\left(K_{i},B_{j-1},T_{m}\right),
\label{def:f}
\end{eqnarray}
and $f$ can be updated inductively from row $j$ to the next
by
\begin{eqnarray}
\label{induct}
f\left(K_{i},B_{j+1},T_{m}\right)=f\left(K_{i},B_{j},T_{m}\right)+\frac{1}{2}\bar{g}\left(K_{i},B_{j},T_{m}\right)\left(\Delta_{B}\left(j\right)+\Delta_{B}\left(j+1\right)\right)\,.
\end{eqnarray}
This sum is then a source term for the $j$-th equation in the barrier direction, defining a right-hand side vector
\[
\mathbf{f}_{.,j}^{m}=\left[f\left(K_{i},B_{j},T_{m}\right)\right]_{i=0,1,...,n_{j}-1},
\]
while the second term in (\ref{def:rhs}) gives a small correction to the diffusion at $B_{j}$ and we can incorporate
it in the discretisation of the corresponding diffusive term of (\ref{eq:Volettera-Type-PIDE}); see
(\ref{eq:Discretised-PIDE-non-uniform}) below.

To approximate the ``boundary derivative'' at $(B,B,T)$ in (\ref{eq:Volettera-Type-PIDE}),
we recall from \cite{Hambly2016} that for any $T>0$
\begin{eqnarray}
\label{thirdder}
\frac{\partial^{3}C}{\partial K^{2}\partial B}\left(B,B,T\right) &=& - \frac{\partial^{3}C}{\partial K^{3}}\left(B,B,T\right) \\
&=&
\DdT \phi(B,B,T), 
\label{density}
\end{eqnarray}
where $\phi(\cdot,\cdot,T)$ is the joint density of $(S_T,M_T)$ at time $T$.
We can compute a second order, five point approximation to the third
derivative on the right-hand side of (\ref{thirdder}) with the algorithm from \cite{Fornberg1988} and denote the 
left-sided difference operator by $\delta^-_{KKK}$ and the discretisation matrix (both obtained by numerical computation) by $\mathbf{\Phi}$. For uniformly spaced grids, the discretisation matrix is given by
\[
\mathbf{\Phi}=\frac{1}{2\Delta_{K}^{3}}\begin{bmatrix}0 & \ldots & 0 & \text{0} & 3 & -14 & 24 & -18 & 5\\
\vdots & \left(0\right) & \vdots & \vdots & \vdots & \vdots & \vdots & \vdots & \vdots\\
0 & \ldots & 0 & 0 & 3 & -14 & 24 & -18 & 5
\end{bmatrix}.
\]

The PIDE algorithm involves solving one-dimensional PDEs for different values of the barrier at every row $j$ of the discretisation. The interconnection between each of these ``layers" is given through the integral term $F$. The first row $j=0$ 
is for the barrier level $B_0=S_{0}\lor K$. As a requirement for stability, we found empirically that the five grid
points used for the approximation $\mathbf{\Phi}$ need to be on the right-hand side
of $S_{0}$. In order to ensure this, we do not compute the solution for $B_j$ with $j\in \left\{1,2,3,4\right\}$, 
and for $j>4$, we start the summation in (\ref{def:f}) at $j'=5$.
No error is introduced if $\frac{\partial\sigma}{\partial b}(K,b,T)=0$
for any $b \in \left[S_{0},B_{5}\right]$ (which will be satisfied by our construction of $\sigma$),
in fact, this allows
to start the induction over $B_j$ in (\ref{induct})
at the barrier level $\Bf = \inf\{b: \frac{\partial\sigma}{\partial b}(\cdot,b,t) \neq 0\}$, since 
\[
\int_{S_{0}\lor K}^{\Bf}\frac{1}{2}K^{2}\frac{\partial^{2}C\left(K,b,T\right)}{\partial K^{2}}\frac{\partial\sigma^{2}\left(K,b,T\right)}{\partial b}\,\db=0.
\]
Please note that $B_\text{first}$ is not linked to any market conventions, but only a numerical convenience.
%
We will refer to the skipped rows 1 to 4 as the ``blank layers'' in the remainder of the section (see Fig.~\ref{fig:Finite-difference-PIDE-hyperbolic-grid} for an illustration). We note that the same effect is obtained if the strike and barrier discretisations are decoupled and that the first barrier level is chosen such that, at least five grid points used for the approximation $\mathbf{\Phi}$ are on the right-hand side of $S_0$. 

Finally, the complete surface of barrier option prices can be retrieved by cubic
spline interpolation in both strike and barrier
(in particular also for
$B<\Bf$ by interpolation between $B=S_{0}$, where $C\left(K,S_{0},T\right)=0$, and $B=\Bf$),
and constant extrapolation
for large 
barriers. The latter is a consistent assumption
since, for any $B>B_{\max}$, the value will be close to that of a vanilla option.

\begin{rem*}
We recall that in \cite{Hambly2016} we were only able to use a first
order accurate approximation of the boundary derivative due to stability issues. 
As this term is present in the discretised equation for all interior
mesh points, the scheme we proposed in \cite{Hambly2016} had a consistency
order reduced to one in $\Delta_{K}$.
We find that
using ``blank layers'' and a second-order BDF time scheme as described below in Section \ref{sub:variable step order-2-BDF} instead of Crank-Nicolson allows
to use a second order accurate approximation and preserve stability.
Overall, we obtain an order two consistent spatial approximation for smooth enough meshes. 
\end{rem*}

\subsection{Pricing vanilla options\label{sub:Pricing-vanilla-options}}


Here, we explain how prices of vanilla contracts can be obtained efficiently as a by-product
of the solution of the forward PIDE (\ref{eq:Volettera-Type-PIDE}) for barrier options.
Note that standard PDE pricing approaches are not directly applicable due to the dependence
of the volatility on the running maximum, such that a two-dimensional
backward PDE would be required  for each strike (see \cite{Hambly2016}).

The straightforward approach to vanillas with the forward PIDE (\ref{eq:Volettera-Type-PIDE})
is to set the maximum up-and-out barrier $B_{\max}$ very high. 
This requires a very large number of mesh rows in the $B$-direction, which increases the computational
time drastically. Therefore, we make and exploit the assumption that the volatility becomes
constant in the running maximum dimension above a given level $\Bl$. Then 
$\partial\sigma(K,b,T)/\partial b=0$ for any $b\geq \Bl$ and all $K, T\ge 0$,
such that (similar to the situation for small $B$ in Section \ref{subset:space})
\begin{equation}
\int_{\Bl}^{\infty}\frac{1}{2}K^{2}\frac{\partial^{2}C\left(K,b,T\right)}{\partial K^{2}}\frac{\partial\sigma^{2}\left(K,b,T\right)}{\partial b} {\, \rm d}b =0 \,.\label{eq:integralAfterBLast=00003D0}
\end{equation}
Moreover, 
it seems reasonable to assume that
\[
\lim_{B\rightarrow\infty} \sigma^{2}\left(B,B,T\right)B^{2}\left(B-K\right)\frac{\partial^{3}C\left(B,B,T\right)}{\partial K^{2}\partial B}=0\,,
\]
since 
by (\ref{density}) the term in the limit 
is proportional to the joint density function of $\left(S_{T},M_{T}\right)$ at $(B,B)$,
and we conjecture here that it goes to 0 faster than $B^{-3}$. 
In other words, 
no error is made by jumping from $B=\Bl$
to a large $B=B_{\max}$ in the solution of (\ref{eq:Volettera-Type-PIDE}).

The PIDE (\ref{eq:Volettera-Type-PIDE}) then becomes \foreignlanguage{british}{{\small{}
\begin{eqnarray}
\frac{\partial C\left(K,B_{\max},T\right)}{\partial T}+\rfT C\left(K,B_{\max},T\right) & = & -\left(\rdT-\rfT\right)K\frac{\partial C\left(K,B_{\max},T\right)}{\partial K}\label{eq:Volettera-Type-PIDE-Van-Layer}\\
 && \hspace{-2 cm} + \frac{1}{2}\sigma^{2}\left(K,B_{\max},T\right)K^{2}\frac{\partial^{2}C\left(K,B_{\max},T\right)}{\partial K^{2}} 
 - \int_{S_{0}\lor K}^{\Bl}\frac{1}{2}K^{2}\frac{\partial^{2}C\left(K,b,T\right)}{\partial K^{2}}\frac{\partial\sigma^{2}\left(K,b,T\right)}{\partial b}\,\db\,,\nonumber 
\end{eqnarray}
with boundary conditions
\begin{eqnarray*}
C\left(K,B_{\max},0\right)=\left(S_{0}-K\right)^{+}, & \quad & T=0,\\
\frac{\partial^{2}C\left(K_{\max},B_{\max},T\right)}{\partial K^{2}}=0, & \quad & K=K_{\max}, 
\end{eqnarray*}
}
for some large enough $K_{\max}\ll B_{\max}$.
One will then compute all mesh rows up to $\Bl$, and
one additional ``vanilla layer'' for $C(K,B_{\max},T)$ for an arbitrarily large level $B_{\max}$
by the PDE (\ref{eq:Volettera-Type-PIDE-Van-Layer}). }

\subsection{BDF2 scheme with variable step size\label{sub:variable step order-2-BDF}}

The main difficulty in the time discretisation of the forward PIDE (\ref{eq:Volettera-Type-PIDE})
arises from the term
\[
\frac{1}{2}\sigma^{2}\left(B,B,T\right)B^{2}\left(B-K\right)\frac{\partial^{3}C\left(B,B,T\right)}{\partial K^{2}\partial B},
\]
which, as per (\ref{density}), contains the joint density $\phi$ of the process $\left(S_{T},M_{T}\right)$
and becomes a Dirac delta point source at $\left(S_{0},S_{0}\right)$
for $T=0$. This potentially causes instabilities for all 
$B$ close to $S_{0}$ for short-term options.

In order to tackle the problem,
we first subdivide the initial time step and perform $4$ fully implicit steps of a quarter step-size.
For the definition of the BDF2 scheme, a single initial fully implicit step would suffice, but for better comparison with the Crank-Nicholson scheme we adopt the 
Rannacher startup with four steps (see \cite{gilescarter}) in both cases.
Also in both cases, we make use of the ``blank layers'' described in Section \ref{subset:space}

\selectlanguage{british}%
If we take into account the finite difference approximations and quadrature
rule for the integral, it is now possible to give a discretised PIDE,
for a given triplet $\left(i,j,m\right)$, $1\le i\le N$, $1\le j\le P$, $1\le m\le M$,
by 
\begin{eqnarray}
\delta_{T}\mathbf{u}_{i,j}^{m}+ \rfTm \mathbf{u}_{i,j}^{m}  +
\left(\rdTm-\rfTm\right)K_{i}\delta_{K}\mathbf{u}_{i,j}^{m}  &&  \nonumber \\
-\frac{1}{2}\left(\sigma^{2}\left(K_{i},B_{j},T_{m}\right)-\frac{1}{2}\frac{\partial\sigma^{2}}{\partial B}\left(K_{i},B_{j},T_{m}\right)\Delta_{B}(j)\right)K_{i}^{2}\delta_{KK}\mathbf{u}_{i,j}^{m} \hspace{-3.5 cm}&&\nonumber \\
- \frac{1}{2}\sigma^{2}\left(B_{j},B_{j},T_{m}\right)B_{j}^{2}\left(B_{j}-K_i \right)^{+}\delta_{KKK}^{-}\mathbf{u}_{n_{j},j}^{m} \hspace{-3.5 cm}& \hspace{3.5 cm} \!=\! &
f\left(K_{i},B_{j},T_{m}\right), \label{eq:Discretised-PIDE-non-uniform}
\end{eqnarray}
where $\delta_{T}$ is a time difference operator, and specify the coefficient matrices 
\begin{eqnarray*}
\mathbf{A}_{.,j}^{m} & = & \left(\rdTm-\rfTm\right) \text{diag}\left(K_{0},...,K_{n_{j}}\right),\\
\mathbf{B}_{.,j}^{m} & = & -\frac{1}{2}\text{diag}\left(\left(\sigma^{2}\left(K_{i},B_{j},T_{m}\right)K_{i}^{2}-\frac{1}{2}K_{i}^{2}\frac{\partial\sigma^{2}\left(K_{i},B_{j},T_{m}\right)}{\partial B}\Delta_{B}\left(j\right)
\right)_{0 \le i \le n_{j}-1}\right),\\
\mathbf{C}_{.,j}^{m} & = & -\frac{1}{2}\text{diag}\left(\left(\sigma^{2}\left(B_{j},B_{j},T_{m}\right)B_{j}^{2}\left(B_{j}-K_{i}\right)^{+}\right)_{0 \le i \le n_{j}-1}\right).
\end{eqnarray*}
Under fully implicit time stepping, the complete scheme can be more
compactly written as 
\begin{eqnarray}
\frac{\mathbf{u}_{.,j}^{m}-\mathbf{u}_{.,j}^{m-1}}{\Delta_{T}\left(m\right)}+\mathbf{L}_{.,j}^{m}\mathbf{u}_{.,j}^{m} & = & \mathbf{f}_{.,j}^{m},\nonumber \\
\mathbf{\mathbf{L}}_{.,j}^{m} & = &  \rfTm \mathbf{I}_{n_{j}} +\mathbf{A}_{.,j}^{m}\mathbf{D}+\mathbf{B}_{.,j}^{m}\mathbf{D_{2}}+\mathbf{C}_{.,j}^{m}\mathbf{\Phi}.\label{eq:PIDE-as-system-of-ODEs}
\end{eqnarray}
To define the BDF scheme for variable step size,
we denote $C\left(K,B,T_{m}\right)$
by $C_{m}$ and write Newton's interpolation polynomial in time as
\[
C\left(T\right)=C_{m}+\left[C_{m,}C_{m-1}\right]\left(T-T_{m}\right)+\left[C_{m,}C_{m-1},C_{m-2}\right]\left(T-T_{m}\right)\left(T-T_{m-1}\right),
\]
where $\left[.,.\right]$ and $\left[.,.,.\right]$ are divided differences. Taking the
derivative with respect to $T$, evaluated at $T_m$, 
\begin{eqnarray*}
\frac{\partial C}{\partial T}(T_m) &=&
\frac{C_{m}-C_{m-1}}{\Delta_{T}} 
 + \frac{\left(\Delta_{T}\left(m\right)\right)^{2}}{\Delta_{T}\left(m\right)+\Delta_{T}\left(m-1\right)}\left[\frac{C_{m}-C_{m-1}}{\Delta_{T}\left(m\right)}-\frac{C_{m-1}-C_{m-2}}{\Delta_{T}\left(m-1\right)}\right]\,.
\end{eqnarray*}
This yields a linear system of equations for each time step,
\[
\left(\mathbf{I}_{n_{j}}+\frac{\Delta_{T}\left(m\right)}{1+\gamma_{m}}\mathbf{L}_{.,j}^{m}\right)\mathbf{u}_{.,j}^{m}=\left(1+\frac{\Delta_{T}\left(m\right)}{\Delta_{T}\left(m-1\right)}\frac{\gamma_{m}}{1+\gamma_{m}}\right)\mathbf{u}_{.,j}^{m-1}-\left(\frac{\Delta_{T}\left(m\right)}{\Delta_{T}\left(m-1\right)}\frac{\gamma_{m}}{1+\gamma_{m}}\right)\mathbf{u}_{.,j}^{m-2}+\frac{\Delta_{T}\left(m\right)}{1+\gamma_{m}}\mathbf{f}_{.,j}^{m},
\]
with 
\[
\gamma_{m}=\frac{\Delta_{T}\left(m\right)}{\Delta_{T}\left(m\right)+\Delta_{T}\left(m-1\right)},
\]
which defines an implicit second-order multi-step method. 
Assuming there exists a smooth bijective mapping between
the non-uniform and a uniform time mesh, the method is consistent of order
2 and stability is preserved if the step-size ratio is bounded as
follows (see \cite{Hairer1993}),
\[
0<\frac{\Delta_T(m)}{\Delta_T(m-1)}<1+\sqrt{2}\,,
\]
which is guaranteed by a smooth change of the step-size.

\subsection{Non-uniform mesh construction and numerical tests}
\label{subsec:nonun}

In order to get the best accuracy, we refine the mesh around $(K,B)=(S_{0},S_{0})$,
for two reasons. First, this will add more barrier mesh rows
where $\frac{\partial C}{\partial B}$ is high and efficiently
capture the change in call prices as well as reduce an eventual error
generated by the ``blank layers''. Second, on the strike
scale, the solution is mainly convex around $K=S_{0}$ as seen in Figure \ref{fig:Up-and-out-call-price-surface}, and therefore higher accuracy
becomes important in that specific zone.

We use a hyperbolic mesh as
in \cite{CMR2016}, where we require $S_{0}$
to be a node and with
$\eta=0.05$ (in their notation) chosen according to our numerical experiments.
In Figure \ref{fig:Finite-difference-PIDE-hyperbolic-grid},
we display the generated mesh with $N=50$ points in each direction, including the initial ``blank layers'' corresponding to the vertical gap in the mesh.

\begin{figure}[h]
\begin{centering}
\includegraphics[scale=0.3]{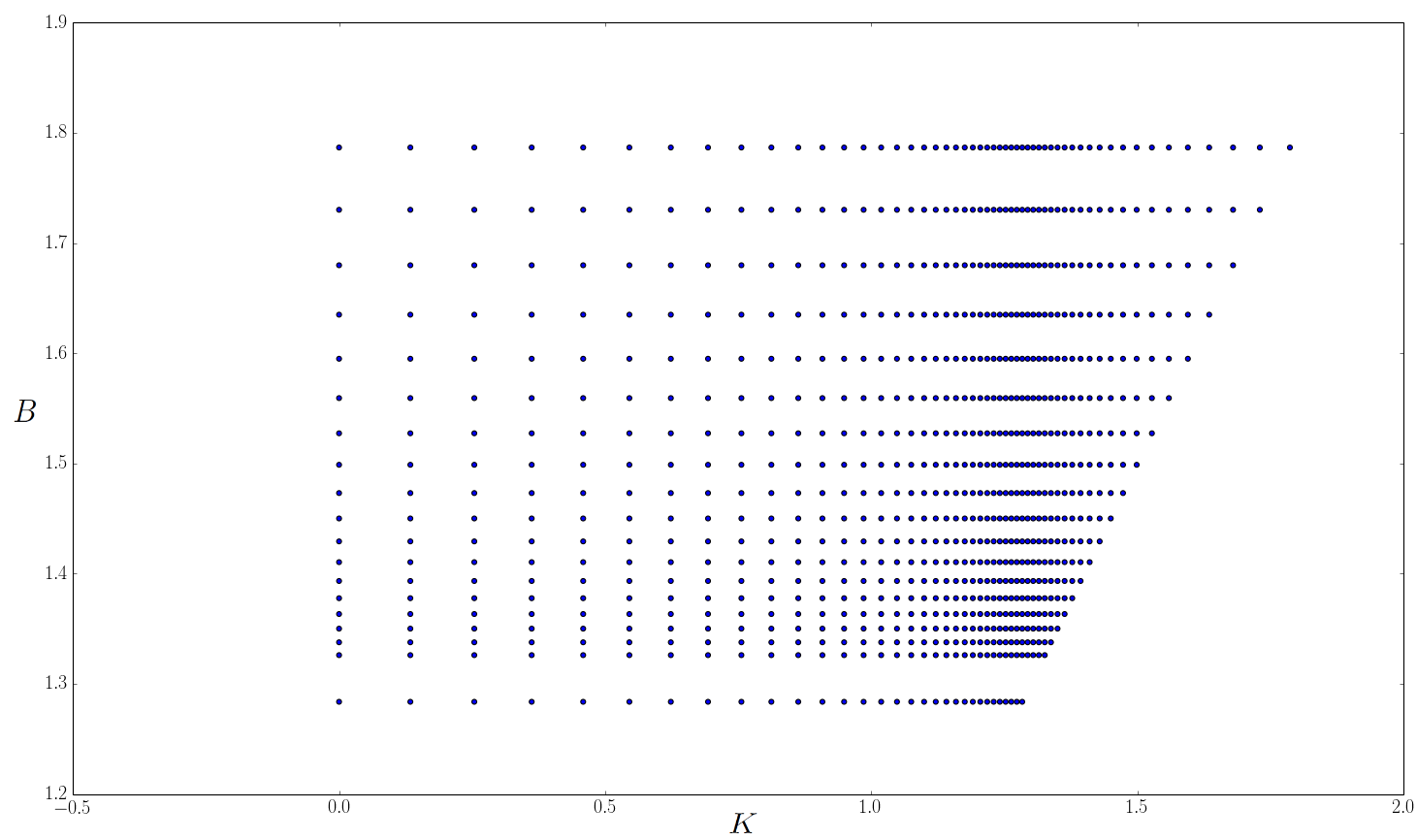}
\par\end{centering}
\caption{Hyperbolic mesh on the domain $\{(K,B) : 0< K \vee S_0 < B\}$, 
initialised with ``blank layers''. \label{fig:Finite-difference-PIDE-hyperbolic-grid}}
\end{figure}

In order to perform numerical tests,
we calibrate a pure local volatility model to the set of vanilla options
presented in Section \ref{sub:Market-Data} and obtain a local volatility
function $\sigma_{\rm LV}$. Our implementation follows a fixed-point
algorithm as described in \cite{CMR2016} based on the work of \cite{Tur2014}. Other methods to retrieve the local volatility function could have been used as well.
From this calibrated local volatility we define a hypothetical 
volatility function of the form 
\[
\sigma\left(s,m,t\right)=\sqrt{\sigma_{\rm LV}\left(s,t\right)\sigma_{\rm LV}\left(m,t\right)} \qquad m\geq s\,,
\]
defined on a mesh of strikes, barriers and maturities $\left(K_{T_{i},j},B_{T_{i},k},T_{i}\right)$,
with $1\le i\le 10$ , $1 \le j \le 5$
and $1\le k\le 5$ such that $B_{T_{i,},k}=K_{T_{i},k}$
and interpolated with cubic splines in space and constant backwards in
time. The initial spot value is $S_0=1.2837$.

For a smooth transition at the boundaries to a constant extrapolation, we propose a smooth transformation by a change of coordinates
in Appendix \ref{sub:Brunick--Shreve-volatility-smooth}.
We plot in Figure \ref{fig:Brunick-Shreve-volatility-slice} the thus assumed 
volatility. 
We emphasise that this volatility surface is in itself not calibrated to any derivatives and used purely as a numerical test example for the discretisation scheme.

\selectlanguage{english}%
In order to demonstrate the importance of ``blank layers'', we plot
the value of  (\ref{density}) obtained
with $N=700$ and $M=100$, as a function of $B$ for
$T=1$ with and without ``blank layers'' in Figure \ref{fig:barrierPriceBlankLayers}
and \ref{fig:barrierPriceNoBlankLayers} respectively. Evaluating
the term $\phi(B,B,T)$ from (\ref{eq:Volettera-Type-PIDE})
accurately is necessary since the value of the foreign no touch option
is directly linked to it by
\[
\frac{\partial C\left(0,B,T\right)}{\partial T}+\rfT C\left(0,B,T\right)=
-\frac{1}{2}\sigma^{2}\left(B,B,T\right)B^{3}
\phi\left(B,B,T\right)
\qquad\forall\left(B,T\right)\in\left(S_{0},+\infty\right)\times\mathbb{R}_{*}^{+}.
\]
 Finally, to highlight the importance of a smoothing scheme for the time stepping, we plot $\phi(B,B,T)$
with ``blank layers'' combined with BDF2 in Figure \ref{fig:barrierPriceBDF2}
and Crank--Nicolson in Figure \ref{fig:barrierPriceCN}, which shows
that Crank--Nicolson can generate instabilities if the number of time
steps is too small, even when using Rannacher initialisation.

\selectlanguage{british}%
\begin{figure}[h]
\begin{minipage}[c]{1\columnwidth}%
\begin{minipage}[c]{0.48\columnwidth}%
\selectlanguage{english}%
\begin{center}
\includegraphics[scale=0.2]{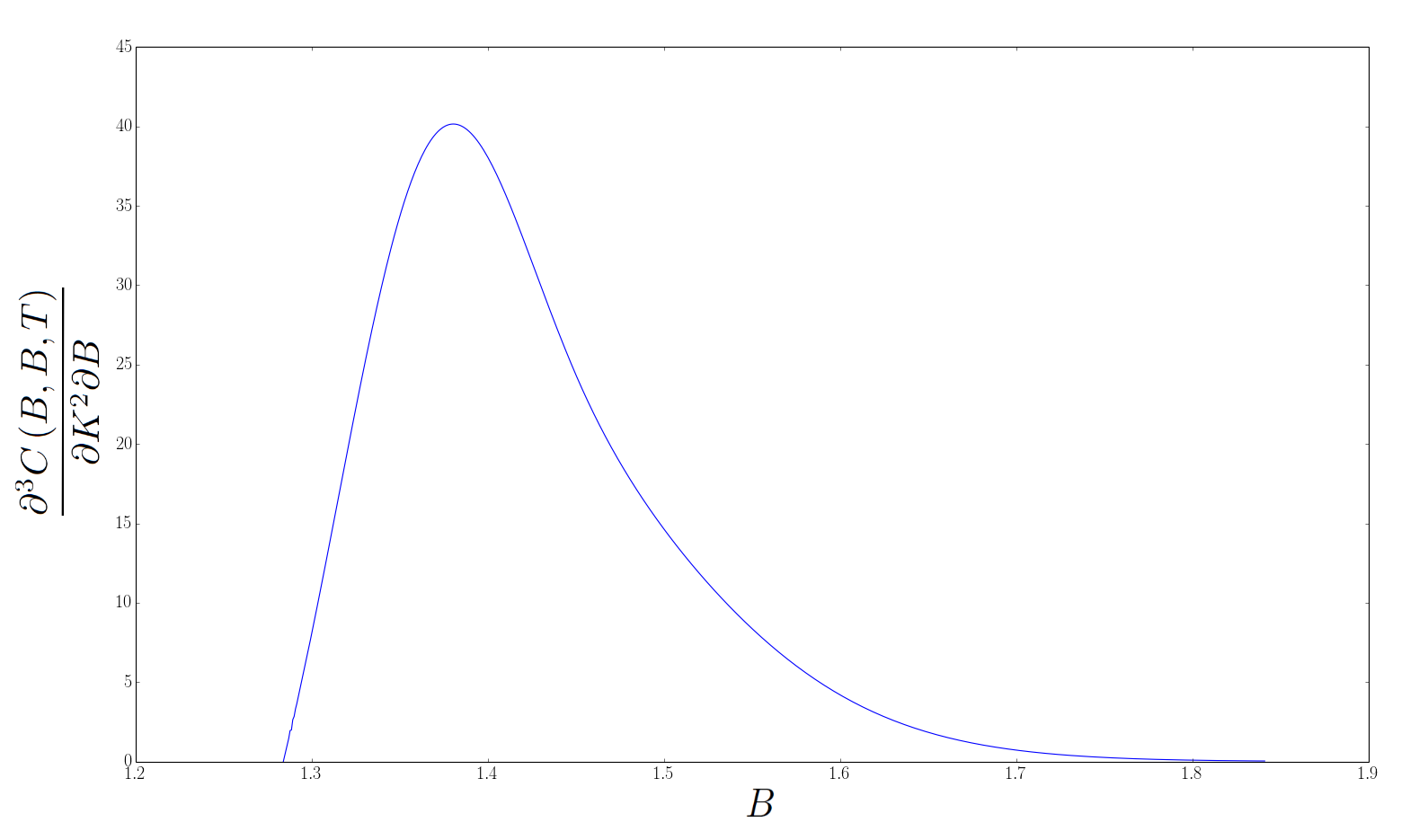}\foreignlanguage{british}{\captionof{figure}{$\partial^3 C/\partial K^2 \partial B$ 
along the diagonal $K=B$ for $T=1$, initialised with ``blank layers". 700 strike steps, 100 BDF2 time steps + Rannacher initialisation.}\label{fig:barrierPriceBlankLayers}}
\par\end{center}\selectlanguage{english}%
\end{minipage}~~~~%
\begin{minipage}[c]{0.48\columnwidth}%
\selectlanguage{english}%
\begin{center}
\includegraphics[scale=0.2]{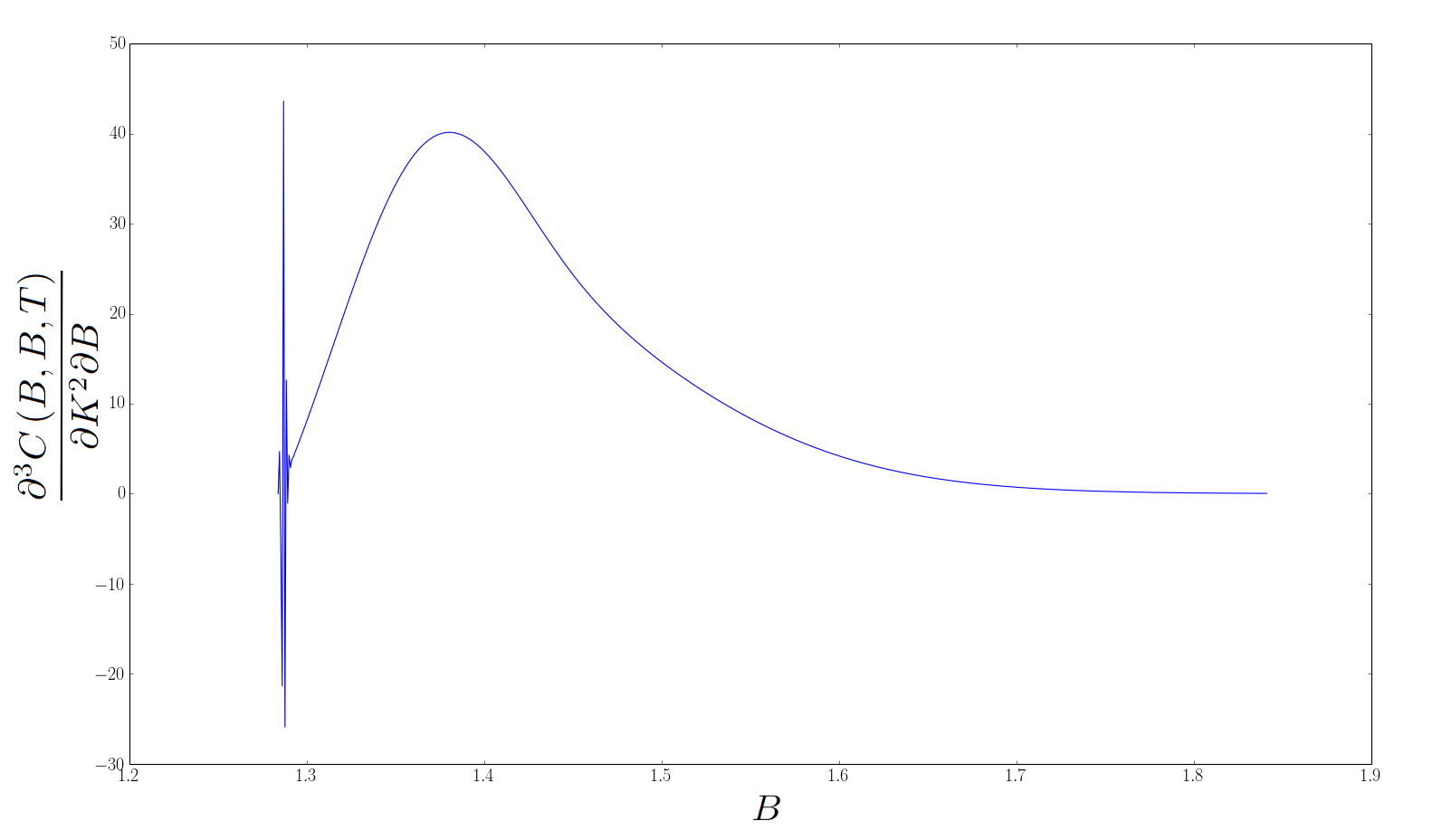}\foreignlanguage{british}{\captionof{figure}{$\partial^3 C/\partial K^2 \partial B$ 
along the diagonal $K=B$ for $T=1$, initialised without ``blank layers". 700 strike steps, 100 BDF2 time steps + Rannacher initialisation.}\label{fig:barrierPriceNoBlankLayers}}
\par\end{center}\selectlanguage{english}%
\end{minipage}%
\end{minipage}
\end{figure}

\begin{figure}[h]
\begin{minipage}[c]{1\columnwidth}%
\begin{minipage}[c]{0.48\columnwidth}%
\selectlanguage{english}%
\begin{center}
\includegraphics[scale=0.16]{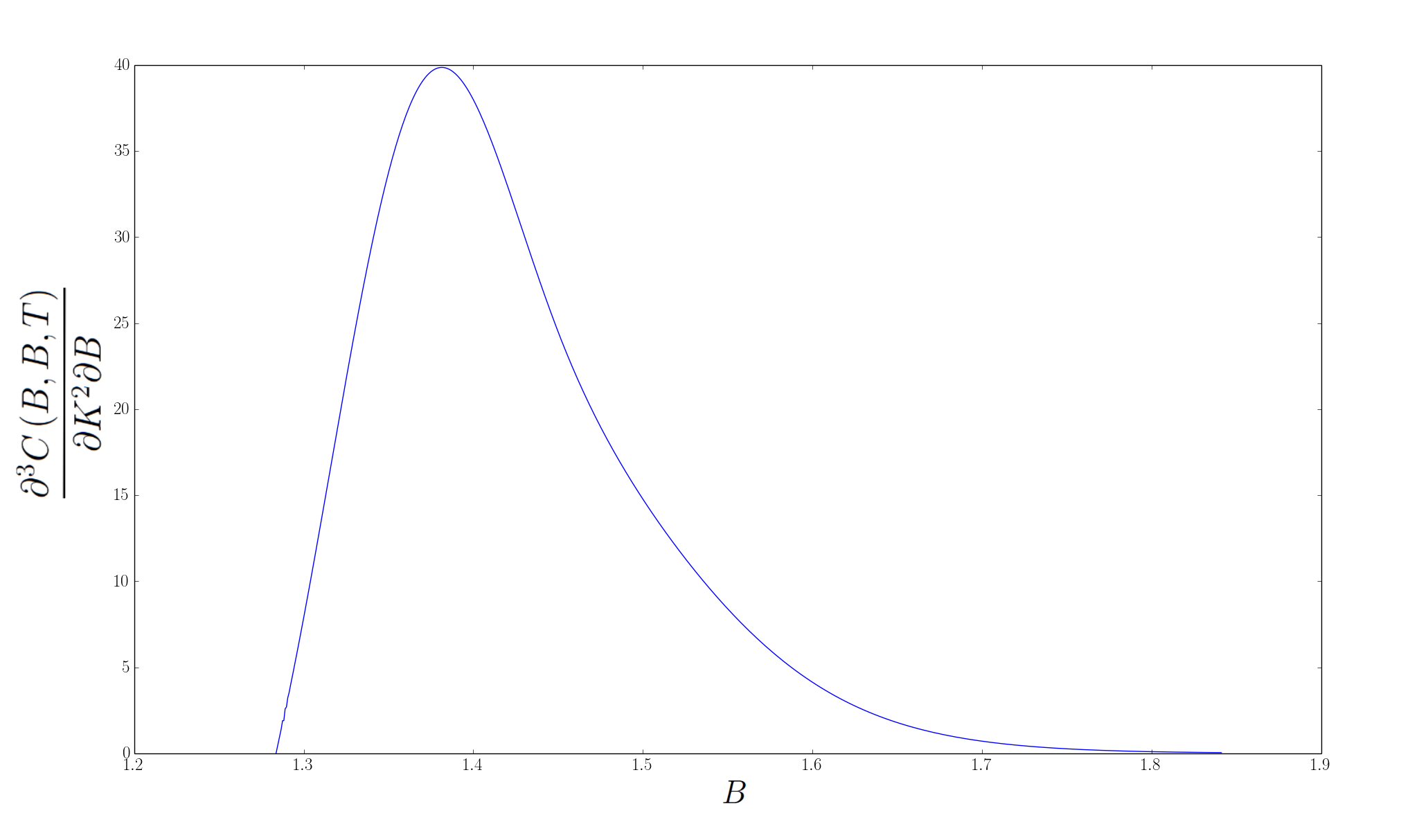}\foreignlanguage{british}{\captionof{figure}{$\partial^3 C/\partial K^2 \partial B$ 
along the diagonal, for $T=1$, initialised with ``blank layers". 700 strike steps, 10 BDF2 time steps + Rannacher initialisation.}\label{fig:barrierPriceBDF2}}
\par\end{center}\selectlanguage{english}%
\end{minipage}~~~~%
\begin{minipage}[c]{0.48\columnwidth}%
\selectlanguage{english}%
\begin{center}
\includegraphics[scale=0.16]{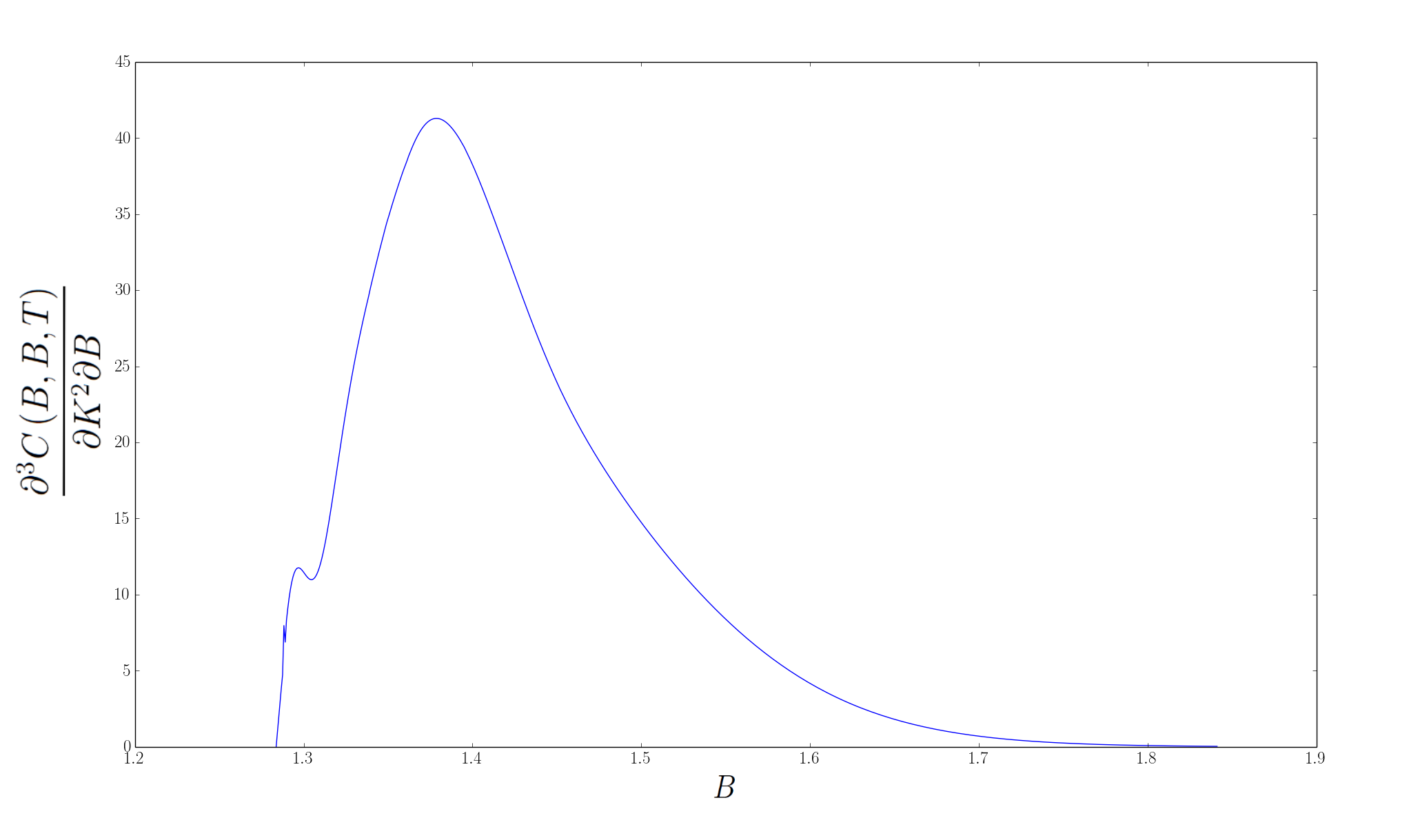}\foreignlanguage{british}{\captionof{figure}{$\partial^3 C/\partial K^2 \partial B$ 
along the diagonal for $T=1$, initialised with ``blank layers". 700 strike steps, 10 Crank--Nicolson time steps + Rannacher initialisation.}\label{fig:barrierPriceCN}}
\par\end{center}\selectlanguage{english}%
\end{minipage}%
\end{minipage}
\end{figure}

\selectlanguage{english}%
In order to numerically verify the PIDE solution, we compute the price
of an up-and-out call option for $K=80\%\times S_{0}$, $B=110\%\times S_{0}$
and $T=1$ with the forward PIDE and
 crude Monte Carlo combined with the
Brownian bridge (BB) technique as in Chapter 6 of \cite{Glasserman2004}.

The results are shown in Table \ref{tab:forwardPIDEvsMC}.

\selectlanguage{british}%
\begin{center}
\begin{table}[h]
\begin{centering}
\begin{tabular}{cc}
\toprule
Forward PIDE  & Monte Carlo (with 95\% conf. int.)\tabularnewline

\midrule

0.15823
 & 
0.15825 (0.15821, 0.15828)
\tabularnewline
\bottomrule 
\end{tabular}
\par\end{centering}

\caption{\selectlanguage{english}%
Price of an up-and-out call option for $K=80\%\times S_{0}$, $B=110\%\times S_{0}$,
computed with the forward PIDE (700 spot
steps and 100 time steps) and Monte Carlo
($5\times10^{7}$ paths and 500 time steps with Brownian bridge interpolation).
\foreignlanguage{british}{\label{tab:forwardPIDEvsMC}}\selectlanguage{english}%
}
\end{table}

\par\end{center}
In Table \ref{tab:PIDECvg-in-K }, we give the convergence order as
a function of the number of strike points. More precisely, the error $e_n$
in row $n$ is the absolute difference between the value
with $2\times N_{K}$ and $N_{K}$ strike points, $N_K = 150\times 2^n$. 
The order displayed
in row $n$ is then $\ln\left(e_{n}/e_{n+1}\right)/\ln 2$.
We notice that while the
convergence order is close to $3$ for a smaller number of strike points,
the asymptotic order is indeed 2. A similar
approach is used for Table \ref{tab:PIDECvg-in-T}, where convergence
of order 2 is obtained after 240 time steps per year. However, even
for a smaller number of time steps, the error is small and the price
is accurate.

\begin{table}[h]
	\begin{minipage}[c]{1\columnwidth}%
		\begin{minipage}[c]{0.48\columnwidth}%
			\selectlanguage{english}%
			\begin{center}
				\begin{tabular}{ccc}
					\toprule 
					$N_{K}$ & Error & Order\tabularnewline
					\midrule 
					300 & $9.43\times10^{-6}$ & 2.93\tabularnewline
					600 & $1.38\times10^{-6}$ & 2.03\tabularnewline
					1200 & $3.23\times10^{-7}$ & 1.98\tabularnewline
					2400 & $6.53\times10^{-8}$ & --\tabularnewline
					\bottomrule
				\end{tabular}\\
\vspace{-0.6 cm}
				~\\
				~\\
				
				\par\end{center}
			
			\selectlanguage{british}%
			\begin{center}
				\captionof{table}{Convergence order in number of strike steps  $N_{K}$ for $N_T=60$ time steps.}\label{tab:PIDECvg-in-K }
				\par\end{center}%
		\end{minipage}~~~~~~~~~~~~%
		\begin{minipage}[c]{0.48\columnwidth}%
			\selectlanguage{english}%
			\begin{center}
				\begin{tabular}{ccc}
					\toprule 
					$N_{T}$ & Error & Order\tabularnewline
					\midrule
					60 & $7.58\times10^{-6}$ & 1.12\tabularnewline
					120 & $3.48\times10^{-6}$ & -0.99\tabularnewline
					240 & $6.91\times10^{-6}$ & 2.07\tabularnewline
					480 & $1.64\times10^{-6}$ & 2.04\tabularnewline
					960 & $3.99\times10^{-7}$ & --\tabularnewline
					\bottomrule
				\end{tabular}
				\vspace{-0.6 cm}
				\par\end{center}
			
			\selectlanguage{british}%
			\begin{center}
				\captionof{table}{Convergence order in number of time steps $N_{T}$ for $N_K=1200$ strike steps.}\label{tab:PIDECvg-in-T}
				\par\end{center}%
		\end{minipage}%
	\end{minipage}
\end{table}

We re-iterate the non-standard nature of the PIDE and that it was only by a careful adaptation of standard techniques that high accuracy and stability was achieved.

Finally, we show in Figure \ref{fig:Up-and-out-call-price-surface}
up-and-out call prices for $T=1$ as a function of $K$ and $B$ where we have $S_0=1.2837$.

\begin{figure}[h]
\begin{minipage}[c]{1\columnwidth}%
\begin{minipage}[c]{0.48\columnwidth}%
\selectlanguage{english}%
\includegraphics[scale=0.25]{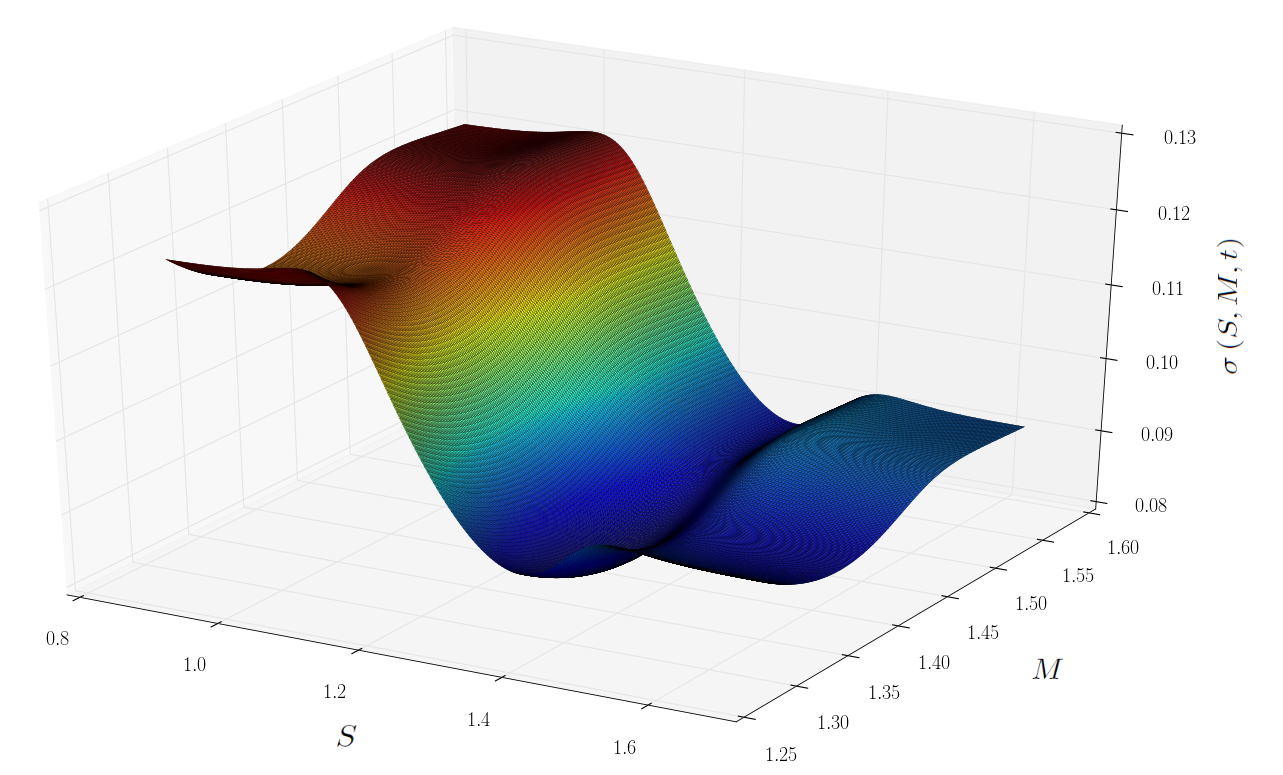}\foreignlanguage{british}{\captionof{figure}{
Volatility function used for numerical test with smooth transition at the boundary. 
The construction is such that  $\sigma\left(s,m,t\right)=\sqrt{\loc\left(s,t\right)\loc\left(m,t\right)}$.}\label{fig:Brunick-Shreve-volatility-slice}}\label{fig:brunick-shreve-test-vol-surface_}
\end{minipage}~~~~%
\begin{minipage}[c]{0.48\columnwidth}%
\selectlanguage{english}%
\begin{center}
\includegraphics[scale=0.25]{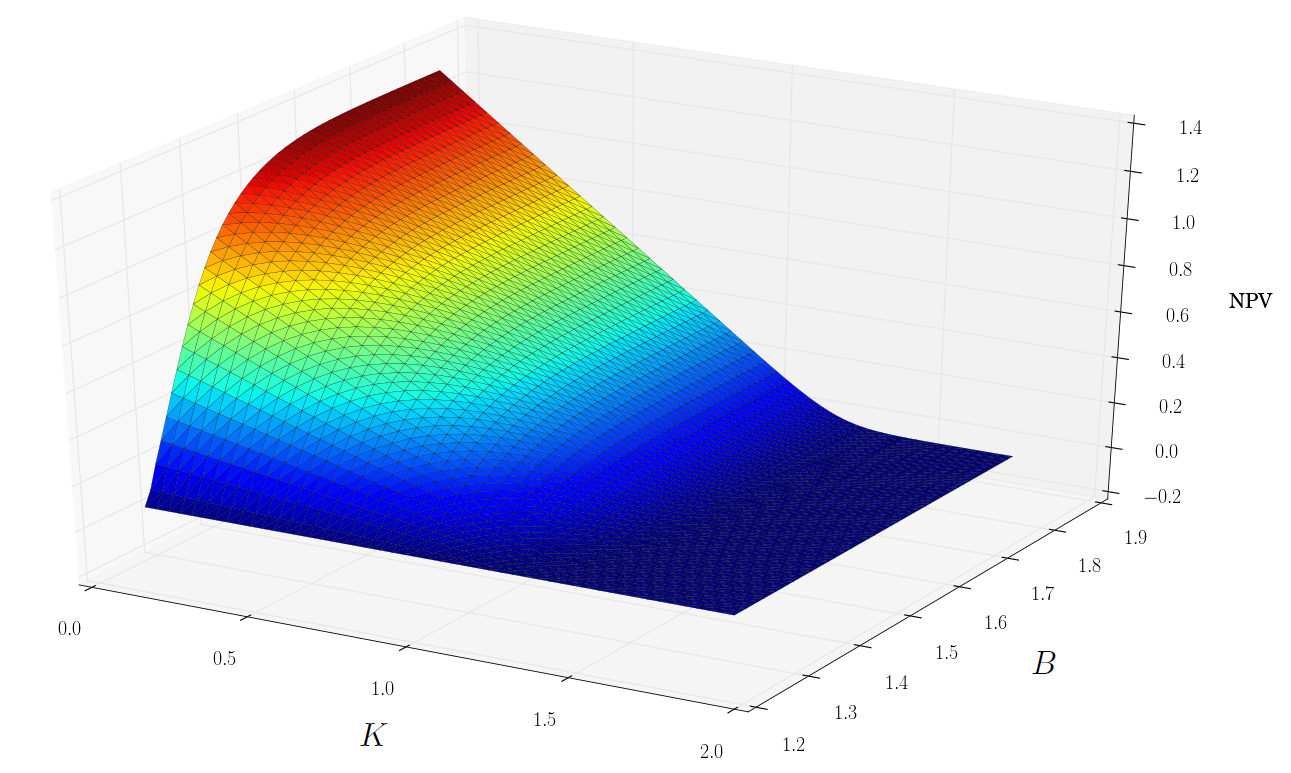}
\par\end{center}
\selectlanguage{british}%
\captionof{figure}{Up-and-out call prices computed with the forward PIDE for different values of strikes and barriers and $T=1$.}\label{fig:Up-and-out-call-price-surface}%
\end{minipage}%
\end{minipage}
\end{figure}

\section{Calibration of path-dependent volatility models \label{sec:Calibration Brunick--Shreve}}

For the calibration to options with barrier features, whose payoff thus depends on the running maximum of the underlying asset,
it seems natural to also consider models where the volatility depends explicitly on the maximum.
This leads to models from the class of path-dependent volatility models proposed in \cite{Guyon2014}.

To this end, let the range of possible $(S_t, M_t)$ values be
\[
\dom=\left\{ \left(s,m\right)\in\mathbb{R}^{2}:\,0\leq s\vee S_{0}\leq m\right\} \,.
\]
First, we define a ``local maximum volatility'' (LMV) model 
by 
\begin{equation}
\begin{cases}
\cfrac{dS_{t}}{S_{t}}=\left(\rdt-\rft\right)\,dt+\sigma_{\mathrm{LMV}}\left(S_{t},M_{t},t\right)\,dW_{t} \\
M_{t}=\underset{0\leq u\leq t}{\max}S_{u}\,,
\end{cases}\label{eq:modelBS}
\end{equation}
where $\sigma_{\mathrm{LMV}}\,:\,\dom \times\left[0,T\right]\rightarrow\mathbb{R}^{+}$
is assumed to be bounded, locally Lipschitz in $S$ and continuously
differentiable in $M$. This implies that $(S_t,M_t)_{t\ge 0}$ is Markovian
(see \cite{BrunickShreve2013}). The construction is motivated by the ability of the model
to mimick the joint distribution of $S_t$ and $M_t$ for any diffusion model, as shown in
\cite{BrunickShreve2013}, and therefore it can fit up-barrier option prices simultaneously for all maturities, strikes, and barrier levels
by construction.

Secondly, we propose a ``local maximum stochastic volatility'' (LMSV) model defined as
\begin{equation}
\begin{array}{c}
\begin{cases}
\cfrac{dS_{t}}{S_{t}}=\left(\rdt-\rft\right)\,dt+\levM\left(S_{t},M_{t},t\right)\sqrt{V_{t}}\,dW_{t}\\
dV_{t}=\kappa\left(\theta-V_{t}\right)\,dt+\xi \sqrt{V_{t}}\,dW_{t}^{V}\\
M_{t}=\underset{0\leq u\leq t}{\max}S_{u}\,,
\end{cases}\end{array}\label{eq:LMSV model}
\end{equation}
with $\levM:\,\dom\times\left[0,T\right]\rightarrow\mathbb{R}^{+}$
a local volatility function which also depends on the running maximum.
This model extends (\ref{eq:modelBS}) in the same way that the LSV model extends the LV model. One might also wish to incorporate a mixing factor $\beta$ as in (\ref{eq:LSV-mixing=00003D0.55}). This would fit naturally into the calibration proposed below.

Note that the class of models in (\ref{eq:Model Definition}) includes (\ref{eq:modelBS}) and (\ref{eq:LMSV model}).

In the following, we discuss the calibration of the path-dependent volatility model (\ref{eq:modelBS})
by forward PIDE (\ref{eq:Volettera-Type-PIDE}) to vanilla and no-touch options.

\subsection{LMV model calibration with regularised best-fit algorithm}
\label{subsec:lmv}

As the LMV model represents
the natural extension of the Dupire local volatility model for European calls to 
up-and-out barriers, the approach taken here is motivated by the literature on calibration and regularisation
of local volatility. Our goal is to encode the market prices of both vanilla options and no-touches in one model.
Moreover, it represents a building block to the calibration of the LMSV model in Section \ref{sec:LMSV calibration} by particle method.

The local maximum volatility function $\locM$ is calibrated directly to these quotes and cannot be expected to be unique,
as the marginal distributions of $S$ (from the vanillas) and $M$ (from the no-touches) do not uniquely identify their joint distribution.
How much they restrict the joint distribution is an interesting reseach question.

Specifically, we will minimise a functional consisting of the least-squares model error for vanillas and no-touches
and a penalty term which steers the optimisation algorithm to a local minimum with certain regularity. 
The optimisation is performed over volatility surfaces which are parameterised with a finite dimensional parameter vector $\Lambda_i$ for each $(T_{i-1},T_{i})$.

In our tests, the 
volatility function is defined by quadratic splines in spot and running maximum and
piecewise constant in time. For each quoted
maturity, we choose  a grid of
points formed by the $N_{K}=5$ quoted strikes in the spot direction
and $N_{B}=4$ nodes $\left(M_{T_{i},k}\right)_{1\le k\le N_{B}}$
in the running maximum direction, uniformly spaced
on the interval 
\[
\left[S_{0}+\frac{\left(B_{T_{i},50\%}-S_{0}\right)}{4},\,B_{T_{i},90\%}\right],
\]
where $B_{T_{i},50\%}$ and $B_{T_{i},90\%}$ are, respectively, the
quoted up-and-out barriers for the corresponding $50\%$ and $90\%$
no-touch probabilities for maturity $T_{i}$.
The optimisation will be performed
over the $\left(N_{K}\times N_{B}\right)$ matrix
\[
\Lambda_{i}\mathbf{=}\left[\sigma_{\mathrm{LMV}}^{i,j,k}\right]_{j,k},
\]
with $\locM^{i,j,k}=\locM\left(K_{T_{i},j},M_{T_{i},k},T_{i}\right)$,
with $1\le i \le N_{\text{Mat}}$, $1\le j\le N_{K}$,
$1 \le k \le N_{B}$. For a given maturity $T_{i}$,
the volatility is extrapolated asymptotically constant as described in Appendix \ref{sub:Brunick--Shreve-volatility-smooth},  outside $\left[K_{T_{i},1},K_{T_{i},N_{K}}\right]\times\left[M_{T_{i},1},K_{T_{i},N_{B}}\right]$.

We define the objective function $\bar{e}$ for $Q_{K}$ quoted strikes,
$Q_{B}$ quoted barrier levels and each maturity $T_{i}$ as
\begin{eqnarray}
\bar{e}\left(\Lambda_{i}\right) & = & e\left(\Lambda_{i}\right)\left(1+\mathcal{P}\left(\Lambda_{i}\right)\right)\nonumber \\
e\left(\Lambda_{i}\right) & = & \sum_{l=1}^{Q_{B}}\left( e_l^{\mathrm{FNT}}\left(\Lambda_{i}\right) \right)^{2}\label{eq:merit_Brunick} + \gamma^{2}\sum_{l=1}^{Q_{K}}\left( e_l^{\Sigma}\left(\Lambda_{i}\right) \right)^{2}, \nonumber \\
e_l^{\mathrm{FNT}}\left(\Lambda_{i}\right) & = & \mathrm{FNT}^{\text{Model}}\left(B_{T_{i},l},T_{i},\Lambda_{i}\right)-\mathrm{FNT}^{\text{Market}}\left(B_{T_{i},l},T_{i}\right) \\
e_l^{\Sigma}\left(\Lambda_{i}\right) & = & \Sigma^{\text{Model}}\left(K_{T_{i},l},T_{i},\Lambda_{i}\right)-\Sigma^{\text{Market}}\left(K_{T_{i},l},T_{i}\right) \,,
\end{eqnarray}
with $\Sigma^{\text{Market}}$ the market Black--Scholes implied volatility,
$\Sigma^{\text{Model}}$ the model implied volatility, $\gamma \in \mathbb{R}$ and
\begin{eqnarray}
\label{penalty}
\hspace{-0.5 cm} \mathcal{P}\left(\Lambda_{i}\right) &\!\! = \!\!& \frac{1}{N_{K}N_{B}}\sum_{l=1}^{N_{K}}\sum_{m=1}^{N_{B}}\left(\frac{\partial^{2}\locM\left(K_{T_{i},l},M_{T_{i},m},T_{i}\right)}{\partial K^{2}}\right)^{2}h\left(-\frac{\partial^{2}\locM\left(K_{T_{i},l},M_{T_{i},m},T_{i}\right)}{\partial K^{2}},0.5,0.5\right)\\
\hspace{-0.5 cm} h\left(x,x_{0},\epsilon\right) &\!\! = \!\!& \frac{1+\text{tanh}(2\frac{\left(x-x_{0}\right)}{\epsilon})}{2}\,,
\nonumber
\end{eqnarray}
where 
the second derivative
is obtained
by differentiation of the interpolant. 
Note that $\bar{e}$, $e$ and $\mathcal{P}$ at $T_i$ are functions of $\left(\Lambda_{i}\right)_{j\le i}$, but in writing $e\left(\Lambda_{i}\right)$ etc, we focus on the dependence
on $\Lambda_{i}$ for the inductive calibration.

The penalisation function $\mathcal{P}$
is a Tikhonov-type regularisation which reduces the number of local
minima and improves the stability of the volatility surface.
For a pure local volatility model, Tikhonov regularisation has been shown to provide well-posedness, under the condition that the local volatility does not depend on time, for the one maturity vanilla calibration problem in \cite{Egger2005}. A similar approach is also used in \cite{Crepey2010} for the pure local volatility model.
The parametric form (\ref{penalty}) for the penalisation was chosen empirically.
We
acknowledge that the penalised error $\bar{e}\left(\Lambda_{i}\right)$
as defined in (\ref{eq:merit_Brunick}) does not prevent over-parametrisation
if the market data fit perfectly, i.e.\ when $e\left(\Lambda_{i}\right)=0$,
as the penalisation is multiplicative.
However, during the iterative optimisation we found $e\left(\Lambda_{i}\right)$ to be always strictly
positive and the penalisation will favour smoother, 
convex shapes of the volatility in the strike direction; see also Section \ref{subsec:reg}.
Here, $h$ acts as a smoothed step function to ensure differentiability
with respect to the parameters $\Lambda_{i}$  (for the BFGS routine used below). Setting, $\epsilon=0.5 \text{ and } x_{0}=0.5$, is such that we get $ h\left(0,0.5,0.5\right) \approx 0.02$, a small positive amount in order to penalise mainly concave solutions while lightly penalising close-to linear solutions as well. In that sense, $h$ helps minimise the impact of small values of the second order derivative.


The
calibration algorithm is described in Algorithm \ref{alg:BFGS_Brunick_Calibration} in Appendix \ref{app:algo}
for $N_{\text{Mat}}$ maturity pillars where we use the calibrated local
volatility (i.e., $\locM(S_t,M_t,t) = \loc(S_t,t)$ independent of $M_t$) as a first guess for the first maturity pillar.


The calibration algorithm uses the bounded L-BFGS routine described in \cite{L-BFGS-B}, 
where the gradient of the objective function $\bar{e}$ needs to be computed. Compared to the gradient-free Nelder--Mead \cite{NelderMead1965} algorithm discussed in Section \ref{sec:LSV local Xi calibration}, the L-BFGS optimisation is considerably faster to converge to a local-minimum if a gradient can be obtained efficiently.
For instance, for parameter $\sigma_{\mathrm{LMV}}^{i,j,k}$, we
can write 
\begin{eqnarray*}
	\frac{\partial e\left(\Lambda_{i}\right)}{\partial\locM^{i,j,k}} & = & 2\sum_{l=1}^{Q_{B}}e_l^{\mathrm{FNT}}\left(\Lambda_{i}\right)\frac{\partial \mathrm{FNT}^{\text{Model}}\left(T_{i},B_{T_{i},l},\Lambda_{i}\right)}{\partial\locM^{i,j,k}} 
	+ 2\gamma^{2}\sum_{l=1}^{Q_{K}}\frac{e_l^{\Sigma}\left(\Lambda_{i}\right)}{\mathcal{V}\left(T_{i},K_{T_{i},l},\locM^{i,j,k}\right)}\frac{\partial {\rm Call}{}^{\text{Model}}\left(T_{i},K_{T_{i},l},\Lambda_{i}\right)}{\partial\locM^{i,j,k}},
\end{eqnarray*}
where $\mathcal{V}\left(T,K,\sigma\right)$ is the standard Black--Scholes
vega for maturity $T$, strike $K$ and volatility $\sigma$.
The computation of the gradient of the model
up-and-out call price $C\left(K,B,T,\Lambda_{i}\right)$ (including calls and no-touches) with respect
to the parameter vector $\Lambda_{i}$ is described in Section \ref{sub:Gradient-PIDE}.

The model prices were computed using the PIDE (\ref{eq:Volettera-Type-PIDE}) discretised with $1200$ strike
steps, $40$ time steps in between each quoted maturity $T_{i}$, which is sufficient to guarantee good accuracy.

\label{subsec:reg}

In order to emphasise the importance of the penalisation function,
we calibrate the LMV model without regularisation, i.e.\
$\mathcal{P}\equiv0$, and plot the resulting LMV function
in Figures \ref{fig:Brunick--Shreve volatility T=00003D1-NoPen}
and \ref{fig:Brunick--Shreve volatility T=00003D5-NoPen},
with the same axis range as for the regularised solution in Figures \ref{fig:Brunick--Shreve volatility T=00003D1}
and \ref{fig:Brunick--Shreve volatility T=00003D5}, which highlights different possible solutions, especially for longer maturities.
We used $\gamma=5$ in (\ref{eq:merit_Brunick}).

The calibration process shows the existence of
a few local minima in the objective function. This is not surprising
as the path-dependent volatility model is, in principle, able to calibrate perfectly
a discrete set of up-and-out call options (which includes vanilla
options), hence by only providing call and no-touch prices, the calibration problem is underdetermined.

In the present setting, with five vanilla and five no-touch quotes per maturity, and
20 parameters, we find many surfaces which fit the data. However, starting the iterative
optimisation procedure from the Dupire volatility (i.e., no dependence on the maximum),
the regularisation steers the approximate minimiser towards a calibrated surface with small penalty term.

\selectlanguage{british}%
\begin{figure}[h]
	\begin{minipage}[c]{1\columnwidth}%
		\begin{minipage}[c]{0.48\columnwidth}%
			\selectlanguage{english}%
			\begin{center}
				\includegraphics[scale=0.37]{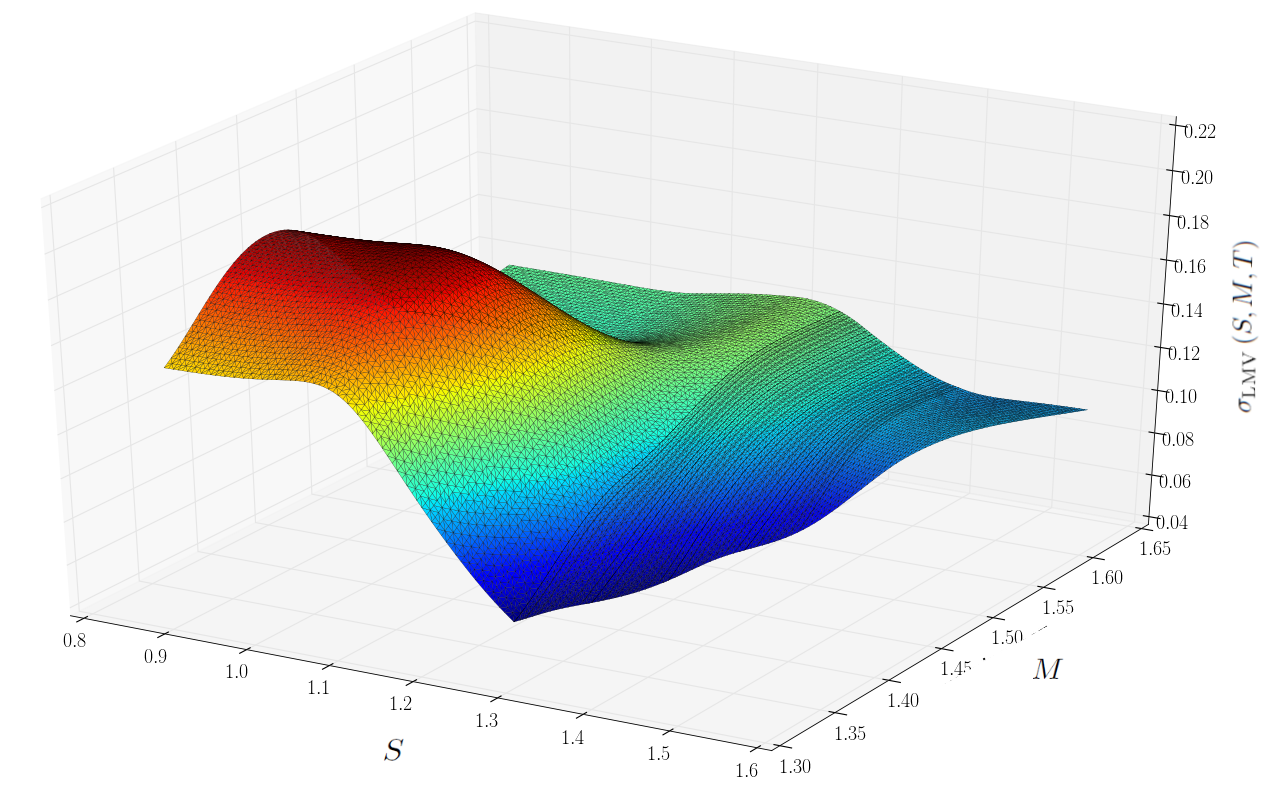}\foreignlanguage{british}{\captionof{figure}{Local maximum volatility function $T=1Y$ with regularisation.}\label{fig:Brunick--Shreve volatility T=00003D1}}
				\par\end{center}\selectlanguage{english}%
		\end{minipage}~~~~~~~%
		\begin{minipage}[c]{0.48\columnwidth}%
			\selectlanguage{english}%
			\begin{center}
				\includegraphics[scale=0.37]{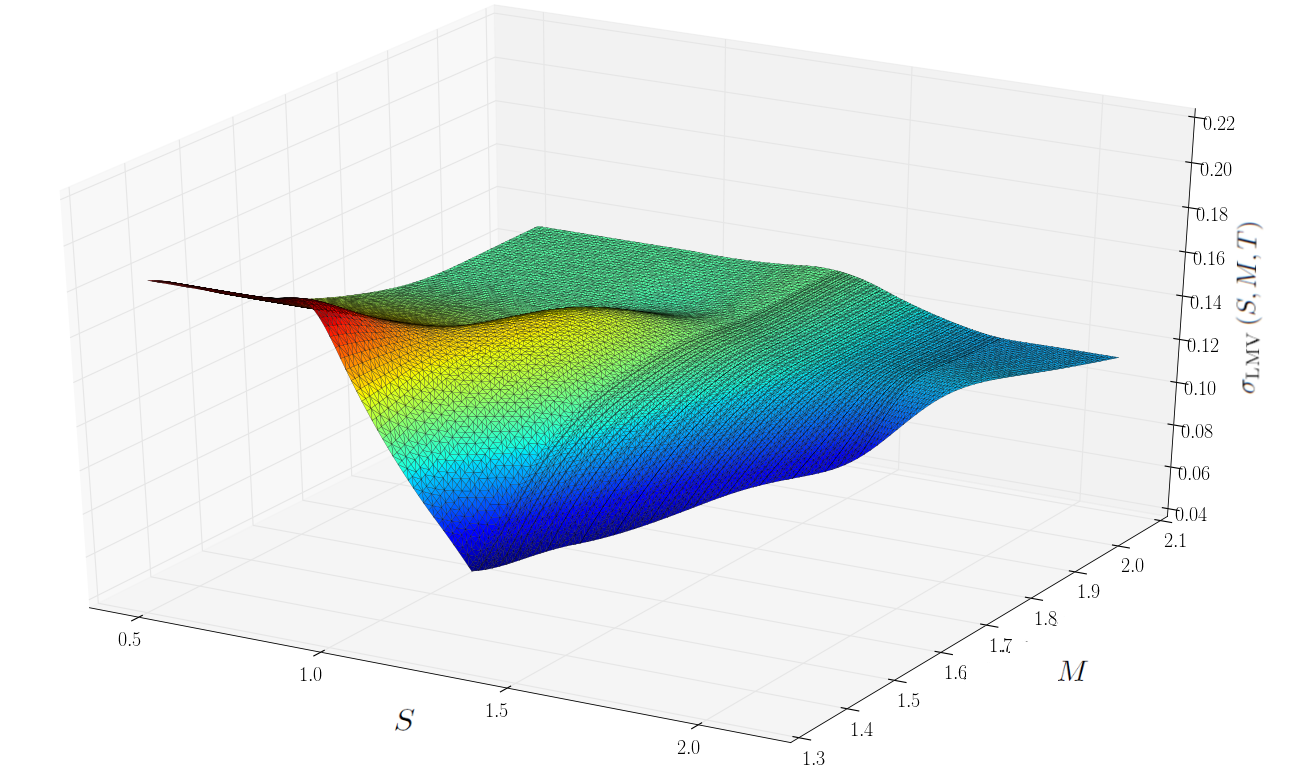}\foreignlanguage{british}{\captionof{figure}{Local maximum volatility function $T=5Y$ with regularisation.}\label{fig:Brunick--Shreve volatility T=00003D5}}
				\par\end{center}\selectlanguage{english}%
		\end{minipage}%
	\end{minipage}
	
	\begin{minipage}[c]{1\columnwidth}%
		\begin{minipage}[c]{0.48\columnwidth}%
			\selectlanguage{english}%
			\begin{center}
				\includegraphics[scale=0.37]{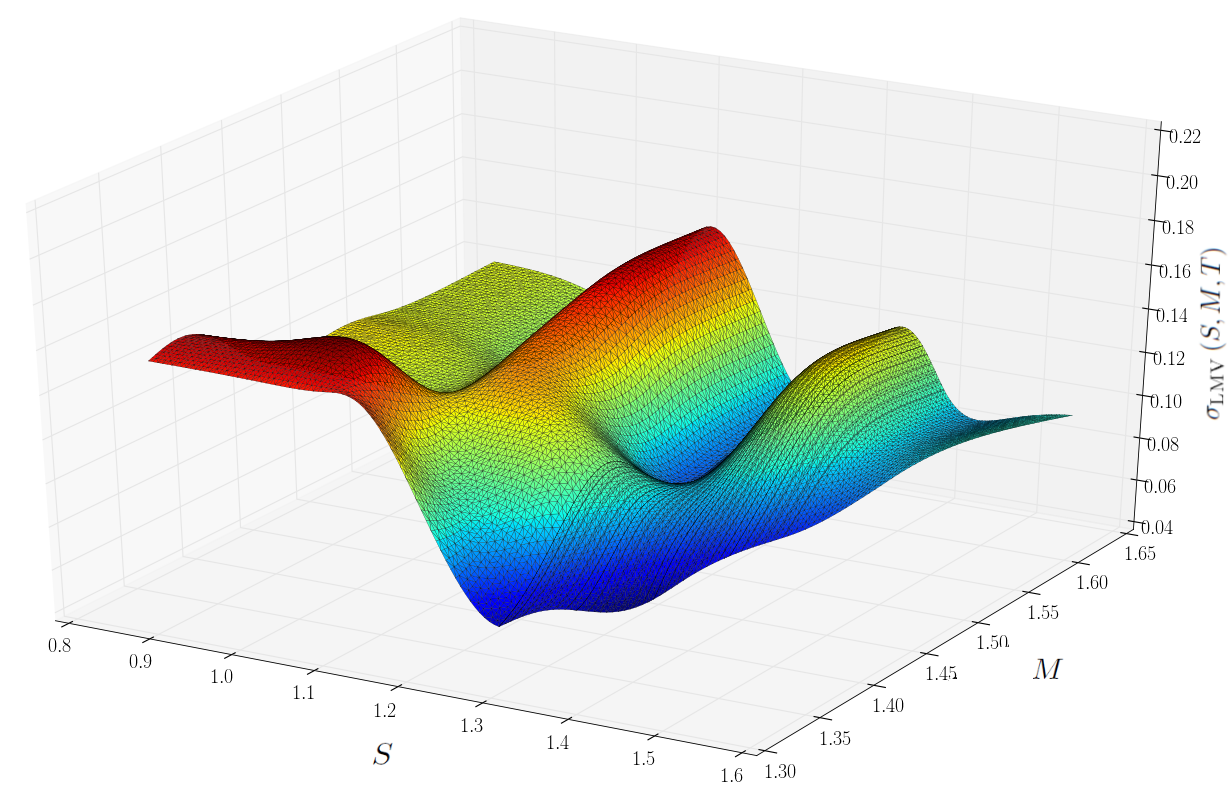}\foreignlanguage{british}{\captionof{figure}{Local maximum volatility function $T=1Y$ with no regularisation.}\label{fig:Brunick--Shreve volatility T=00003D1-NoPen}}
				\par\end{center}\selectlanguage{english}%
		\end{minipage}~~~~~~~%
		\begin{minipage}[c]{0.48\columnwidth}%
			\selectlanguage{english}%
			\begin{center}
				\includegraphics[scale=0.37]{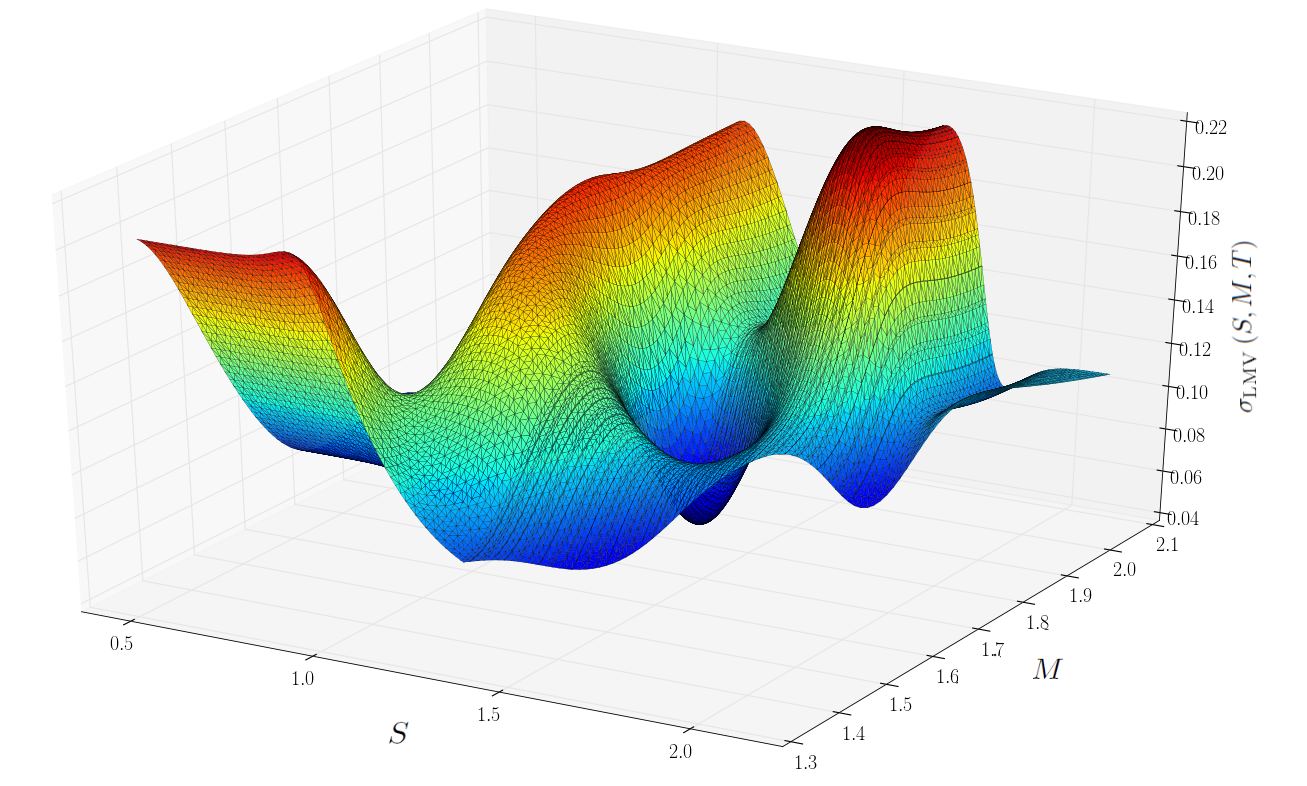}\foreignlanguage{british}{\captionof{figure}{Local maximum volatility function $T=5Y$ with no regularisation.}\label{fig:Brunick--Shreve volatility T=00003D5-NoPen}}
				\par\end{center}\selectlanguage{english}%
		\end{minipage}%
	\end{minipage}
\end{figure}

We will see in Section \ref{sec:Calibration-results}, specifically the first column of Table \ref{tab:Error No Touch}, that the calibration is very precise, with an absolute error for no-touches never higher than $0.03\%$ in price.

\subsection{Gradient operator with respect to the volatility parameters\label{sub:Gradient-PIDE}}

In order to perform a best-fit algorithm, knowledge of the gradient with
respect to the model parameters is required for the chosen
(gradient-based) optimisation process. Assume that the
volatility in $(T_i,T_{i+1})$ is a function of $N$ parameters $\Lambda_{i}=\left(\sigma_{i,1},...,\sigma_{i,N}\right)$, and constant between quoted maturities, where we drop the subscript `LMV' for brevity.

So, we need to compute
$
\nabla C\left(\sigma_{i,1},...,\sigma_{i,N}\right),
$
where $\nabla$ is the gradient operator with respect to $\Lambda_{i}$.

Equation (\ref{eq:Volettera-Type-PIDE}) can be written
as
{\small{}
	\begin{eqnarray}
	\frac{\partial C\left(\Lambda_{i}\right)}{\partial T}+\rfT C\left(\Lambda_{i}\right)+\left(\rdT-\rfT\right)K\frac{\partial C\left(\Lambda_{i}\right)}{\partial K}-\frac{1}{2}\sig^{2}\left(\Lambda_{i}\right)K^{2}\frac{\partial^{2}C\left(\Lambda_{i}\right)}{\partial K^{2}}=\hspace{3.5cm}
	\nonumber
	\\
	-\frac{1}{2}\left.\sig^{2}\left(\Lambda_{i}\right)\right\rfloor _{K=B}B^{2}\left(B-K\right)\left.\frac{\partial^{3}C\left(\Lambda_{i}\right)}{\partial K^{2}\partial B}\right\rfloor _{K=B}-\int_{S_{0}\lor K}^{B}K^{2}\frac{\partial^{2}C\left(\Lambda_{i}\right)}{\partial K^{2}}\sig\left(\Lambda_{i}\right)\frac{\partial\sig\left(\Lambda_{i}\right)}{\partial b}\,\db\,.\hspace{-1cm}\nonumber 
	\end{eqnarray}
}
We can differentiate with respect to each of the parameter vectors 
$\Lambda_{i}$,
which gives
{\small{}
	\begin{eqnarray}
	\hspace{0.5cm}
	\frac{\partial\nabla C\left(\Lambda_{i}\right)}{\partial T}+\rfT\nabla C\left(\Lambda_{i}\right)+\left(\rdT-\rfT \right)K\frac{\partial\nabla C\left(\Lambda_{i}\right)}{\partial K}-\frac{1}{2}\sig^{2}\left(\Lambda_{i}\right)K^{2}\frac{\partial^{2}\nabla C\left(\Lambda_{i}\right)}{\partial K^{2}}=\hspace{3.5cm} 
	\nonumber 
	\\
	-\frac{1}{2}\left.\sig^{2}\left(\Lambda_{i}\right)\right\rfloor _{K=B}B^{2}\left(B-K\right)\left.\frac{\partial^{3}\nabla C\left(\Lambda_{i}\right)}{\partial K^{2}\partial B}\right\rfloor _{K=B}-\int_{S_{0}\lor K}^{B}K^{2}\frac{\partial^{2}\nabla C\left(\Lambda_{i}\right)}{\partial K^{2}}\sig\left(\Lambda_{i}\right)\frac{\partial\sig\left(\Lambda_{i}\right)}{\partial b}\,\db+R\left(\Lambda_{i}\right)\,,\nonumber 
	\end{eqnarray}
	with
	\begin{eqnarray*}
		R\left(\Lambda_{i}\right) & = & \sig\left(\Lambda_{i}\right)\left(\nabla\sig\left(\Lambda_{i}\right)\right)K^{2}\frac{\partial^{2}C\left(\Lambda_{i}\right)}{\partial K^{2}} 
		-\left.\left(\sig\left(\Lambda_{i}\right)\left(\nabla\sig\left(\Lambda_{i}\right)\right)\right)\right\rfloor _{K=B}B^{2}\left(B-K\right)\left.\frac{\partial^{3}C\left(\Lambda_{i}\right)}{\partial K^{2}\partial B}\right\rfloor _{K=B}\\
		&  & -\int_{S_{0}\lor K}^{B}K^{2}\frac{\partial^{2}C\left(\Lambda_{i}\right)}{\partial K^{2}}\left(\nabla\left(\sig\left(\Lambda_{i}\right)\right)\frac{\partial\sig\left(\Lambda_{i}\right)}{\partial b}+\sig\left(\Lambda_{i}\right)\nabla\left(\frac{\partial\sig\left(\Lambda_{i}\right)}{\partial b}\right)\right)\,\db\,.
	\end{eqnarray*}
}Hence, $\nabla C\left(\Lambda_{i}\right)$ follows the same PDE as
$C$, but with an inhomogeneous term which is a function of $C$ and
its spatial derivatives, with initial condition 
\[
\nabla C\left(\Lambda_{i}\right)\left(K,B,0\right)=0\qquad 0\le K, \quad  K \vee S_0 \le B,
\]
and with boundary conditions{\small{}
	\begin{eqnarray*}
		\nabla C\left(\Lambda_{i}\right)\left(B,B,T\right)=0, & \quad & S_0 \le B,\\
		\nabla C\left(\Lambda_{i}\right)\left(K,S_{0},T\right)=0, & \quad & K\le S_{0}, 
	\end{eqnarray*}
}which match the Dirichlet boundary conditions for $C\left(K,B,T\right)$.
This useful property confirms that we can use the same discretised
linear operator for both $C$ and $\nabla C\left(\Lambda_{i}\right)$.
The additional source term $R$ on the right-hand side is fully known
since the solution for $C$ is computed beforehand. Only one costly
LU factorisation is needed to compute both $C$ and $\nabla C\left(\Lambda_{i}\right)$
at each implicit time step. Solving the linear systems of \foreignlanguage{english}{$N+1$} equations is then fast by forward and backward substitution.
Additionally, since the volatility is assumed piecewise constant in maturity,
the set of parameters $\Lambda_{i+1}$ has no impact on the values
of $C\left(K,B,T\right)$ for any $T\leq T_{i}$. Hence we also have
\[
\nabla C\left(\Lambda_{i}\right)\left(K,B,T_j \right)=0,\, j<i\,.
\]

Some numerical experiments for a volatility defined on a grid of $5\times 5$
points, i.e.\ $25$ parameters, showed that the additional evaluation
of the gradient (with respect to each of the parameters) requires
only twice the time needed to solve the PDE. As a comparison, the
powerful adjoint algorithmic differentiation (AAD) technique (see \cite{giles2006smoking} for applications in derivative pricing)
can achieve the same task for
a computational time between three to four times the time needed
to solve the original PDE independent of the number of parameters (see Section 4.6 in \cite{griewank2008evaluating}).
Therefore, the chosen approach
leads to a competitive computational time in the present setting. The gradient
components can be computed in parallel which would further reduce
the computational cost. 

\subsection{Calibration of the LMSV model by 2D particle method}
\label{sec:LMSV calibration}

In this section, we discuss a possible calibration algorithm for the LMSV
model (\ref{eq:LMSV model}). We assume that a calibrated LMV
volatility function $\locM$ is at our disposal, e.g.\ obtained as in Section \ref{subsec:lmv}.

With the Heston parameters $(\kappa, \xi, \theta)$ and $\locM$ fixed,
$\levM$ in (\ref{eq:LMSV model}) can be found from the calibration condition (see (\ref{proj}) and thereafter)
\begin{eqnarray}
\label{projectLMSV}
\levM^2(K,B,T) \, \EQd \left[V_{T}^{2}\,|\,S_{T}=K,\,M_{T}=B\right]
&=& \locM^{2}\left(K,B,T\right)\,.
\end{eqnarray}

Through the conditional expectation, the function
$\levM$ in (\ref{projectLMSV}) depends on the distribution of the joint process $X=(X_{t})_{t\ge 0}=\left(S_{t},M_{t},V_{t}\right)_{t\ge 0}$.
If we insert $\levM$ expressed from (\ref{projectLMSV}) in (\ref{eq:LMSV model}) for a model calibrated to vanilla and barrier quotes, via $\locM$,
the resulting process thus falls in the class of McKean-Vlasov processes \cite{McKean1966}.

The particle method for the estimation of conditional expectations was introduced in \cite{McKean1966},
and is discussed in detail in \cite{Sznitman1991}; it was
applied to LSV model calibration in \cite{guyon2012being,GuyonLabordere2013}. More details about stochastic filtering problems, as well as a literature review, can also be found in \cite{Bain2008}.



We consider $N$-sample paths $\left(X_{t}^{i}\right)_{1 \le i\le N}=\left(S_{t}^{i},M_{t}^{i}, V_{t}^{i} \right)_{1\le i\le N}$, $t\ge 0$,
i.e.\ $N$ independent realisations of $X$,
and write for brevity $\X_{\cdot} = \left(X_{\cdot}^{i}\right)_{1 \le i\le N}$.

The ($3\times N$)-dimensional SDE driving the system $\X$ 
in the case of the LMSV model can be approximated by
\[
\begin{cases}
\cfrac{d\hat{S}_{t}^{i}}{\hat{S}_{t}^{i}}=\left(\rdt-\rft\right)\,dt+\hat{\levM}_{N}\!\left(\hat{S}_{t}^{i},\hat{M}_{t}^{i},t;\X \right)\sqrt{V_{t}^{i}}\,dW_{t}^{i}\\
dV_{t}^{i}=\kappa\left(\theta-V_{t}^{i}\right)\,dt+\xi\sqrt{V_{t}^{i}}\,dW_{t}^{V,i} \\
\hat{M}_{t}^{i}=\underset{0\leq u\leq t}{\max} \hat{S}_{u}^{i}\,,
\end{cases}
\]
where $(W_{\cdot}^{i},W_{\cdot}^{V,i})$, $1\le i\le N$,
are independent samples of the two correlated driving Brownian motions,
$\hat{\sigma}_N$ is an estimator for $\levM$ to be defined below,
and $\hat{\X}_{t} = \left(\hat{X}_{t}^{i}\right)_{1 \le i\le N}=\left(\hat{S}_{t}^{i},\hat{M}_{t}^{i}, V_{t}^{i} \right)_{1\le i\le N}$, $t\ge 0$,
with $\hat{M}_{t}^{i} = \sup_{s\le t} \hat{S}_{t}^{i}$.

We use an extension of the QE-scheme \cite{Andersen2008} (see also Section \ref{subsec:particle}) where the volatility now depends on
the running maximum as well as the spot.
The Brownian increments are generated with a pseudorandom number generator.s
The running maximum 
is sampled approximately
with a Brownian bridge technique as described in Chapter 6 of \cite{Glasserman2004},
with $\lev$ kept constant in time between timesteps,
\begin{eqnarray*}
G_t & = & \frac{S_{t}+S_{t+\Delta t}+\sqrt{\left(S_{t}+S_{t+\Delta t}\right)^{2}-2\left(S_{t}\sigma\left(S_{t},t\right)\sqrt{V_{t}}\right)^{2}\,\Delta t\,\log\left(U_{t}\right)}}{2}\\
M_{t+\Delta t} & = & \max\left(G_t,M_{t}\right)\,,
\end{eqnarray*}
where $U_{t}$ is an independent draw from the uniform distribution $\mathcal{U}\left(0,1\right)$, i.i.d.\ across $t$.

The accuracy of the integration of the SDE would be of lesser importance if the same scheme were used in calibration and pricing, 
since the calibration will be to the conditional law of the (approximate) model. Here, we discuss calibration by PIDE and pricing by MC and hence use an accurate timestepping scheme.

We refer to \cite{guyon2012being} and \cite{Sznitman1991}
for more extensive details about the particle method and conditions for its convergence
(which, to the best of our knowledge, are not proven for the present case).
%

For the construction of the LMSV by particle method, we estimate the Markovian projection (\ref{proj}) as
\begin{eqnarray}
\label{proj2d}
\hat{p}_{N}\left(K,B,T; \X_T \right)=\frac{\frac{1}{N}\sum_{i=1}^{N}V_{T}^{i}\delta_{N}\left(S_{T}^{i}-K,\,M_{T}^{i}-B,T\right)+2\theta\xi\epsilon}{\frac{1}{N}\sum_{i=1}^{N}\delta_{N}\left(S_{T}^{i}-K,\,M_{T}^{i}-B,T\right)+\xi\epsilon},
\end{eqnarray}
with $\delta_{N}$ 
an anisotropic bi-variate Gaussian kernel 
\begin{eqnarray}
\label{kernelDef}
\delta_{N}\left(x,y,T\right) & = & \frac{\text{exp}\left(-\frac{1}{2}\frac{\zeta\left(T\right)}{1-\rho_{xy}^{2}\left(T\right)}\right)}{\gamma\left(T\right)}\,\\ \nonumber
\zeta\left(T\right) & = &  \frac{x^{2}}{h_{x}^{2}\left(T\right)}+\frac{y^{2}}{h_{y}^{2}\left(T\right)}-2\frac{\rho_{xy}\left(T\right)xy}{h_{x}\left(T\right)h_{y}\left(T\right)}\\ \nonumber
\gamma\left(T\right) & = & 2\pi h_{x}\left(T\right)h_{y}\left(T\right)\sqrt{1-\rho_{xy}^{2}\left(T\right)}\,,\nonumber
\end{eqnarray}
where $h_x$, $h_y$, $\rho_{xy}$ as well as the specific bandwidth details are again given in Appendix \ref{subsec:kernel-construction}.
When we used the normal Silverman rule bandwidth with $\rho_{xy}\left(T\right)=0$,
the accuracy was found to drop drastically and became unsatisfactory for the considered number of particles in the range $100\,000-2\,000\,000$.
We note that cheaper to evaluate kernels, such as uniform or triangle kernels, could be considered for better performance. Additionally, we used a high number of particles for our numerical study, however, a smaller sample size will likely suffice for most calibrations.

The extra terms $2\theta\xi\epsilon$ and $\xi\epsilon$ in (\ref{proj2d}) serve
as a smooth extrapolation rule for areas containing only a few particles.
for all $t$, then the calibration algorithm recovers, as expected,
the calibrated path-dependent (mimicking) volatility $\locM$. Similarly, if $\xi=0$, then $V$ is deterministic and a single particle is sufficient to calibrate the LMSV model.

Then, $\levM$ can be estimated by 
\[
\hat{\levM}_{N}\left(K,B,T;\X\right)=\sqrt{\frac{\locM^{2}\left(K,B,T\right)}{\hat{p}_{N}\left(K,B,T;\X\right)}}\,,
\]
with $\hat{p}_{N}\left(K,B,T;\X\right)$ given in (\ref{proj2d}) and  ${\X}_\cdot = ({S}^i_\cdot,V^i_\cdot,{M}^i_\cdot)_{1\le i\le N}$.

The step-by-step calibration is detailed in Algorithm \ref{alg:ParticleMethod} in Appendix \ref{app:algo}.

The computational complexity of a direct evaluation of $\sigma_{\text{LMV}}$ by $\hat{p}_{N}\left(S_t^i,M_t^i,t; \X_t \right)$ for all particles is quadratic in $N$.
To reduce the cost, we first approximate $\sigma_{\text{LMV}}$ on a mesh in $(S,M)$ and then interpolate by splines for the actual evaluation;
see Appendix \ref{subsec:spline-interpolation} for details.
We do not use any regularisation for $\sig$ here.

Moreover, for a given point $(K,B)$ in (\ref{proj2d}), only a small number of particles in the vicinity contributes significantly to the sum due to the fast decay of the kernel function.
We perform an efficient search for those particles on a $k$-d tree as described in Appendix \ref{subsec:kernel-construction},
which reduces the complexity per time step from $\mathcal{O}(N^2)$ to $\mathcal{O}( N_S N_M \log N)$, if the 2D spline has $N_S N_M$ nodes.

Finally, we note that alternative approaches could be used in order to estimate the conditional expectation. One example would be
to bucket the particles and then locally linearly regress on the variables, e.g., $S$ and $M$. This bucketing approach is used frequently for applications such as  Monte Carlo valuation of American options and for CVA computations, to avoid having to use higher order polynomial terms as in the Longstaff--Schwartz algorithm \cite{Longstaff01valuingamerican}.

%

We calibrated the model to the vanilla and no-touch quotes from Section \ref{sub:Market-Data}.
We plot in Figure \ref{fig:LMSV leverage} the calibrated
local volatility function for $T=1$ with 350 time steps per year and $2\,000\,000$
particles.
The calibration fit is compared to other models in Section \ref{sec:Calibration-results}.

\begin{figure}[h]
	\begin{minipage}[c]{1\columnwidth}%
		\begin{minipage}[c]{0.48\columnwidth}%
			\selectlanguage{english}%
			\begin{center}
				\includegraphics[scale=0.4]{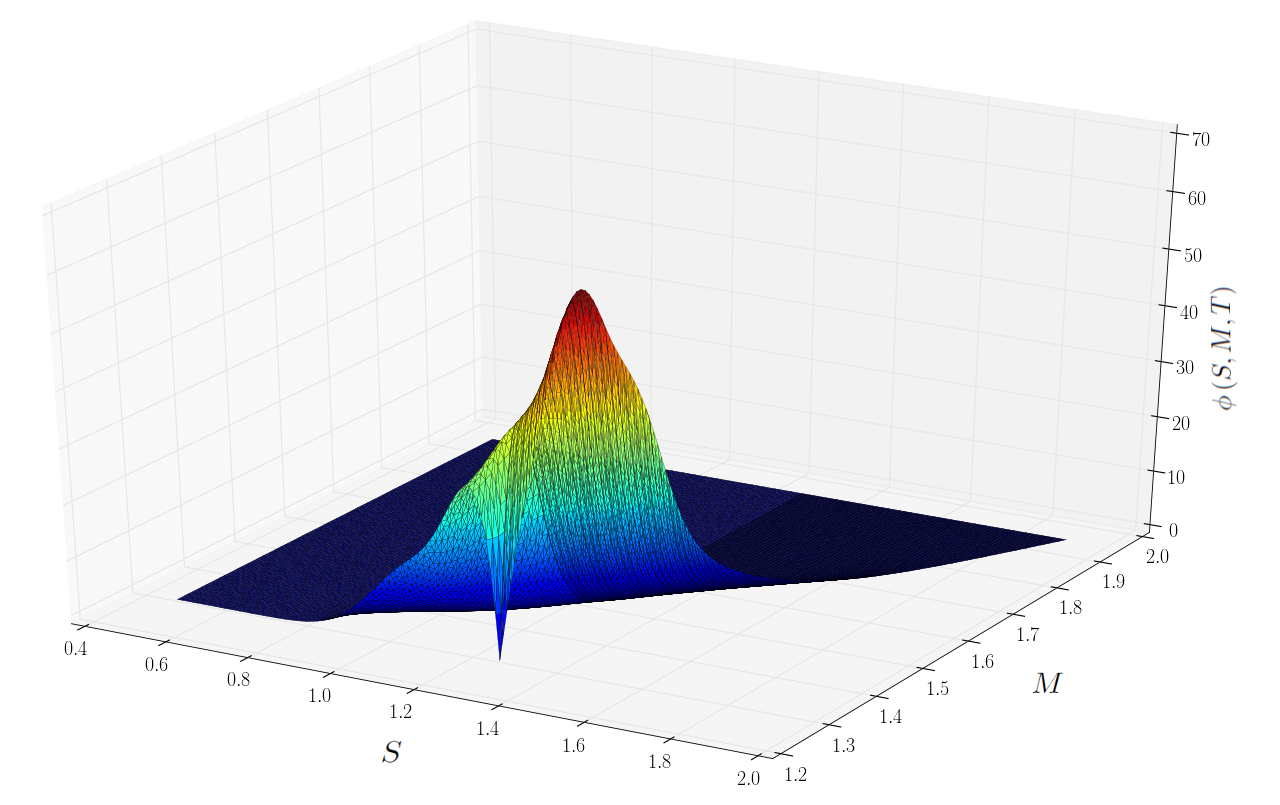}\foreignlanguage{british}{\captionof{figure}{Joint density $\phi\left(S,M,T\right)$ of the spot and running maximum after calibration at time $T=1$.}\label{fig:SMJointCalibratedDensity1Y}}
				\par\end{center}\selectlanguage{english}%
		\end{minipage}~~~~~~~~~~~~%
		\begin{minipage}[c]{0.48\columnwidth}%
			\selectlanguage{english}%
			\begin{center}
				\smallskip{}
				\includegraphics[scale=0.4]{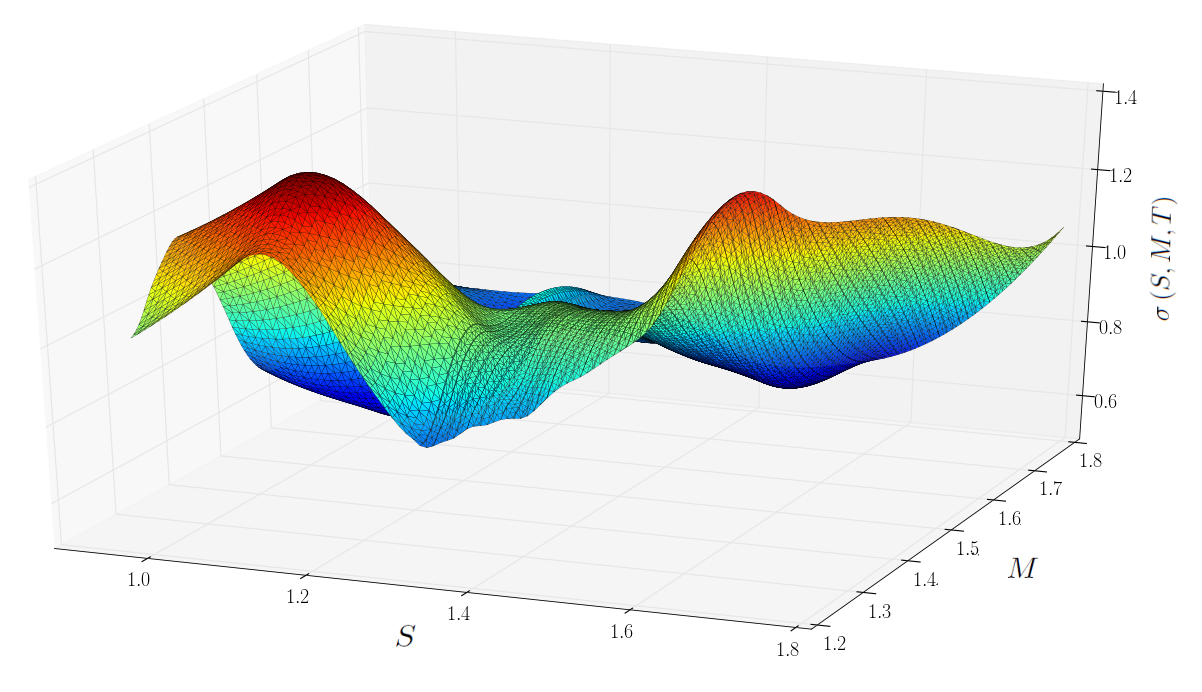}
				\par\end{center}
			
			\selectlanguage{british}%
			\begin{center}
				\captionof{figure}{Calibrated local volatility function $\levM\left(S,M,T\right)$ for the LMSV model at time $T=1$.}\label{fig:LMSV leverage}
				\par\end{center}%
		\end{minipage}%
	\end{minipage}
\end{figure}

\section{Calibration of the Heston-type LSV-LVV model \label{sec:LSV local Xi calibration}}

In this section, we describe the calibration of the LSV-LVV model (\ref{eq:Heston LSV model})
to both vanilla and no-touch options. 
Model (\ref{eq:Heston LSV model}) generalises the Heston-type LSV model (\ref{eq:LSV-mixing=00003D0.55}) and we briefly discuss the prevalent approach to the calibration of this model first.

The calibration can be based on two different methods. On the one hand, if prices of exotic products are computed by Monte Carlo, it is possible to rely on a full Monte Carlo calibration approach. Accurate re-pricing of vanilla options will be ensured by computation of the particle estimator (\ref{proj1d}), while keeping track of the running maximum for each particle will allow to compute no-touch prices for all maturities and barrier levels. An optimisation algorithm can then be used to calibrate no-touch options. However, this approach leads to inaccuracies and parameter instabilities for longer maturities and higher barrier levels. Therefore, if one wishes to use PDE techniques in order to price a set of derivative products, a full Monte Carlo calibration becomes far less suitable. On the other hand, and in order to provide consistent and stable calibration for both PDE and Monte Carlo pricing, we propose a calibration method where no-touch prices are computed with PIDE (\ref{eq:Volettera-Type-PIDE}), coupled with the LMV volatility calculated by a particle estimator we describe hereafter. This PIDE based calibration approach is described in the remainder of this section.


For any set of parameters $v_{0},\kappa,\theta,\xi,\rho$ in the Heston-LSV model ($\beta=1$), 
a sufficient condition on the local volatility function $\lev$ (see, e.g., \cite{GuyonLabordere2013})
such that the model gives a perfect fit to arbitrage-free vanilla quotes is 
\begin{eqnarray}
\lev\left(K,T\right) & = & \frac{\loc\left(K,T\right)}{\sqrt{\EQd \left[V_{T}\,|\,S_{T}=K\right]}}\,, \label{eq:LSV--LocalXi Leverge}
\end{eqnarray}
where $ \loc$ is a local volatility function (i.e., calibrated to vanilla quotes). 
Notice here that the right-hand side depends on $\lev(\cdot,t)$ for $t<T$ through the conditional expectation.


One approach to the calibration is to consider the case $\sigma(S_t,t) = 1$ independent of $S_t$ and $t$ and calibrate a Heston type model with $\beta=1$ to the vanilla quotes by choice of $v_0, \theta, \xi, \kappa, \rho$. Then, having established this choice of these parameters, $\beta$ is adjusted 
and for each choice the local volatility $\sigma(S,t)$ is recalibrated. The choice of $\beta$ is made for the best match to the barrier option prices (while by construction maintaining the calibration to the vanilla options).

The parameter $\beta$ is commonly within the range $[0, 1]$ and called the ``mixing factor'' (see \cite{Clark2010}):
the market is believed to stand in between pure local volatility models, i.e.\ $\beta=0$, and full Heston LSV models, i.e.\ $\beta=1$.
We show calibration results which support this claim in Section \ref{sec:Calibration-results}.

To improve the calibration accuracy to no-touch options, whose payoff depends on the running maximum, the vol-of-vol ``local volatility function'' is made spot- and time-dependent in our model (\ref{eq:Heston LSV model}).

\subsection{Particle method and parametrisation of $\boldsymbol{\sigma}$}
\label{subsec:particle}

We satisfy the calibration condition (\ref{eq:LSV--LocalXi Leverge}) by a modificaiton of the particle method from Section \ref{sec:LMSV calibration}.

We consider $N$-sample paths $\left(X_{t}^{i}\right)_{1 \le i\le N}=\left(S_{t}^{i},M_{t}^{i}, V_{t}^{i} \right)_{1\le i\le N}$, $t\ge 0$
of $X$, and write for brevity $\X_{\cdot} = \left(X_{\cdot}^{i}\right)_{1 \le i\le N}$. 
Then $\lev$ can be estimated by 
\[
\hat{\lev}_{N}\left(K,T;\X_T\right)=\frac{{\loc}\left(K,T\right)}{\sqrt{\hat{p}_{N}\left(K,T;\X_T\right)}}\,,
\]
with
\begin{eqnarray}
\label{proj1d}
\hat{p}_{N}\left(K,T;\X_T\right)&=&\frac{\frac{1}{N}\sum_{i=1}^{N}V_{T}^{i}\delta_{N}^S\left(S_{T}^{i}-K,T\right)+2\theta\xi\epsilon}{\frac{1}{N}\sum_{i=1}^{N}\delta^S_{N}\left(S_{T}^{i}-K,T\right)+\xi\epsilon},
\end{eqnarray}
with $\delta^S_{N}$ a one-dimensional kernel function, specifically,
\begin{eqnarray}
\label{kernelDef1D}
\delta_{N}^S\left(x,T\right) & = & \frac{1}{\sqrt{2\pi} h_{x}\left(T\right)} \text{exp}\left(-\frac{1}{2}\frac{x^2}{h_x^2\left(T\right)}\right),
\end{eqnarray}
where $h_x$ as well as the specific bandwidth details, constructed heuristically, are given in Appendix \ref{subsec:kernel-construction}.
In our tests we pick $\epsilon=10^{-4}$.

The ($2\times N$)-dimensional SDE approximating the system $\X$ 
is in the case of the LSV model
\begin{eqnarray*}
\begin{cases}
\cfrac{d\hat{S}_{t}^{i}}{\hat{S}_{t}^{i}}=\left(\rdt-\rft\right)\,dt+\hat{\lev}_{N}\!\left(\hat{S}_{t}^{i},t;\X_t \right)\sqrt{V_{t}^{i}}\,dW_{t}^{i}\\
dV_{t}^{i}=\kappa\left(\theta-V_{t}^{i}\right)\,dt+\xi\sqrt{V_{t}^{i}}\,dW_{t}^{V,i}\,, \\
\end{cases}
\end{eqnarray*}
where $(W_{\cdot}^{i},W_{\cdot}^{V,i})_{1\le i\le N}$
are $N$ independent samples of the two correlated driving Brownian motions.
For the LSV-LVV model, $\xi \equiv \xi(S_t^i,t)$, where the function $\xi$ is assumed as given for now.

For application of the forward PIDE (\ref{eq:Volettera-Type-PIDE}), we also require the Markovian projection (\ref{proj}), and we estimate this again as
\begin{eqnarray*}
\hat{p}_{N}\left(K,B,T; \X_T \right)=\frac{\frac{1}{N}\sum_{i=1}^{N}V_{T}^{i}\delta_{N}\left(S_{T}^{i}-K,\,M_{T}^{i}-B,T\right)+2\theta\xi\epsilon}{\frac{1}{N}\sum_{i=1}^{N}\delta_{N}\left(S_{T}^{i}-K,\,M_{T}^{i}-B,T\right)+\xi\epsilon},
\end{eqnarray*}
with $\delta_{N}$ 
an anisotropic bi-variate Gaussian kernel as earlier and bandwidth details given in Appendix \ref{subsec:kernel-construction}.

\subsection{Parametrisation of $\xi$ \label{sub:xi-param}}

We recall that the available data is described in Section \ref{sub:Market-Data} as they inform the parametric form of $\xi$.
Given the scarcity of the data, and to avoid over-fitting, we will consider two simple parametric vol-of-vol functions: one which is constant in the spot variable and piecewise constant in time,
and one which is linear for a range of spot values (but capped
above and below, i.e., piecewise linear in the spot) and constant in time between quoted maturities.
More precisely, we write
\[
\begin{cases}
\xi\left(S,T\right)= & \bar{\xi}\left(q_{0}\left(S\right),T\right) \\
\bar{\xi}\left(S,T\right)= & \max\left(\left(a_{n+1}\left(S-S_{0}\right)+b_{n+1}\right),\xi_{\text{low}}\right),\quad
T \in \left[T_{n},T_{n+1}\right),
\end{cases}
\]
where we set $T_{0}=0$, $\xi_{\text{low}}=0.01$ and with $q_{0}$ 
defined as in (\ref{eq:h_x smooth}).
The construction performs a smooth, asymptotically constant extrapolation of $\bar{\xi}$ outside the
interval $\left[S_{0},B_{\max}\right]$,
where $B_{\max}$ is the largest quoted no-touch barrier for the last quoted maturity $T_{N_{\text{Mat}}}$.

Note that there are (only) two parameters per maturity, compared to one global vol-of-vol parameter
for the Heston model, and one parameter per maturity for the purely time-dependent case.
For the following discussion of the calibration, we focus on the piecewise  linear example as the other one is a special case.

\subsection{Overall calibration methodology}
\label{subsec:overall}

First, we calibrate a pure Heston model to vanilla options only 
by minimising
the mean square error for the difference between market and Heston
implied volatilities. For research purposes only, we use a Basin-Hopping
global optimisation \cite{Basin-Hopping1997}\footnote{For equity options, a variance swap-based calibration may be preferable
 (see \cite{Guillaume2014} for details).}.

As shown by 
the results in \cite{Clark2010}, the Heston-type LSV model with
the Heston parameters calibrated to vanilla options overestimates the
no-touch prices. A rule-of-thumb \cite{Clark2010} suggests that dividing $\xi$
calibrated to vanillas
by two, i.e.\ taking the so-called ``mixing factor'' in (\ref{eq:LSV-mixing=00003D0.55})
to be $\beta=0.5$, provides a good starting point to fitting no-touch options; see also the beginning of Section \ref{sec:LSV local Xi calibration}.
The parameters used are thus as displayed in Table \ref{tab: heston params} where the vol-of-vol calibrated to vanilla options, has been scaled by $0.5$ and set as $\xi$.

\begin{table}[h]
\begin{centering}
\caption{The calibrated Heston parameters.}
\label{tab: heston params} 
\par\end{centering}

\centering{}%
\begin{tabular}{ccccc}
\toprule 
$v_{0}$ & $\theta$ & $\kappa$ & $\rho$ & $\xi$\tabularnewline
\midrule

0.00827 & 0.01564 & 0.7147 & -0.4429 & 0.0947\tabularnewline
\bottomrule
\end{tabular}
\end{table}

A local volatility function $\loc$ is calibrated with the procedure presented in the Appendix of \cite{CMR2016}, but any
stable method, e.g., based on Dupire's formula or a regularisation approach would be adequate.
We then minimise an error measure $\bar{e}$ over the parameters $a$ and $b$ for the fit to no-touch quotes with the model's local volatility component $\sigma$ chosen to accurately fit vanilla options.
To this end, we use the particle method from Section \ref{subsec:particle} for the calibration to vanillas with ``outer'' iterations over the parameters $a$ and $b$ for
the best-fit to no-touch quotes.
For a proof of concept, we use
the Nelder--Mead gradient free optimisation algorithm \cite{NelderMead1965}, for which sufficient convergence
is obtained in 10 to 20 iterations in our tests. 
There is clearly room for improvement by a faster optimisation procedure.


\begin{rem*}[Mixing factor]
Applying this approach to (\ref{eq:LSV-mixing=00003D0.55}),
i.e.\ best-fitting the (constant) mixing factor $\beta$ to no-touch options
with all other parameters best-fitted to vanilla options (see Table \ref{tab: heston params}, and with $\xi=0.1894$),
the obtained mixing factor is $\beta=0.5528$. The associated
model will be denoted ``LSV mixing=0.55'' in the remainder of
the article.
\end{rem*}

The model prices of barrier options are computed by the PIDE (\ref{eq:Volettera-Type-PIDE}), where the local maximum mimicking volatility of the LSV model is estimated by particle method as (\ref{proj2d}).
We refer to Section \ref{sec:NumericalPIDE} for the details of the finite difference solution of the PIDE.

We briefly contrast this approach against two alternatives, namely a purely PDE-based approach and a purely Monte Carlo-based approach.
For the present two-factor ($S$ and $V$),
three-state ($S$, $V$ and $M$) model, it would be possible to
numerically solve the Kolmogorov forward PDE 
for $\phi$, the joint density of the spot, stochastic variance and running-maximum, and compute the Markovian projection by quadrature as
\[
\EQd\left[V_{T}\,|\,S_{T}=K,\,M_{T}=B\right]=\frac{\int_{0}^{\infty}v\phi\left(K,B,v,T\right)\,dv}{\int_{0}^{\infty}\phi\left(K,B,v,T\right)\,dv}\,.
\]
The main reason why we estimate the conditional expectation by a particle method 
is that it allows a more straightforward extension to
higher-dimensional models, such as those with stochastic
rates (or stochastic vol-of-vol or stochastic correlation).

As discussed in the introduction of Section \ref{sec:LSV local Xi calibration}, another possible approach is
to compute the no-touch prices directly with the simulated
paths and perform the optimisation process.
The particle estimator to compute (\ref{proj1d}) is then still needed to compute the local volatility function $\lev$. 
We will refer to this approach as ``full
Monte Carlo'' as the forward PIDE is not required anymore. 
%
We denote by $T_{n}^{\text{Mat}}$, $1\leq n\leq N_{\text{Mat}}$,
the quoted maturities, by
$T_{m}$, $m\leq N_{T}$, the time grid, which is constructed to contain
all quoted maturities $T_{n}^{\text{Mat}}$, and by $N$ the number of particles.
The step-by-step calibration is detailed in Algorithm \ref{alg:ParticleMethod-LSVLocalXi} in Appendix \ref{app:algo}.
For ``pure Monte Carlo'', one can remove
line \ref{algoline: mprojsm} and line \ref{algoline: pide step}
of Algorithm \algref{ParticleMethod-LSVLocalXi} and replace line
\algolineref{ compute price pide} by ``compute model foreign no-touch
price $\mathrm{FNT}^{\text{Model}}$ for maturity $T_{n}^{\text{Mat}}$ from the simulated
Monte Carlo particles''.

\subsection{Performance}

We carry out three ``full Monte Carlo" calibrations as detailed at the end of Section \ref{subsec:overall},
with $100\,000$ and $500\,000$ particles, both
with $100$ time steps per year, and one with 1~000~000 particles and 300
time steps per year, and one calibration using the forward
PIDE with $100\,000$ particles, 100 time steps per year and 200 strike
points. We pick $\epsilon=10^{-4}$ in (\ref{proj1d}) and (\ref{proj2d}).
We found $15\times10$ spline nodes for the estimation of $\sigma_{\text{LMV}}$
to provide a good trade-off between accuracy and smoothness.
While having $N_{S}$ and $N_{M}$ too
small will lead to accuracy problems, choosing them too large will make
the surface rougher due to over-fitting.

The error measure we use for this comparison is the relative
error for one-touch prices,
\begin{eqnarray*}
rel & = & 100\times\sum_{l=1}^{Q_{B}}\left|\frac{\mathrm{FOT}^{\text{Model}}\left(B_{T_{n}^{\text{Mat}},l},T_{n}^{\text{Mat}}\right)-\mathrm{FOT}^{\text{Market}}\left(B_{T_{n}^{\text{Mat}},l},T_{n}^{\text{Mat}}\right)}{\mathrm{FOT}^{\text{Market}}\left(B_{T_{n}^{\text{Mat}},l},T_{n}^{\text{Mat}}\right)}\right|,
\end{eqnarray*}
where the foreign one-touch price $\mathrm{FOT}$ can be computed from the
foreign no-touch price $\mathrm{FNT}$ with (\ref{eq:foreign no touch definition}).
A relative error is best suited in order to compare numerical methods for different levels of barriers.\footnote{From a practitioner perspective, the absolute difference is more relevant. For calibration results expressed in terms of absolute difference, see Section \ref{sec:Calibration-results}.}

Let $\tau_\text{\text{PIDE}}$
be the computational time needed to perform the PIDE calibration with
$100\,000$ particles, 200 strike space points and 100 time steps
per year.
If we denote by $\tau_\text{MC}$ the timing for a full Monte Carlo
calibration, 
the ``timing factor'' $\Theta$ is defined as
\[
\Theta=\frac{\tau_\text{MC}}{\tau_\text{PIDE}}\,.
\]
In Table \ref{tab:Full-Monte-Carlo-vs-PIDE}, we display the average
relative error over all barrier levels and maturities as well as the
relative computational times $\Theta$ with respect to the PIDE approach.
Table \ref{tab:Full-Monte-Carlo-vs-PIDE}
allows us to conclude that the full Monte Carlo calibration error becomes
comparable to the PIDE calibration error only with more than $1\,000\,000$
particles and $300$ time steps per year. This makes the full Monte Carlo method ten times slower than the PIDE approach.

\begin{table}
\begin{centering}
\begin{tabular}{ccccc}
\toprule 
 & PIDE $N=100\,000$ & MC $N=100\,000$ & MC $N=500\,000$ & MC $N=1\,000\,000$\tabularnewline
 & $N_{K}=200$, $N_{T}=100$ & $N_{T}=100$ & $N_{T}=100$ & $N_{T}=300$\tabularnewline
\midrule
Average relative error & 0.474\% & 1.437\% & 0.797\% & 0.621\%\tabularnewline
Timing factor $\Theta$ & 1 & 0.3 & 1.5 & 10.6\tabularnewline
\bottomrule
\end{tabular}
\par\end{centering}

\caption{Average relative error for one-touch prices in comparison Full Monte Carlo vs PIDE.
``Timing factor'' is the time needed for the calibration, relative
to the PIDE approach computational time. $N_{T}$ is the number of
time steps per year, $N$ the number of particles and $N_{K}$ the
number of strike points in the PIDE scheme.\label{tab:Full-Monte-Carlo-vs-PIDE}}

\end{table}

According to our numerical experiments, the two-dimensional particle
method to compute $\lev$ and $\sigma_{\text{LMV}}$
takes approximately
two to three times as long as the one-dimensional particle method
used for vanilla calibration, i.e.\ the computation of $\lev$
alone (including the computational time for the particle scheme evolution).
Additionally, we also need to solve the forward PIDE at each time
step, a task that has a comparable computational time as solving the
two-dimensional Heston pricing PDE.

The results plotted in Figure \ref{fig:Calibration error LSVLocalXI PIDE vs MC}
show the relative error for one short (left) and one long maturity (right), where
we notice that the full Monte Carlo approach, even with $1\,000\,000$
particles and 300 times steps, has larger relative error for the longer maturity.
Monte Carlo pricing of barrier options is numerically challenging
as it translates into integrating a discontinuous payoff function,
a problem that becomes more pronounced for higher levels of no-touch
barriers as only a few particles will breach the barrier. Estimating $\sigma_{\text{LMV}}$
and pricing with the forward PIDE does not suffer from this problem in the same way
as the knock-out feature is simply treated as a Dirichlet boundary condition.

Additionally, we plot
the calibrated local vol-of-vol functions for both the forward PIDE
approach with $100\,000$ particles in Figure \ref{fig:local xi PIDEvsMC PIDE}
and the full Monte Carlo approach with $500\,000$ particles in Figure
\ref{fig:local xi PIDEvsMC MC}, both with 100 time steps per year,
and notice that the use of the forward PIDE leads to more stable parameters.

Hence, we conclude that combining the forward PIDE with
the particle method provides a more efficient solution for the calibration
problem compared to the full Monte Carlo technique, as seen by comparing to the ``fully'' converged
surface in Figure \ref{fig:LSV--localXi Xi} with more points and particles. However, the full Monte Carlo approach is a good alternative if one is not willing to implement
the finite difference discretisation of the forward PIDE.
A key benefit of the full Monte Carlo calibration is that one calibrates exactly the model simulated for pricing. Additionally, the calibration can be done at the same time as the pricing. Thus, the accuracy of the numerical discretisation of the model becomes less of a concern. The calibrated model is therefore the chosen discretisation scheme. This  means, however, that it becomes important to perform tests on the implementation to be sure that the discretisation has properties close to the desired model, for example by pricing moments of the variance process.
Additionally, if one wishes to price by PDE methods, a Monte Carlo calibration becomes far less suitable.

Finally, we plot in Figure \ref{fig:LSV--localXi Leverage _} the
calibrated local volatility function $\lev$ and in Figure \ref{fig:LSV--localXi Xi}
the calibrated local vol-of-vol $\xi$ from 3 months onward, obtained
with $100$ time steps per year, $500\,000$ particles and $900$
strike points for the PIDE. For these numerical parameters, the calibration error on the implied volatility is on average smaller than 2bps in absolute volatility.
The fit to market data, especially regarding no-touch options, is discussed in detail in Section \ref{sec:Calibration-results}.

In our tests, $a_{n}$
is negative and lies inside $\left[-1.5,0\right]$ and $b_{n}$ is
usually in the interval $\left[\xiH /2,3 \, \xiH/2\right]$,
where $\xiH$ is the vol-of-vol of a pure Heston model calibrated
to vanilla prices.

As seen in Figure \ref{fig:LSV--localXi Xi} from the resulting shape of $\xi$,
both parameters are stable from one maturity to
the next and make thus a good first guess for the next quoted pillar.
Hence, for the shortest quoted maturity, we start
the optimisation with $a_{1}=-1$ and $b_{1}=\xiH$ and then
use the calibrated $\left(a_{n},b_{n}\right)$ of $T_{n}^{\text{Mat}}$ as
a first guess for the iterative solver in the calibration to no-touch quotes at $T_{n+1}^{\text{Mat}}$.

\begin{figure}[h]
\begin{minipage}[c]{1\columnwidth}%
\begin{minipage}[c]{0.48\columnwidth}%
\begin{center}
\includegraphics[scale=0.37]{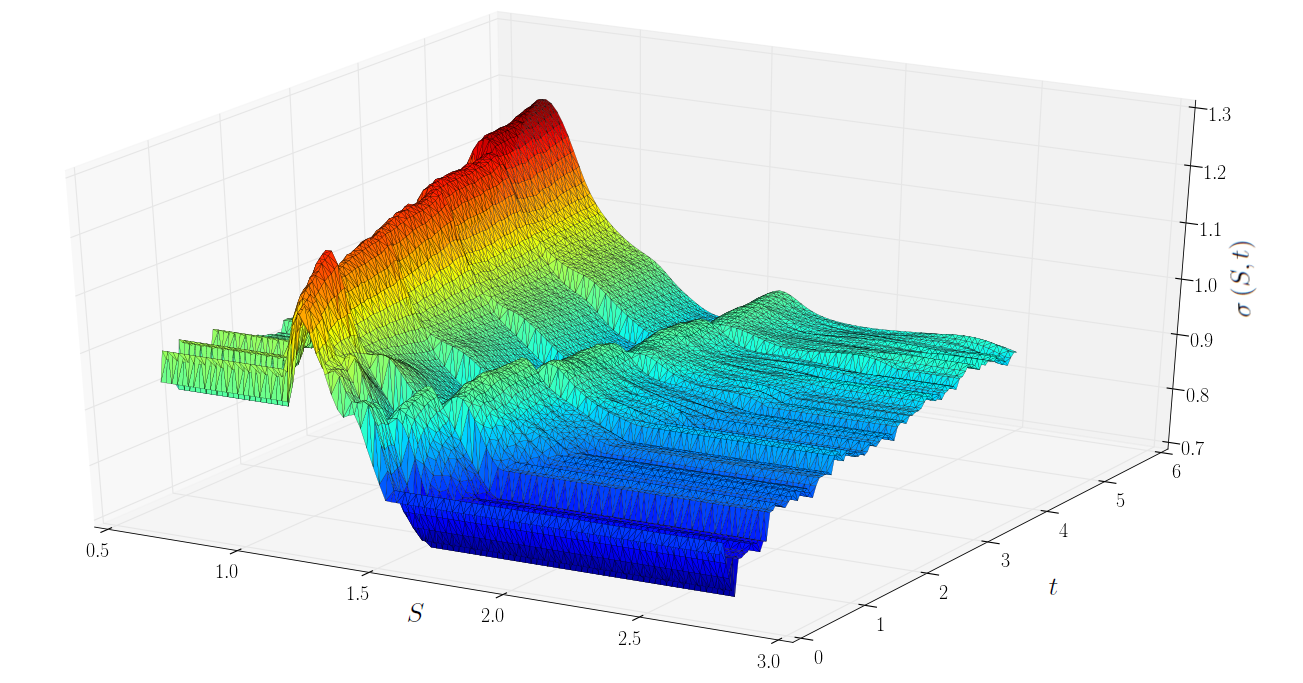}{\captionof{figure}{Calibrated local volatility function $\lev$ for the LSV-LVV model (\ref{eq:Heston LSV model}).}\label{fig:LSV--localXi Leverage _}}
\par\end{center}
\end{minipage}~~~~~~~~~~~~%
\begin{minipage}[c]{0.48\columnwidth}%
\begin{center}
\includegraphics[scale=0.37]{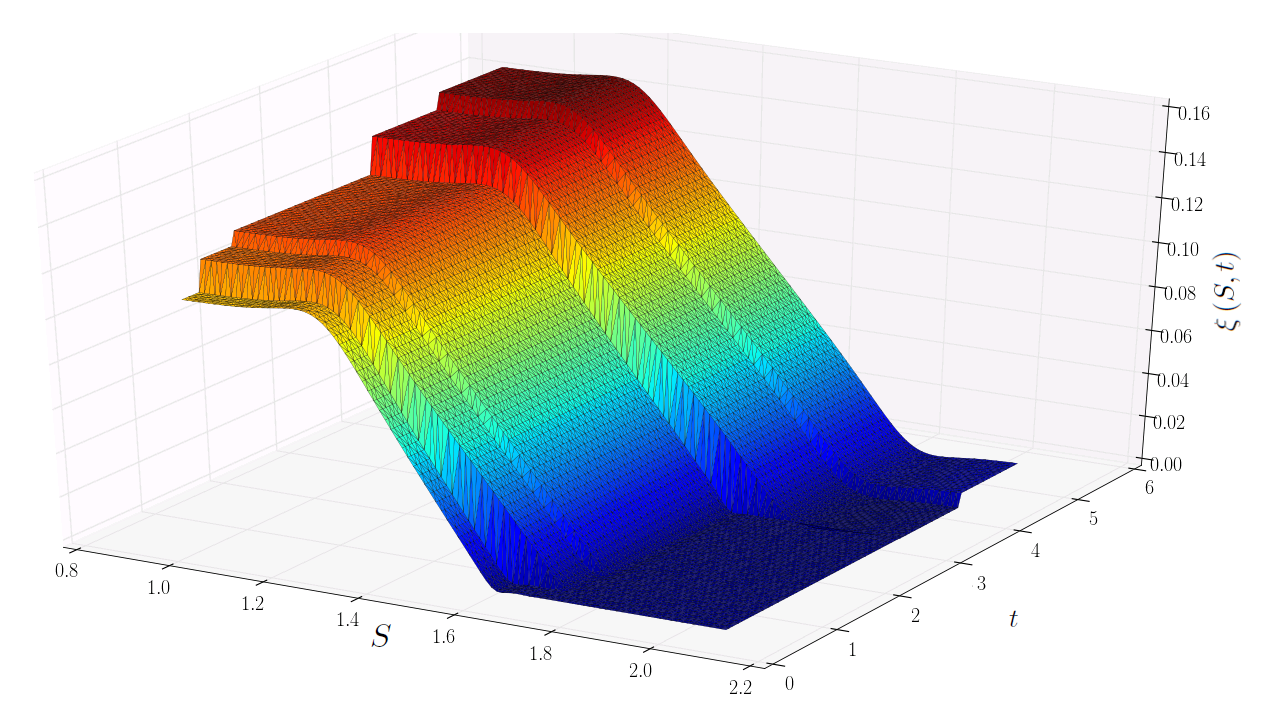}{\captionof{figure}{Calibrated local vol-of-vol $\xi$ for the LSV-LVV model (\ref{eq:Heston LSV model}).}\label{fig:LSV--localXi Xi}}
\par\end{center}
\end{minipage}%
\end{minipage}
\end{figure}


%




\begin{figure}[h]
\begin{centering}
\includegraphics[scale=0.3]{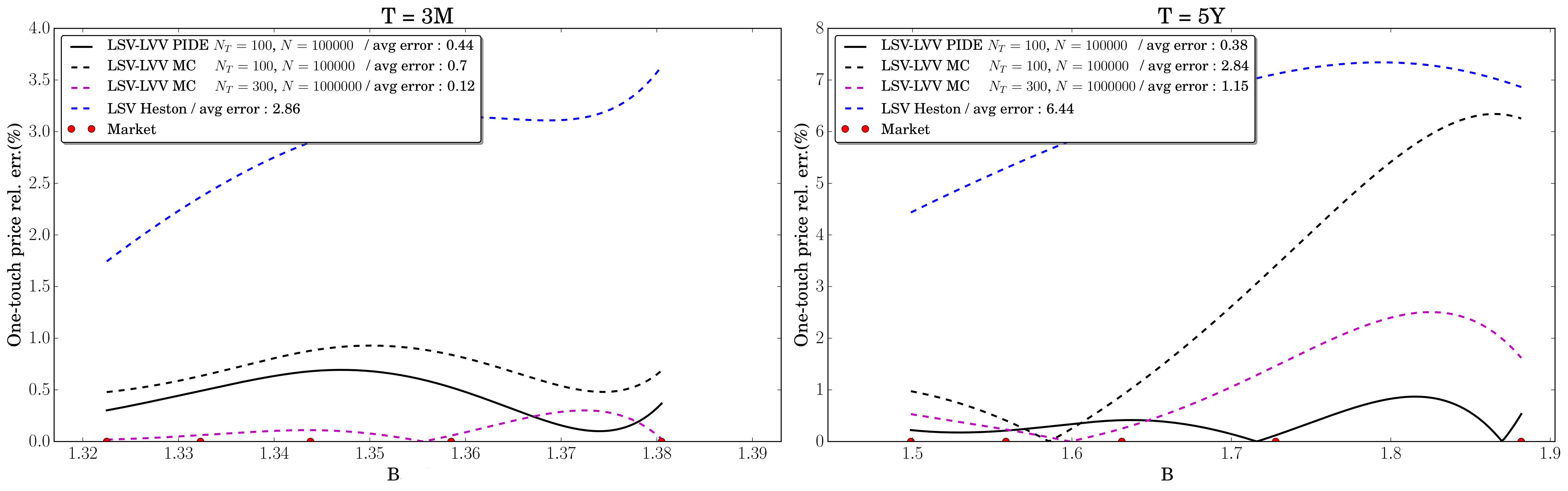}
\par\end{centering}

\caption{Full Monte Carlo vs forward PIDE calibration comparison for a short-
and long-term maturity pillar, $N_{T}$ is the number of time steps
per year and $N$ the number of particles. 
\label{fig:Calibration error LSVLocalXI PIDE vs MC}}
\end{figure}

\selectlanguage{british}%
\begin{figure}[h]
\begin{minipage}[c]{1\columnwidth}%
\begin{minipage}[c]{0.43\columnwidth}%
\selectlanguage{english}%
\begin{center}
\includegraphics[scale=0.39]{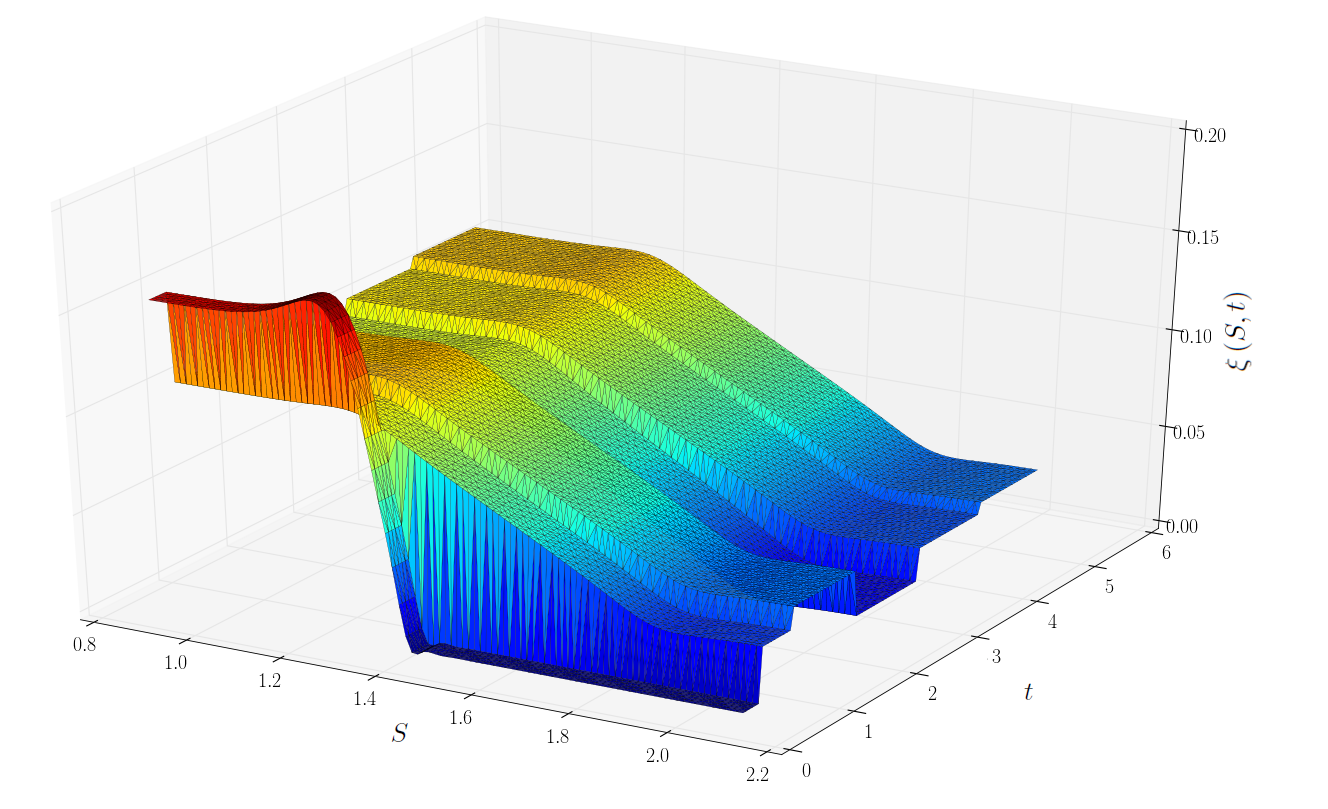}\foreignlanguage{british}{\captionof{figure}{Local vol-of-vol $\xi\left(S,t\right)$ function calibrated by forward PIDE 200 strike points, 100 000 particles and 100 time steps per year. }\label{fig:local xi PIDEvsMC PIDE}}
\par\end{center}\selectlanguage{english}%
\end{minipage}~~~~~~~%
\begin{minipage}[c]{0.48\columnwidth}%
\selectlanguage{english}%
\begin{center}
\includegraphics[scale=0.39]{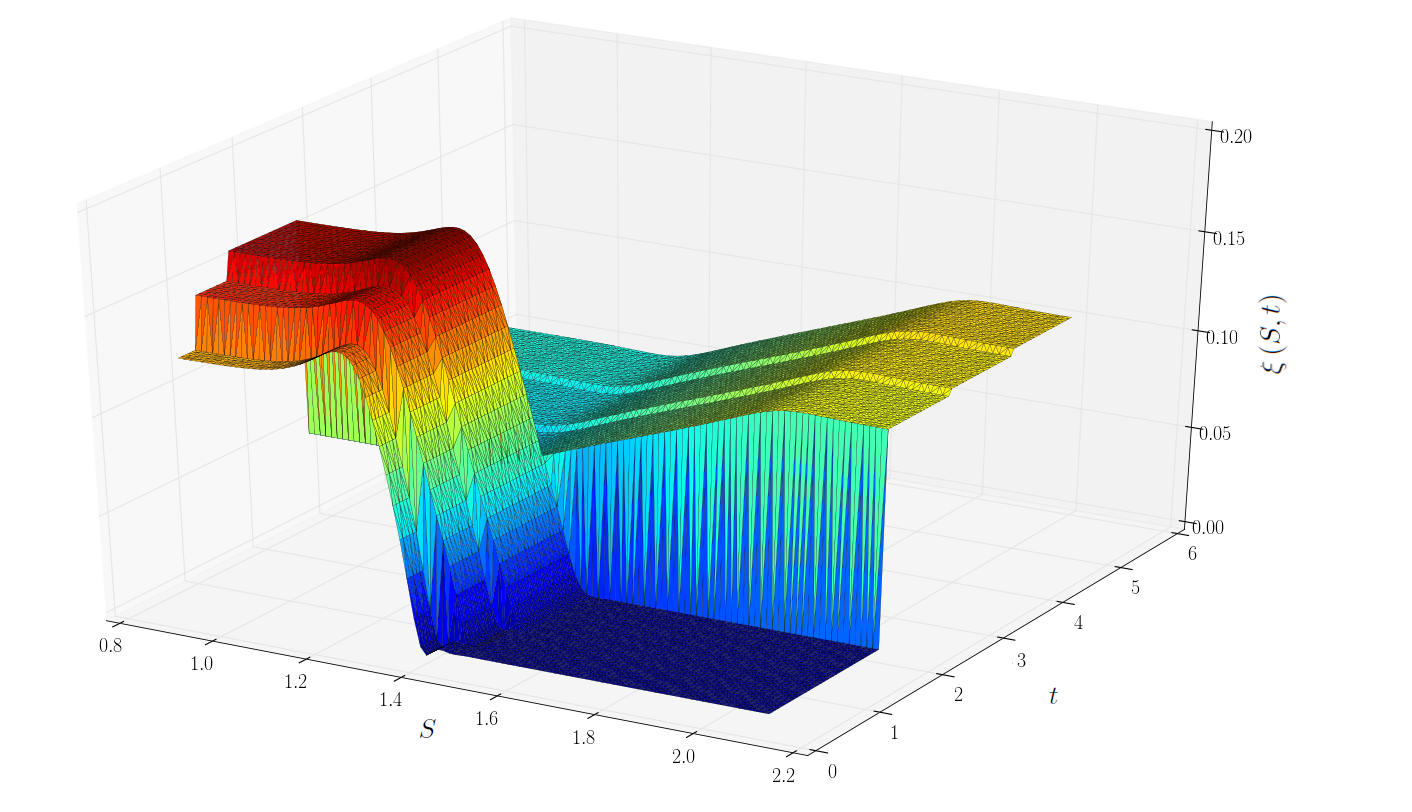}\foreignlanguage{british}{\captionof{figure}{Local vol-of-vol function $\xi\left(S,t\right)$ calibrated by full  Monte Carlo, 500 000 particles and 100 time steps per year.}\label{fig:local xi PIDEvsMC MC}}
\par\end{center}\selectlanguage{english}%
\end{minipage}%
\end{minipage}
\end{figure}

\selectlanguage{english}%

\section{Calibration results and model comparison \label{sec:Calibration-results}}

In this section, we present the calibration fit for all models in this paper, benchmarked against some widely used models.
The
calibration error for
vanilla options is given in Table \ref{tab:Error Vanilla},
and for foreign no-touch options in Table \ref{tab:Error No Touch},
for both path-dependent models, i.e.\ the LMV model (\ref{eq:modelBS})
and LMSV model (\ref{eq:LMSV model}), as well as the LSV-LVV model
(\ref{eq:Heston LSV model}) and the standard LSV model (\ref{eq:LSV-mixing=00003D0.55}) with mixing factor $\beta$.
As a benchmark, we also include the pure local volatility model, i.e. (\ref{eq:LSV-mixing=00003D0.55}) with $\beta=0$, and
the LSV Heston model, i.e. (\ref{eq:LSV-mixing=00003D0.55}) with $\beta=1$, calibrated to vanilla options and where the Heston parameters are also calibrated to call options.

The average
error in absolute implied volatility for vanillas is $0.005\%$ for the LMV model,
$0.009\%$ for the LMSV model and $0.014\%$ for the LSV-LVV model. 

\begin{table}[h]
	\begin{centering}
		\begin{tabular}{c|cccccc}
			\toprule 
			\textbf{T} & \textbf{~~~~LMV~~~~} & \textbf{~~~~LMSV~~~~} & \textbf{~~~LSV-LVV~~~} & \textbf{LSV mix.=0.55} & \textbf{~~~~LV~~~~} & \textbf{LSV Heston}\tabularnewline
			\midrule 
			\textbf{0.26} & 0.001 & 0.008 & 0.017 & 0.055 & 0.001 & 0.001\tabularnewline
			
			\textbf{0.51} & 0.003 & 0.007 & 0.014 & 0.040 & 0.001 & 0.006\tabularnewline
			 
			\textbf{1.01} & 0.006 & 0.005 & 0.016 & 0.027 & 0.001 & 0.005\tabularnewline
			 
			\textbf{2.01} & 0.006 & 0.005 & 0.016 & 0.016 & 0.000 & 0.003\tabularnewline
			 
			\textbf{3.01} & 0.005 & 0.007 & 0.014 & 0.011 & 0.000 & 0.003\tabularnewline
			
			\textbf{4.01} & 0.006 & 0.012 & 0.013 & 0.009 & 0.001 & 0.005\tabularnewline
			 
			\textbf{5} & 0.006 & 0.016 & 0.012 & 0.005 & 0.001 & 0.003\tabularnewline
			
			\bottomrule
		\end{tabular}
		\par\end{centering}
	
	\caption{Average absolute error of implied volatilities in \% for the vanilla options for all models. 
		\label{tab:Error Vanilla}}
\end{table}

For the no-touch prices, we display the average absolute error in
\% for each maturity as 
\[
e=\frac{1}{Q_{B}}\sum_{l=1}^{Q_{B}}\left|\mathrm{FNT}^{\text{Model}}\left(B_{T^{\text{Mat}},l},T^{\text{Mat}}\right)-\mathrm{FNT}^{\text{Market}}\left(B_{T^{\text{Mat}},l},T^{\text{Mat}}\right)\right|
\]
in Table \ref{tab:Error No Touch}, where $Q_{B}$ is the number of
quoted barriers and $B_{T^{\text{Mat}},l}$, $1\leq l\leq Q_{B}$,
the set of quoted barriers (e.g., for an average absolute error
of $0.1\%$ and for a market no-touch probability of $70\%$, the
model will price it at $70\%\pm0.1\%$ on average). The fit for the
two path-dependent models should theoretically be perfect for all
vanilla and barrier contracts, and any mismatches consist in numerical errors
and penalisation, while the LSV-LVV is by construction calibrated
to vanilla options but has only two further free parameters per maturity
to fit five no-touch prices.

In Figure \ref{fig:Calibration-fit-NoTouch}, we plot as a function of $B$, for fixed $T$, the error
\[
e\left(B,T\right)=\left(\mathrm{FNT}^{\text{Model}}\left(B,T\right)-\mathrm{FNT}^{\text{Market}}\left(B,T\right)\right)\,.
\]

All the model prices are computed with
$1\,000\,000$ Monte Carlo paths and $365$ time steps per year, where
Brownian increments are generated with Sobol sequences and Brownian
bridge construction \cite{SobolBB1997}. The running maximum is sampled with the Brownian bridge
technique as in Chapter 6 of \cite{Glasserman2004}. We note that to reach faster convergence for pricing, Sobol sequences and Brownian bridge construction can be used naturally as each path is simulated independently. 

\begin{table}[h]
\begin{centering}
\begin{tabular}{c|cccc|cc}
\toprule 
\textbf{T} & \textbf{~~~~LMV~~~~} & \textbf{~~~~LMSV~~~~} & \textbf{~~~LSV-LVV~~~} & \textbf{LSV mix.=0.55} & \textbf{~~~~LV~~~~} & \textbf{LSV Heston}\tabularnewline
\midrule 
\textbf{0.26} & 0.012 & 0.060 & 0.079 & 0.834 & 1.198 & 0.706\tabularnewline
\textbf{0.51} & 0.008 & 0.043 & 0.140 & 0.675 & 1.455 & 1.454\tabularnewline
\textbf{1.01} & 0.028 & 0.047 & 0.120 & 0.435 & 1.605 & 1.894\tabularnewline
\textbf{2.01} & 0.029 & 0.055 & 0.157 & 0.123 & 1.590 & 2.097\tabularnewline
\textbf{3.01} & 0.024 & 0.043 & 0.148 & 0.124 & 1.439 & 2.186\tabularnewline
\textbf{4.01} & 0.026 & 0.055 & 0.122 & 0.147 & 1.420 & 2.062\tabularnewline
\textbf{5} & 0.019 & 0.063 & 0.166 & 0.187 & 1.380 & 1.960\tabularnewline
\bottomrule
\end{tabular}
\par\end{centering}

\caption{Average absolute error in \% for foreign no-touch quotes for all models;
see Fig.~\ref{fig:Calibration-fit-NoTouch} for error plots. On the left hand-side are models calibrated to no-touches (LMV, LMSV, LSV-LVV); on the right-hand side models not calibrated on no-touches (LSV mix.=0.55, LV, LSV Heston).
\label{tab:Error No Touch}}
\end{table}

%

\begin{figure}[h]
\begin{centering}
\includegraphics[scale=0.3]{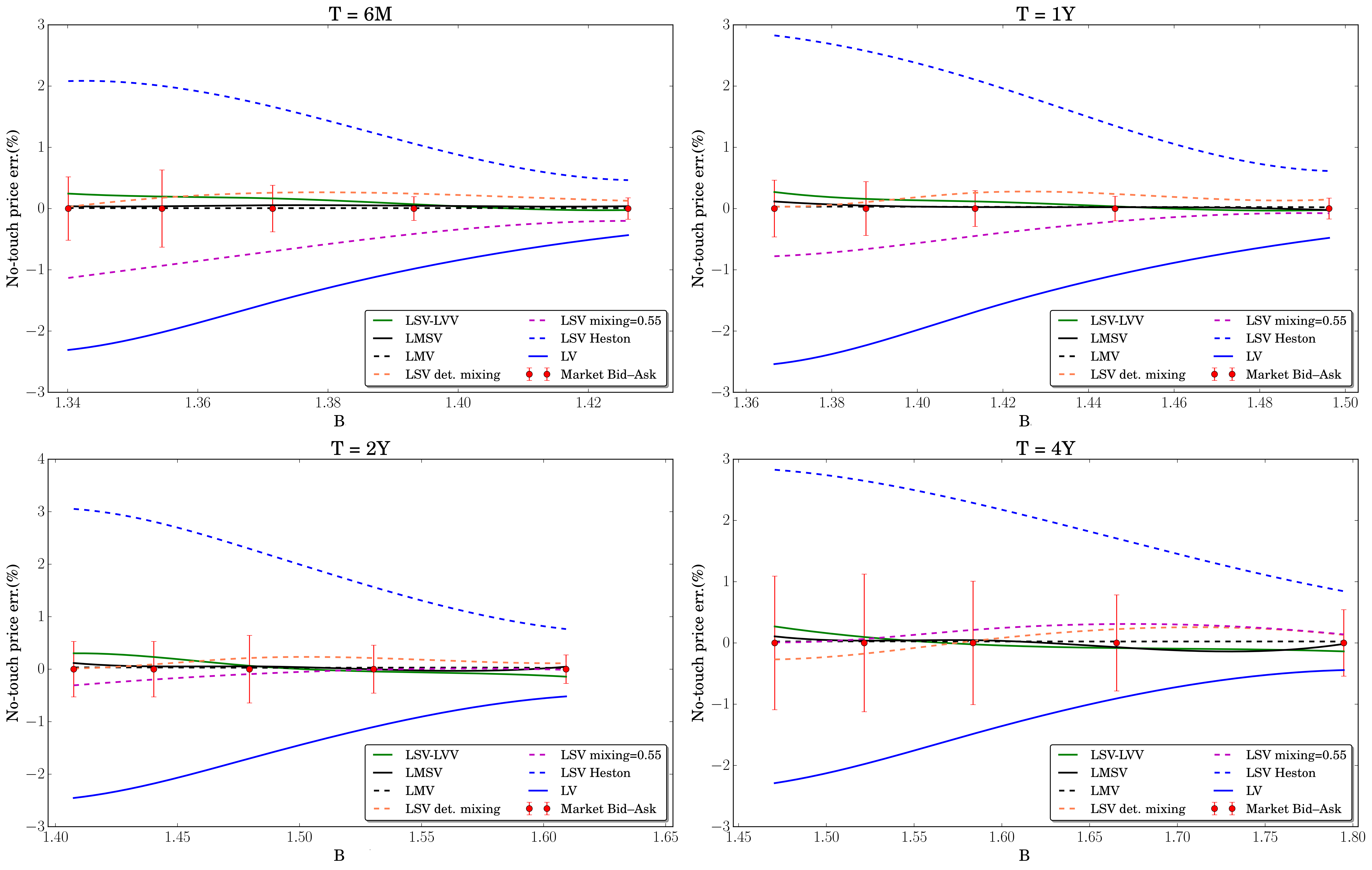}
\par\end{centering}

\caption{Calibration fit to foreign no-touch options for all models, as a function of barrier level $B$;
see Table \ref{tab:Error No Touch} for tabulated errors. \label{fig:Calibration-fit-NoTouch}}
\end{figure}

As Figure \ref{fig:Calibration-fit-NoTouch}
suggests, if no-touch options are not included in the set of calibration
instruments, more classical models like the Heston LSV model can largely mis-price
the no-touch probability (in fact, a mis-pricing significantly higher than $3\%$ is common).
These results show that the
calibration of no-touch options is of paramount importance in order
to incorporate the information about the distribution of the running
maximum process provided by the market.

The LSV-LVV, LMV and LMSV models, calibrated
to both vanilla and foreign no-touch options, perform significantly
better for the valuation of no-touch options than the LV or the Heston
LSV model calibrated to vanilla options only. The inclusion of a constant mixing factor improves the fit for longer maturities, but still does not allow calibration within the bid--ask spread.
This is almost achieved by a time-dependent mixing factor, and fully achieved with a time-dependent vol-of-vol which is also a linear function of the spot FX rate (LSV-LVV model of Section \ref{sec:LSV local Xi calibration}).


\section{Conclusion\label{sec:Conclusion}}

In this work, we demonstrated on the example of three volatility models the calibration to two traded
product classes, namely, vanilla and no-touch options. 

We introduced a new LSV-LVV model, an extension of the classic
Heston-type LSV model with a local vol-of-vol.
Due to the small number of degrees-of-freedom, 
the chosen LSV-LVV parametrisation cannot
match no-touch mid-prices perfectly, however, the fit is very satisfactory
as the model price lies well within the market spread across all quoted
barrier levels and maturities.

We also studied a model based directly on the  local maximum mimicking diffusion
as the natural extension to the Dupire local volatility
framework, namely the LMV model.
Then, the addition of a Heston-type stochastic volatility on top of the maximum-dependent volatility
leads to a new LMSV model with
a potentially more interesting spot-vol dynamics.

Two approaches were proposed for the calibration; one based a two-dimensional particle method to compute the Markovian projection onto the two-dimensional state space $\left(S,M\right)$, and the other using the numerical solution of a forward PIDE for barrier option prices.

An interesting extension will be to compare the volatility dynamics
implied by the three models through the pricing of forward start options.
One would then be able to understand to which extent the calibration
to touch options, and barriers in general, is compatible with the market
smile dynamics.

\section*{Acknowledgements}
Simon McNamara (now at \textsc{UBS London}) and the first author originally proposed the LSV-LVV model in workshop sessions held at \textsc{BNP Paribas London}.
The authors thank Marek Musiela from the \textsc{Oxford-Man Institute} 
for insightful comments.

\appendix

\section{Implementation details}

\subsection{Construction of kernel for particle method \label{subsec:kernel-construction}}

The bandwidth of the one-dimensional Gaussian kernel in (\ref{kernelDef1D}) is given by
\begin{eqnarray}
\label{eqn:hx}
h_{x}\left(T\right) & = & \eta S_{0}{ \loc\left(S_{0},T\right)} \sqrt{\max\left(T,T_{\min}\right)}N^{-\frac{1}{6}},
\end{eqnarray}
where $\loc\left(S_{0},T\right)$ is replaced by $\locM\left(S_{0},S_0,T\right)$ for the LMV model, and 
\[
\eta=1.5,\,\quad T_{\min}=0.25,\,N_{0}=180.
\]

In the two-dimensional case of (\ref{proj2d}), (\ref{kernelDef}),
\begin{eqnarray*}
h_{y}\left(T\right) & = & h_{x}(T) \\ 
\rho_{xy}\left(T\right) & = & \left(\rho_{\max}-\rhos\left(T\right)\right)\text{\,exp}\left(-k\frac{N_{T}}{T_{\max}}\right)+\rhos\left(T\right)\\
k & = & \frac{2T_{\max}\ln\left(2\right)}{N_{0}},
\end{eqnarray*}
where $T_{\max}$ is the last quoted maturity, $\rhos \left(T\right)$
the correlation 
between $S_T$ and $M_T$ estimated 
with the sampled particles at time $T$, and $\rho_{\max}=0.98$.
To speed up the computation, it is enough to update $\rhos\left(T\right)$
once every year or half-year. The use of an appropriate bandwidth
was found to have a significant impact on the accuracy of the method. This bandwidth
is inspired by a Silverman-type rule (see \cite{Silverman1986}, \cite{guyon2012being})
for the values of $h_{x}$ and $h_{y}$ and an experimental definition
of the correlation part

 A heuristic analysis suggests that if the number of time steps is low, the running maximum with the Brownian
bridge technique, as described in Chapter 6 of \cite{Glasserman2004}, tends to be underestimated in the important area where spot and its running maximum are
around the initial values $\left(S_{0},S_{0}\right)$, since, as seen in Figure \ref{fig:LMSV leverage},
the local volatility function is an increasing function of the running maximum
in this region. Indeed, between two time steps $t$ and $t+\Delta t$,
the Brownian bridge technique freezes the value of the volatility function $\sigma_{\text{LMV}}$ from (\ref{eq:LMSV model})for a
running maximum $M$ at time $t$ smaller than the value of $M$ along
$\left[t,t+\Delta t\right]$, and therefore, $\sigma_{\text{LMV}}$ is underestimated
on average such that $M_{t+\Delta t}$ computed by Brownian bridge
is smaller than its exact value.

The intuition behind the specification
of the correlation $\rho_{xy}\left(T\right)$ is as follows: if the
number of time steps is low, the running maximum will tend to be underestimated,
hence, we give more importance to the particles where the running
maximum is higher (which happens when the spot increases); if the
number of time steps is large enough, i.e.\ more than one per day,
the running maximum bias is lower and we rely on the sample correlation.
It is important to mention that in practice, $\rhos\left(T\right)$
is around $80\%\pm5\%$ and does not change significantly over time. Also, the closer
we get to the boundary $S=M$, the higher the correlation, which can
easily reach $95\%$. This is due to the fact that for particles where
$M$ is large, it is highly probable that $S$ is large as well, therefore,
most of the particles will gather around the diagonal boundary. Conversely,
if $M$ is around $S_{0}$, most particles are for spots going downwards.
This is easily observed from the $\left(S_{t},M_{t}\right)$ joint
density where the mass aggregates around the area $\left(S_{0},S_{0}\right)$,
the line $S=M$ and the line $M=S_{0}$ as displayed in Figure \ref{fig:SMJointCalibratedDensity1Y}
where we plot the density computed with a finite element method.


\subsection{Particle search on tree} \label{subsec:tree-search}

For a given level of spot and running maximum, $K$ and $B$, we only
want to compute the kernel function in (\ref{proj2d}) for particles which give a
significant contribution to the sum, i.e.\ particles close enough
with respect to the metric implied by $\delta_N$. We measure this by $\delta_{N}\left(x,y,T\right)\leq\epsilon_{0}$ for some $\epsilon_{0}>0$, i.e.\
all the points contained in the ellipse $\mathcal{E}$
defined by
\begin{eqnarray*}
x^{2}+y^{2}-2\rho_{xy}\left(T\right)xy-H\left(T\right) & = & 0\,,
\end{eqnarray*}
with
\[
H\left(T\right)=-2h_{x}^{2}\left(T\right)\ln\left(\gamma\left(T\right)\epsilon_{0}\right)\,.
\] where $\gamma$ is defined in (\ref{kernelDef}). 
The canonical form of ellipse $\mathcal{E}$ expressed in the coordinate
system defined by its principal axes is (see \cite{Conic1993})
\[
\frac{x^{2}}{-D/\left(\lambda_{1}^{2}\lambda_{2}\right)}+\frac{y^{2}}{-D/\left(\lambda_{1}\lambda_{2}^{2}\right)}=1\,,
\]
with $D=-H\left(T\right)\left(1-\rho_{xy}^{2}\left(T\right)\right)$,
$\lambda_{1}=1-\rho_{xy}\left(T\right)$ and $\lambda_{2}=1+\rho_{xy}\left(T\right)$, the
roots of $\lambda\rightarrow\lambda^{2}-2\lambda+\left(1-\rho_{xy}^{2}\left(T\right)\right)$.
Since $\rho_{xy}\left(T\right)\geq0$ , the semi-major axis length
is 
\[
R\left(T\right)=\sqrt{-D/\left(\lambda_{1}^{2}\lambda_{2}\right)}=\sqrt{\frac{H\left(T\right)}{1-\rho_{xy}\left(T\right)}}.
\]

Working for simplicity with the Euclidean distance,
the particles with a significant
contribution to the value of $\hat{p}_{N}\left(K,B,T\right)$ are contained
in the ball with center $\left(K,B\right)$ and radius $R\left(T\right)$.
We use $\epsilon_{0}=\left({\xi\epsilon}\right)/{10}$ in our tests. A large
value of $\epsilon_{0}$ leads to a fast computation of $\hat{p}_{N}$
but also reduces the accuracy of the result.

In order to perform a
distance query efficiently, we build a $k$-d tree as described in
\cite{Brown2015kdtree}, with a worst case complexity of $\mathcal{O}\left(2N\log N\right)$
and in which we can perform a binary search with complexity $\mathcal{O}\left(\log N\right)$
on average, to find all particles with distance less than $\epsilon_0$. 
This provides fast access to the nearest neighbours, and,
as a consequence, allows to list the particles contained in the ball
of centre $\left(K,B\right)$ and radius $R\left(T\right)$.


\begin{rem*}
	In order to further improve the performance, we could define a distance function 
	\[
	d\left(\mathbf{x}_{1},\mathbf{x}_{2}\right)=\sqrt{\left(s_{1}-s_{2}\right)^{2}+\left(m_{1}-m_{2}\right)^{2}-2\rho\left(s_{1}-s_{2}\right)\left(m_{1}-m_{2}\right)}\,,
	\]
	where the coordinates of a given particle $i$ are given by $\mathbf{x}_{i}=\left[s_{i},\,m_{i}\right]^{T}$,
	with $i \in \{1,2\}$ in this example. Then
	$d$ can be combined with a metric tree algorithm \cite{Yianilos93} for an
	optimal nearest neighbours search, by only keeping particles within
	a distance $H\left(T\right)$ to the point $\left(K,B\right)$ of
	interest. 
It is straightforward to check that $d$ is in fact a metric.
	
\end{rem*}

\subsection{Spline interpolation of the volatility surface} \label{subsec:spline-interpolation}

We describe here the spline parameterisation of the volatility surfaces for a given time.
Let $N_{T}$ be the number of time steps.
For a given time $T_{m}$, the function $\sig$ is approximated
on a rectangle $\left[S_{\min}^{m},\,S_{\max}^{m}\right]\times\left[S_{0},\,S_{\max}^{m}\right]$,
with $S_{\min}^{m} < S_0 < S_{\max}^{m}$, by
bi-variate quadratic
splines in the spot and running maximum directions (and piecewise constant in time),
and is extrapolated outside these bounds as detailed in Appendix \ref{sub:Brunick--Shreve-volatility-smooth}.
There are $N_{T}+1$ volatility ``slices'' in total such
that we denote the $m$-th time slice, i.e., $(x,y)\rightarrow\sig\left(x,y,T_{m}\right)$,
by $\sig_{m}$.

The surface construction
starts by defining a spot grid where we need more grid points around
the forward value and less around $S_{\min}^{m}$ and $S_{\max}^{m}$.
We then use a hyperbolic grid (see \cite{CMR2016}
for more details) refined around the forward value 
\[
F_{m}=S_{0}e^{\int_{0}^{T_{m}}\left(\rdt-\rft\right)dt}
\]
with 
\[
S_{\min}^{m}=F_{m}e^{-\frac{6}{2}{ \sigma_{F}\left(T_{m}\right)} \sqrt{T_{m}}},\quad 
S_{\max}^{m}=F_{m}e^{\frac{6}{2}{ \sigma_{F}\left(T_{m}\right)}\sqrt{T_{m}}}\,,
\] where $\sigma_{F}\left(T_{m}\right)$ is the at-the-money forward volatility of the market for maturity $T_{m}$.
The spot grid is denoted by $\left(S_{m,j}\right)_{m\leq N_{T},\,j\leq N_{S}}$.
The creation of the running maximum grid is done selecting the nodes
of the spot grid above the initial spot $S_{0}$, which leads to $N_{B}\left(m\right)\leq N_{S}$,
where $N_{B}$ is now time dependent 
This particular construction is crucial for accuracy as it ensures that
the diagonal where $S=M$ is part of the grid,
with the associated maximum grid points $\left(M_{m,k}\right)_{m\leq N_{T},\,k\leq N_{M}\left(m\right)}$.
Each of the grid values
can be seen as a parameter and we denote them by $\left(\sig_{m,j,k}\right)_{m\leq N_{T},\,j\leq N_{S},k\leq N_{M}\left(m\right)}$.

\subsection{Smooth volatility extrapolation\label{sub:Brunick--Shreve-volatility-smooth}}

Here, we describe how we extrapolate volatility functions smoothly to be asymptotically constant in the spatial coordinates from
a rectangle $\left[x_{\min},x_{\max}\right]\times\left[y_{\min},y_{\max}\right]$.
For $\sig(\cdot,\cdot, t)$ defined on $\left[x_{\min},x_{\max}\right]\times\left[y_{\min},y_{\max}\right]$, we first extend the function to $\mathbb{R}^2 $ by constant extrapolation,
\[
\sig\left(x,y,t\right)=\sig\left(\bar{q}_{0}\left(x\right),\bar{q}_{1}\left(y\right),t\right), \qquad (x,y) \in \mathbb{R}^2,
\]
with 
\begin{eqnarray*}
\bar{q}_{0}\left(x\right) & = & x_{\max}\mathbf{1}_{x\geq x_{\max}}+\mathbf{1}_{x<x_{\max}}\left[x_{\min}\mathbf{1}_{x\leq x_{\min}}+x\mathbf{1}_{x>x_{\min}}\right]\\
\bar{q}_{1}\left(y\right) & = & y_{\max}\mathbf{1}_{x\geq y_{\max}}+\mathbf{1}_{y<y_{\max}}\left[y_{\min}\mathbf{1}_{y\leq y_{\min}}+y\mathbf{1}_{y>y_{\min}}\right].
\end{eqnarray*}
From this, we define a linear extrapolation $\sig$ as 
\begin{eqnarray*}
\bar{\sig}\left(x,y,t\right) & = & \sig\left(\bar{q}_{0}\left(x\right),\bar{q}_{1}\left(y\right),t\right)\\
 & + & \mathbf{1}_{x>x_{\max}}\frac{\partial\sig\left(x_{\max},\bar{q}_{1}\left(y\right),t\right)}{\partial x}\left(x-x_{\max}\right)+\mathbf{1}_{x<x_{\min}}\frac{\partial\sig\left(x_{\min},\bar{q}_{1}\left(y\right),t\right)}{\partial x}\left(x-x_{\min}\right)\\
 & + & \mathbf{1}_{y>y_{\max}}\frac{\partial\sig\left(\bar{q}_{0}\left(x\right),y_{\max},t\right)}{\partial y}\left(y-y_{\max}\right)+\mathbf{1}_{y<y_{\min}}\frac{\partial\sig\left(\bar{q}_{0}\left(x\right),y_{\min},t\right)}{\partial y}\left(y-y_{\min}\right).
\end{eqnarray*}
We then introduce a smoothed transition of the coordinate $x$ at both
$x_{\max}$ and $x_{\min}$, 
\begin{equation}
q_{0}\left(x\right)=x_{\max}w\left(x,x_{\max},\eta_{0}\right)+\left(1-w\left(x,x_{\max},\eta_{0}\right)\right)\left[x_{\min}\left(1-w\left(x,x_{\min},-\eta_{0}\right)\right)+xw\left(x,x_{\min},-\eta_{0}\right)\right]\label{eq:h_x smooth}
\end{equation}
with
\[
\begin{cases}
w\left(x,x_{0},\eta_{0}\right) & =\frac{1}{2}(1+\tanh\left(\frac{2x_{0}}{\epsilon}\frac{\left(x-\bar{x}\left(\eta_{0}\right)\right)}{\bar{x}\left(\eta_{0}\right)}\right)\\
\bar{x}\left(\eta_{0}\right) & =\frac{2x_{0}^{2}}{2x_{0}+\epsilon \, \text{arctanh}\left(\eta_{0}(1-\frac{\epsilon}{2})\right)}\\
\epsilon & =\frac{S_{0}}{10}\,,
\end{cases}
\]
and similar for
$y$ at $y_{\max}$ and $y_{\min}$, which we denote $q_{1}$. The idea
of the smoothing is to be able to control the impact of the transition
on the inside of the domain $\left(x_{\min},x_{\max}\right)$ by means
of the parameter $\eta_{0}$. If $\eta_{0}=1$, most of the transition
happens outside of the domain, which allows to match the values of
the original function inside the domain. In contrast, if
$\eta_{0}=-1$, most of the transition will happen inside the domain.
This behaviour can seem attractive at first, however, it will give
rise to issues if the function needs to take a specific shape inside
the domain, a local volatility for instance. In both cases, we have
$q_{0}\left(x_{\max}\right)\approx x_{\max}$ and $q_{0}\left(x_{\min}\right)\approx x_{\min}$.
Finally, we reach a trade-off when $\eta=0$. This is the value we
pick in our implementation. The new volatility is then
\[
\bar{\sig}\left(q_{0}\left(x\right),q_{1}\left(y\right),t\right)\,.
\]
An example is shown in Figure \ref{fig:Brunick-Shreve-volatility-slice}.

\section{Algorithms}
\label{app:algo}

\begin{algorithm}[h]
	\begin{algor}
		\item [{set}] $\locM\left(x,\cdot,T_{1}\right)=\sigma_{\text{LV}}\left(x,T_{1}\right), \forall x $
		\selectlanguage{british}%
		\item [{for}] ( $i=1\,;\,i\leq N_{\text{Mat}}\,;\,i{++}$) 
		
		\begin{algor}
			\item [{while}] $\bar{e}>10^{-10}$ and $\left\Vert \nabla\bar{e}\right\Vert >10^{-8}$
			
			\selectlanguage{english}%
			\begin{algor}
				\item [{{*}}] \textbf{solve} forward PIDE (\ref{eq:Volettera-Type-PIDE})
				on $\left[T_{i-1},T_{i}\right]$ (with $T_0=0$)
				\item [{{*}}] \textbf{compute} model implied vol $\Sigma^{\text{Model}}$
				for 
				$T_{i}$ from computed up-and-out call prices $C\left(K,B_{\max},T_{i}\right)$
				\item [{{*}}] \textbf{compute} model foreign no-touch price $\mathrm{FNT}^{\text{Model}}$
				for maturity $T_{i}$ from the computed up-and-out call prices $C\left(0,B,T_{i}\right)/S_{0}$
				\item [{{*}}] \textbf{compute} the objective function 
				\[
				\bar{e}\left(\Lambda_{i}\right)=e\left(\Lambda_{i}\right)\left(1+\mathcal{P}\left(\Lambda_{i}\right)\right)
				\]
				as in (\ref{eq:merit_Brunick})
				\item [{{*}}] \textbf{compute }the objective function gradient $\nabla\bar{e}\left(\Lambda_{i}\right)$
				using the forward PIDE solution for $\nabla C\left(\Lambda_{i}\right)$
				\item [{{*}}] \textbf{update} volatility surface points $\Lambda_{i}$
				with the L-BFGS-B algorithm
			\end{algor}
			\item [{endwhile}]~
			\item [{{*}}] \textbf{set} $\locM\left(x,y,T_{i+1}\right)=\locM\left(x,y,T_{i}\right), \forall\left(x,y\right)$
		\end{algor}
		\selectlanguage{english}%
		\item [{endfor}]~
	\end{algor}
	\selectlanguage{british}%
	\caption{\selectlanguage{english}%
		Calibration of the local maximum volatility model \label{alg:BFGS_Brunick_Calibration}\selectlanguage{english}%
	}
\end{algorithm}

\begin{algorithm}[h]
	\caption{Calibration of path-dependent $\lev$ with 2D particle method\label{alg:ParticleMethod}}
	
	\begin{algor}[1]
		\item [{{*}}] $\lev\left(S,M,T_{0}=0\right)=\frac{\locM\left(S,M,0\right)}{\sqrt{v_{0}}}$
		\item [{for}]\textbf{each time point} ( $m=0\,;\,m\leq N_{T}-1\,;\,m{++}$) 
		
		\begin{algor}[1]
			\item [{{*}}] \begin{raggedright}
				\textbf{generate} $\left(Z,Z_{v}\right)_{i\leq N}$ and $\left(U_{v},U_{\max}\right)_{i\leq N}$,
				i.e. $2\times N$ independent draws from $\mathcal{N}\left(0,1\right)$
				and $2\times N$ draws from $\mathcal{U}\left(\left[0,1\right]\right)$,
				respectively\\
				
				\par\end{raggedright}
			\item [{{*}}] \textbf{evolve }the \foreignlanguage{english}{2-factor 3-state
				particle system} from $T_{m}$ to $T_{m+1}$ with the $QE-$Scheme
			($Z,Z_{v},U_{v}$) where the running maximum is computed by Brownian
			bridge ($U_{\max}$) and where $\lev\left(S,M,\left[T_{m},T_{m+1}\right[\right)=\lev\left(S,M,T_{m}\right)$
			\vspace{-1em}
			\item [{{*}}] \textbf{build} the particles $k$-d tree partitioning on
			the position values $\left(S^{i}\,,M^{i}\right)$ as described in Appendix \ref{subsec:kernel-construction}.
			\vspace{-1em}
			\item [{{*}}] \textbf{set }$T=T_{m+1}$
			\item [{for}]\textbf{each maximum level}( $k=1\,;\,k\leq N_{M}\,;\,k{++}$)\foreignlanguage{english}{ }
			
			\begin{algor}[1]
				\item [{for}]\textbf{each spot level} ( $j=1\,;\,j\leq N_{S}\,;\,j{++}$)
				
				\selectlanguage{english}%
				\begin{algor}[1]
					\item [{{*}}] \textbf{set} $K=S_{m+1,j}$; $B=M_{m+1,k}$
					\item [{{*}}] \textbf{select} using the $k$-d tree, a set of selected significant particles $I(K, B)$
					\item [{if}] ($K\leq B$)
					\item [{{*}}] \textbf{compute and update 
						\[
						\hat{p}_{N}=\frac{\frac{1}{\left|I\left(K,B\right)\right|}\sum_{i\in I\left(K,B\right)}V_{T}^{i}\delta_{N}\left(S_{T}^{i}-K,\,M_{T}^{i}-B,T\right)+2\theta\xi\epsilon}{\frac{1}{\left|I\left(K,B\right)\right|}\sum_{i\in I\left(K,B\right)}\delta_{N}\left(S_{T}^{i}-K,\,M_{T}^{i}-B,T\right)+\xi\epsilon}
						\]
					} 
					\item [{endif}]~
					\selectlanguage{british}%
					\item [{{*}}] \textbf{compute} 
					\[
					\lev_{m+1,j,k}=\frac{\locM\left(K,B,T\right)}{\sqrt{\hat{p}_{N}}}\,
					\]
					
				\end{algor}
				\item [{endfor}]~
			\end{algor}
			\item [{endfor}]~
		\end{algor}
		\item [{endfor}]~\end{algor}
\end{algorithm}

\begin{algorithm}[h]
\caption{Calibration of LSV-LVV $\left(\lev,\xi\right)$ with 2D particle method,
forward PIDE and inner iterations\label{alg:ParticleMethod-LSVLocalXi}}

\begin{algor}[1]
\item [{{*}}] $\lev\left(S,0\right)=\frac{\loc\left(S,0\right)}{\sqrt{v_{0}}}$
\item [{{*}}] $\sigma_{\text{LMV}}\left(S,M,0\right)=\loc\left(S,0\right)$
\item [{{*}}] \foreignlanguage{english}{\textbf{$N_{T}^{\text{down}}=0$}}

\item [{{*}}] \textbf{set}$ \left(a_{1},b_{1}\right)=\left(-1,\xiH \right)$
\item [{for}]\textbf{each maturity}( $n=1\,;\,n\leq N_{T}^{\text{Mat}}\,;\,n{++}$)
\selectlanguage{english}%
\begin{algor}[1]
\item [{{*}}] \textbf{set} $\left(a_{n},b_{n}\right)=\left(a_{n-1},b_{n-1}\right)$
\item [{{*}}] \textbf{find $N_{T}^{\rm up}$ }such that $T_{N_{T}^{\rm up}}=T_{n}^{\text{Mat}}$
\item [{{*}}] \textbf{set} the optimisation variable $\text{\it convergence} = \text{false}$
\item [{while}] $\text{\it convergence} \text{ is false}$

\selectlanguage{british}%
\begin{algor}[1]
\item [{for}] \textbf{each time point}( $m=N_{T}^{\text{down}}\,;\,m<N_{T}^{\rm up}\,;\,m{++}$) 

\begin{algor}[1]
\item [{{*}}] \begin{raggedright}
\textbf{generate} $\left(Z,Z_{v}\right)_{i\leq N}$ and $\left(U_{v},U_{\max}\right)_{i\leq N}$,
i.e., $2\times N$ independent draws from $\mathcal{N}\left(0,1\right)$
and $\mathcal{U}\left(\left[0,1\right]\right)$
\\

\par\end{raggedright}
\item [{{*}}] \textbf{evolve }the \foreignlanguage{english}{2-factor 3-state
particle system} of model (\ref{eq:Heston LSV model}) from $T_{m}$
to $T_{m+1}$ with the QE-Scheme, 
 where the maximum is computed by Brownian bridge ($U_{\max}$) and $\lev\left(S,\left[T_{m},T_{m+1}\right[\right)=\lev\left(S,T_{m}\right)$
\selectlanguage{english}%
\item [{{*}}] \textbf{build}, as in Appendix \ref{subsec:tree-search},
the $k$-d tree of the particles by their positions $\left(S^{i},\,M^{i}\right)$. This allows to define a set of significant particles $I\left(K,B\right)$ to use in (\ref{eq:MarkovProjSM_LocalXi}).
\selectlanguage{british}%
\item [{{*}}] \textbf{set }$T=T_{m+1}$
\item [{{*}}] \textbf{for each} $K$ on a grid, compute as in \cite{guyon2012being,CMR2016}
for a set of selected significant particles $I\left(K\right)$
\[
\lev\left(K,T\right)=\frac{\loc\left(K,T\right)\sqrt{\sum_{i\in I\left(K\right)}\delta_{N}^{S}\left(S_{T}^{i}-K,T\right) +2\theta\xi\epsilon}}{\sqrt{\sum_{i\in I\left(K\right)}V_{T}^{i}\delta_{N}^{S}\left(S_{T}^{i}-K,T\right)+\xi\epsilon}}\,,
\]
with $\delta_{N}^{S}$ a one-dimensional kernel function
\item [{{*}}] \textbf{for each} $\left(K,B\right)$ on a grid, compute
as in Section \ref{sec:LMSV calibration} for a set of selected
significant particles $I\left(K,B\right)$\label{algoline: mprojsm} as in Section \ref{sec:LMSV calibration}
\begin{equation}
\sigma_{\text{LMV}}\left(K,B,T\right)=\frac{\lev\left(K,T\right)\sqrt{\frac{1}{\left|I\left(K,B\right)\right|}\sum_{i\in I\left(K,B\right)}V_{T}^{i}\delta_{N}\left(S_{T}^{i}-K,\,M_{T}^{i}-B,T\right)+2\theta\xi\epsilon}}{\sqrt{\frac{1}{\left|I\left(K,B\right)\right|}\sum_{i\in I\left(K,B\right)}\delta_{N}\left(S_{T}^{i}-K,\,M_{T}^{i}-B,T\right)+\xi\epsilon}},\label{eq:MarkovProjSM_LocalXi}
\end{equation}
where $\delta_{N}$ is a two-dimensional kernel function
\item [{{*}}] \foreignlanguage{english}{\textbf{solve} the forward PIDE
for barriers (\ref{eq:Volettera-Type-PIDE}) by BDF2 implicit step
on $\left[T_{m},T_{m+1}\right]$ with a volatility of $\sigma_{\text{LMV}}\left(K,B,T_{m+1}\right)$\label{algoline: pide step}}
\end{algor}
\selectlanguage{english}%
\item [{endfor}]~
\item [{{*}}] \textbf{compute} model foreign no-touch price $\mathrm{FNT}^{\text{Model}}$
for maturity $T_{n}^{\text{Mat}}$ from the up-and-out call prices
$C\left(0,B,T_{n}^{\text{Mat}}\right)$ computed with the PIDE \label{algoline: compute price pide}
\item [{{*}}] \textbf{compute} the objective function 
\[
\bar{e}\left(a_{n},b_{n}\right)=\sum_{l=1}^{Q_{B}}\left|\mathrm{FNT}^{\text{Model}}\left(B_{T_{n}^{\text{Mat}},l},T_{n}^{\text{Mat}},a_{n},b_{n}\right)-\mathrm{FNT}^{\text{Market}}\left(B_{T_{n}^{\text{Mat}},l},T_{n}^{\text{Mat}}\right)\right|
\]

\item [{{*}}] \textbf{update} $\left(a_{n},b_{n}\right)$ guess with the
Nelder--Mead algorithm \cite{NelderMead1965}
\item [{{*}}] \textbf{check} difference
with the previous iteration and set $\text{\it convergence} = {\text{true}}$ if converged
\end{algor}
\item [{endwhile}]~
\end{algor}
\selectlanguage{british}%
\item [{endfor}]~\end{algor}
\end{algorithm}

\FloatBarrier


\bibliographystyle{abbrv}
\bibliography{GeneralBiblio}

\end{document}